\documentclass[fleqn,usenatbib]{rasti}

\usepackage[T1]{fontenc}

\newcommand{\cm}{cm$^{-1}$}
\usepackage{multirow}
\usepackage{geometry}
\usepackage{pdflscape}
\usepackage{color}
\usepackage{url}
\usepackage{lineno}

\usepackage{booktabs}

\usepackage{amssymb, xcolor}

\usepackage{newtxtext,newtxmath}

\newcommand{\red}[1]{{\color{red} #1}}

\usepackage[version=4]{mhchem}

\DeclareRobustCommand{\VAN}[3]{#2}
\let\VANthebibliography\thebibliography
\def\thebibliography{\DeclareRobustCommand{\VAN}[3]{##3}\VANthebibliography}

\usepackage{graphicx}   
\usepackage{amsmath}    
\title[Ariel databases WG: White paper]{Data availability and requirements relevant for the Ariel space mission and other exoplanet atmosphere applications}

\title[Data for the Ariel space mission and other exoplanet applications]{Data availability and requirements relevant for the Ariel space mission and other exoplanet atmosphere applications}

\author[Katy L. Chubb {\em et al.\/}]{
\newauthor Katy L. Chubb$^{1,2}$\footnotemark{},
S\'{e}verine Robert$^{3}$\footnotemark{},
Clara Sousa-Silva$^{4,5}$\footnotemark{},
Sergei N. Yurchenko$^{6}$\footnotemark{},
\newauthor Nicole F. Allard$^{7}$,
Vincent Boudon$^{8}$,
Jeanna Buldyreva$^{9}$,
Benjamin Bultel$^{10}$,
Athena Coustenis$^{11}$,
\newauthor Aleksandra Foltynowicz$^{12}$,
Iouli E. Gordon$^{13}$,
Robert J. Hargreaves$^{13}$,
Christiane Helling$^{14,15}$,
\newauthor Christian Hill$^{16}$,
Helgi Rafn Hrodmarsson$^{17}$,
Tijs Karman$^{18}$,
Helena Lecoq-Molinos$^{14,15,31}$,
\newauthor Alessandra Migliorini$^{19}$,
Michaël Rey$^{20}$,
Cyril Richard$^{8}$,
Ibrahim Sadiek$^{21}$,
Frédéric Schmidt$^{10}$,
\newauthor Andrei Sokolov$^{6}$,
Stefania Stefani$^{19}$,
Jonathan Tennyson$^{6}$,
Olivia Venot$^{17}$,
Sam O. M. Wright$^{6}$,
\newauthor Rosa Arenales-Lope$^{22}$,
Joanna K. Barstow$^{23}$,
Andrea Bocchieri$^{24}$,
Nathalie Carrasco$^{25}$,
\newauthor Dwaipayan Dubey$^{22}$,
Oleg Egorov$^{26}$,
Antonio García Muñoz$^{27}$,
Ehsan (Sam) Gharib-Nezhad$^{28}$,
\newauthor Leonardos Gkouvelis$^{22}$,
Fabian Grübel$^{22}$,
Patrick Gerard Joseph Irwin$^{29}$,
Antonín Knížek$^{30}$,
\newauthor David A. Lewis$^{14}$,
Matt G. Lodge$^{1}$,
Sushuang Ma$^{6}$,
Zita Martins$^{32}$,
Karan Molaverdikhani$^{22}$,
\newauthor Giuseppe Morello$^{33}$,
Andrei Nikitin$^{26}$,
Emilie Panek$^{34}$,
Miriam Rengel$^{35}$,
Giovanna Rinaldi$^{19}$,
\newauthor Jack W. Skinner$^{36,40}$,
Giovanna Tinetti$^{6}$,
Tim A. van Kempen$^{37}$,
Jingxuan Yang$^{29}$,
\newauthor Tiziano Zingales$^{38,39}$ 
\\
$^{1}$University of Bristol, School of Physics, HH Wills Physics Laboratory, Tyndall Avenue, Bristol BS8 1TL, UK\\ 
$^{2}$Centre for Exoplanet Science, University of St Andrews, North Haugh, St Andrews, KY16 9SS, UK\\ 
$^{3}$Royal Belgian Institute for Space Aeronomy (BIRA-IASB), Av. Circulaire 3, 1180 Uccle, Belgium\\ 
$^{4}$Bard College, Institute of Astrophysics and Space Sciences, 30 Campus Rd, Annandale-On-Hudson, NY 12504\\ 
$^{5}$Institute of Astrophysics and Space Sciences, Rua das Estrelas, 4150-762 Porto, Portugal\\ 
$^{6}$Department of Physics and Astronomy, University College London, London WC1E 6BT, UK\\ 
$^{7}$Observatoire de Paris, PSL Research University, 61, Avenue de l’Observatoire, F-75014 Paris, France\\ 
$^{8}$Lab. ICB, UMR 6303 CNRS / Université de Bourgogne, 9 Av. A. Savary, BP 47870, F-21078 Dijon Cedex, France\\
$^{9}$Institut UTINAM, UMR CNRS 6213, Universit\'{e} de Franche-Comt\'{e}, 16 Route de Gray, 25030 Besan\c{c}on cedex, France\\ 
$^{10}$Université Paris-Saclay, CNRS, GEOPS, 91405, Orsay, France\\ 
$^{11}$LESIA, Paris Observatory, CNRS, PSL Univ., 5 place Jules Janssen, 92190 Meudon Cedex, France\\ 
$^{12}$Umeå University, Department of Physics, 901 87 Umeå, Sweden\\ 
$^{13}$Center for Astrophysics | Harvard \& Smithsonian, 60 Garden Street, Cambridge, MA 02138, USA\\ 
$^{14}$Space Research Institute, Austrian Academy of Sciences, Schmiedlstrasse 6, 8042 Graz, Austria\\ 
$^{15}$TU Graz, Fakultät für Mathematik, Physik und Geodäsie, Petersgasse 16, A-8010 Graz, Austria\\ 
$^{16}$International Atomic Energy Agency, Wagramer Strasse 5, Vienna A-1400, Austria\\ 
$^{17}$Université Paris Cité and Univ Paris Est Creteil, CNRS, LISA, F-75013 Paris, France\\ 
$^{18}$Institute for Molecules and Materials, Radboud University, Heyendaalseweg 135, 6525 AJ Nijmegen, Netherlands\\ 
$^{19}$IAPS-INAF, Via Fosso del Cavaliere, 100, 00133, Rome, Italy\\ 
$^{20}$Université de Reims Champagne-Ardenne, UMR CNRS 7331, BP 1039, F-51687, Reims Cedex 2, France\\ 
$^{21}$Leibniz Institute for Plasma Science and Technology (INP), Felix-Hausdorff-Straße 2, 17489 Greifswald, Germany\\ 
$^{22}$Fakultät für Physik, Ludwig-Maximilians-Universität München, Scheinerstraße 1, D-81679 München, Germany\\ 
$^{23}$School of Physical Sciences, The Open University, Walton Hall, Milton Keynes, MK7 6AA, UK\\ 
$^{24}$Dipartimento di Fisica, La Sapienza Università di Roma, Piazzale Aldo Moro 5, Roma, 00185, Italy\\ 
$^{25}$University of Paris-Saclay and ENS Paris-Saclay, LATMOS, 11 bourlevard d'Alembert, 78280 Guyancourt, France\\ 
$^{26}$V.E. Zuev Institute of Atmospheric Optics SB RAS, 1, Akademician Zuev Sq., Tomsk, 634055 Russia\\ 
$^{27}$Universit\'e Paris-Saclay, Universit'e Paris Cit\'e, CEA, CNRS, AIM, 91191, Gif-sur-Yvette, France\\ 
$^{28}$Space Science and Astrobiology Division, NASA Ames Research Center, Moffett Field, CA, 94035 USA\\ 
$^{29}$University of Oxford, Atmospheric, Oceanic and Planetary Physics, Department of Physics, Parks Rd, Oxford OX1 3PU, UK\\ 
$^{30}$J. Heyrovský Institute of Physical Chemistry, Czech Academy of Sciences, Dolejškova 2155/3, 18223, Prague, Czech Republic\\ 
$^{31}$Institute of Astronomy, KU Leuven, Celestijnenlaan 200D bus 2401, 3001 Leuven\\ 
$^{32}$CQE, IMS, Department of Chemical Engineering, Instituto Superior Técnico, Universidade de Lisboa, Portugal\\ 
$^{33}$Instituto de Astrofísica de Andalucía (IAA-CSIC), Glorieta de la Astronomía s/n, 18008, Granada, Spain\\ 
$^{34}$Institut d’Astrophysique de Paris (CNRS, Sorbonne Université), 98bis Bd Arago, 75014 Paris, France\\ 
$^{35}$Max-Planck-Institut für Sonnensystemforschung, Justus-von-Liebig-Weg 3, 37077 Göttingen, Germany\\ 
$^{36}$Division of Geological and Planetary Sciences, California Institute of Technology, 1200 E. California Blvd., Pasadena, CA, 91125, USA\\ 
$^{37}$SRON Netherlands Institute for Space Research, Niels Bohrweg 4, 2333 CA, Leiden, the Netherlands\\ 
$^{38}$Dipartimento di Fisica e Astronomia, Università degli studi di Padova , Vicolo dell’Osservatorio 3, 35122 Padova, Italy\\ 
$^{39}$INAF, Osservatorio Astronomico di Padova, Vicolo dell’Osservatorio 5, 35122 Padova, Italy\\
$^{40}$Martin A. Fisher School of Physics, Brandeis University, 415 South St, Waltham, MA, 02453, USA
}

\begin{document}

\maketitle

\begin{abstract}

The goal of this white paper is to provide a snapshot of the data availability and data needs primarily for the Ariel space mission, but also for related atmospheric studies of exoplanets and cool stars. It covers the following data-related topics: molecular and atomic line lists, line profiles, computed cross-sections and opacities, collision-induced absorption and other continuum data, optical properties of aerosols and surfaces, atmospheric chemistry, UV photodissociation and photoabsorption cross-sections, and standards in the description and format of such data. 
These data aspects are discussed by addressing the following questions for each topic, based on the experience of the ``data-provider'' and ``data-user'' communities: (1) what are the types and sources of currently available data, (2) what work is currently in progress, and (3) what are the current and anticipated data needs. 
We present a GitHub platform for Ariel-related data, with the goal to provide a go-to place for both data-users and data-providers, for the users to make requests for their data needs and for the data-providers to link to their available data.
Our aim throughout the paper is to provide practical information on existing sources of data whether in databases, theoretical, or literature sources.

\end{abstract}

\addtocounter{footnote}{-1}
\footnotetext[1]{katy.chubb@bristol.ac.uk}
\footnotetext[2]{severine.robert@aeronomie.be}
\footnotetext[3]{csousasilva@bard.edu}
\footnotetext[4]{s.yurchenko@ucl.ac.uk}

\section{Introduction}

\subsection{The Ariel space mission}

The Ariel space mission is due to launch in 2029. It will measure spectra of exoplanet atmospheres between 0.5~-~7.8~$\mu$m ($\sim$~1280~-~20,000~cm$^{-1}$), with the highest resolving power (R~=~100) between 1.95~-~3.9~$\mu$m ($\sim$~2500~-~5100~cm$^{-1}$), undertaking a chemical survey of around 1000 of the planets in our galaxy~\citep{18ZiTiPi,18VeDrMi,18TiDrEc.exo}. Ariel will be used to constrain elemental abundances in hot gas giant atmospheres~\citep{23WaChTi}, as well as allowing for detections of rocky Super-Earth atmospheric signatures~\citep{22ItChEd.exo} and the observation of 
temperate exoplanet atmospheres~\citep{22EnCoGi}. Ariel has inspired various data challenges, such as in retrievals~\citep{22BaChCh} and machine learning~\citep{23ChYi,23NiWaTs}. As highlighted in \cite{tinetti2021ariel}, there is a synergy between Ariel and other exoplanet atmosphere missions; both high-resolution ground-based spectroscopic observations such as the Telescopio Nazionale Galileo (TNG)~\citep{22GuSoGi.exo}, and space-based missions such as PLATO (PLAnetary Transits and Oscillations of stars) and TESS (Transiting Exoplanets Survey Satellite)~\citep{23KaSzBo}, and JWST~\citep{22ChEdAl.exo}.

The Hubble Space Telescope (HST) has been widely used in exoplanet science; in particular the coverage of optical and UV regions, for example by the Wide Field Camera 3 WFC3/UVIS and Space Telescope Imaging Spectrograph (STIS) instruments, are important for understanding aerosol properties in exoplanet atmospheres~\citep{20WaSiSt,24FaWaMa}. The WFC3 infrared (IR) instrument has been extensively used for IR observations of hot exoplanet atmospheres, particularly for characterising H$_2$O spectral features which have been detected in abundance in such atmospheres~\citep[see, for example,][]{16IySwZe}. The wider spectral coverage and improvement in spectral resolution of both Ariel and JWST compared to HST/WFC3 allows for the improved detection and characterisation of several molecular species in the optical and IR regions~\citep{22GuSoGi.exo}. While JWST has a larger telescope than Ariel, allowing for observations of smaller cooler exoplanets using less transits~\citep{22ChEdAl.exo}, JWST was designed to make observations for many more astronomy fields than only exoplanets. Ariel, on the other hand, will observe simultaneously over a wide wavelength range; it is the first space mission fully dedicated to the study of exoplanet atmospheres and has the unique ability to probe the atmosphere of warm and hot exoplanets with a statistical approach, aiming to observe $\sim$1000 exoplanet atmospheres during its first tier~\citep{tinetti2021ariel}. Ariel is optimised to observe transiting exoplanets, and can be equally or more efficient than JWST for characterising exoplanets orbiting bright stars~\citep[see Figure 8-2 of][]{tinetti2021ariel}. Meanwhile, some of the exoplanets observed by JWST will be used by Ariel for calibration purposes in the first months of observation, and will likely be used to refine the Ariel target list~\citep{19EdMuTi}, highlighting the strong synergy between these two missions.

High-resolution spectroscopy utilises cross-correlation techniques to allow for the detection of individual species present at lower abundances than can be observed using the lower resolution observations of spectroscopic missions such as Ariel~\citep{22GuSoGi.exo}. The two types of observation together allow for a more complete understanding of an atmosphere~\citep[see, for example,][]{19BrLixx.exo}.

Overviews of potential exoplanet targets for Ariel can be found in \cite{19EdMuTi,2022AJ....164...15E}. The mission's instrument wavelength coverage is summarised in Table~\ref{tab:instruments}. Although the main spectral coverage (besides photometric bands) of Ariel is from 1.1~-~7.8~$\mu$m, we consider data which cover the wavelength region from the UV to the far infrared, because of their use in atmospheric models relevant to the Ariel mission and other exoplanet atmosphere applications, such as climate models and analysing observed spectra from complementary exoplanet atmosphere telescopes. To characterise different exoplanet atmospheres, spectra are typically modelled using radiative transfer calculations. These calculations rely on a priori knowledge, often based on assumptions, of a variety of atmospheric processes, such as: molecular absorption and emission, scattering, chemistry, atmospheric dynamics, cloud formation, and atmosphere-surface interaction. 

\begin{table*}
    \centering
    \caption{Summary of Ariel instruments and their wavelength coverage and resolving power R=$\frac{\lambda}{\Delta\lambda}$, from \citet{19EdMuTi}. AIRS is the Ariel InfraRed Spectrometer, NIRSpec the Near-InfraRed Spectrometer, and VISPhot, FGS1 and FGS2 are photometers~\citep{22SzKaSz}.}
    \label{tab:instruments}
    \begin{tabular}{lccc}
    \hline 
\rule{0pt}{3ex}Instrument & Wavelength range ($\mu$m) & Wavenumber range (cm$^{-1}$) &R=$\frac{\lambda}{\Delta\lambda}$\\
    \hline 
\rule{0pt}{3ex}VISPhot & 0.5~-~0.6 & 16,667~-~20,0000 &  \\
\rule{0pt}{3ex}FGS 1 & 0.6~-~0.81 & 12,346~-~16,667 & Photometric bands \\
\rule{0pt}{3ex}FGS 2 & 0.81~-~1.1 &  9091~-~12,346 & \\
\hline
\rule{0pt}{3ex}NIRSpec & 1.1~-~1.95 & 5128~-~9091 &  20 \\
\hline
\rule{0pt}{3ex}AIRS Ch0 & 1.95~-~3.9 & 2564~-~5128 & 100\\
\hline
\rule{0pt}{3ex}AIRS Ch1 & 3.9~-~7.8 & 1282~-~2564 &  30\\
\hline
    \end{tabular}

\end{table*}

\subsection{Atmospheric modelling and retrieval codes}\label{sec:retrieval}

A comprehensive list of exoplanetary retrieval codes (as of March 2023) can be found in \citet{23MaBa}, with updates online\footnote{\url{https://doi.org/10.5281/zenodo.7675743}}. Another detailed overview of radiative transfer and retrieval codes (as of May 2023) can be found in \cite{23ReAd}. In the Ariel science team, several radiative transfer retrieval packages are available, including ARCiS \citep{18OrMi.arcis,20MiOrCh.arcis}, NEMESIS \citep{NEMESIS}, Pyrat Bay \citep{Pyrat-Bay},  POSEIDON \citep{POSEIDON} and  TauREx3 \citep{21AlChWa}. They have been compared in the frame of a retrieval challenge for the Ariel mission \citep{22BaChCh}. Table~\ref{tab:retrieval_codes} lists a non-exhaustive selection of atmospheric modelling and retrieval codes and their respective repositories used in exoplanetary studies. 
Whilst these multiple retrieval codes all perform the same basic function -- comparing a range of typically one-dimensional, parametric radiative transfer simulations to observed data via a sampling algorithm such as MCMC or nested sampling -- they each have slightly different methods for parameterising often complex atmospheric processes. Previous retrieval comparison efforts by \citet{20BaChQu} and \citet{22BaChCh} showcase the agreement between them for simple cases where all codes are able to implement the same parameterisations. However, \citet{20BaChQu} compared a range of cloud parameterisations and showed that whilst the overall conclusions using each method were not in disagreement with each other, the different ways of representing cloud in each model meant that the sensitivity to different aspects of the cloud varied between approaches. Therefore, a multiplicity of retrieval codes is considered to be an asset for the Ariel mission, since each has been written with different priorities in mind. Some codes include more complex chemistry than others \citep[see e.g.][]{22AlChVe} and others have focused particularly on representation of cloud physics \citep[see e.g.][]{18OrMi.arcis}. Applying multiple retrieval codes to a single dataset is rapidly becoming accepted as field best-practice as it allows these different strengths to be captured and also provides a useful cross-check for cases where solutions may be model dependent \citep[see e.g.][]{18KiCuSt,20LeWaMa.UV,24Welbanks_TBS}. Regardless of any differences in parameterisations in these codes used to model exoplanet atmospheres and interpret observations, they all rely on various data inputs crucial for successful analyses of atmospheres, as will be discussed in this paper.

\begin{table*}
    \centering
    \caption{Examples of typical atmospheric retrieval and modelling codes used by the exoplanetary atmosphere community.}
    \label{tab:retrieval_codes}
    \begin{tabular}{lll}
    \hline 
\rule{0pt}{3ex} Code  & Reference & Link \\
    \hline 
    \rule{0pt}{3ex}ARCiS & \cite{20MiOrCh.arcis}  & \url{https://www.exoclouds.com} \\
    \rule{0pt}{3ex}ATMO & \cite{ATMO}  & \url{https://www.erc-atmo.eu}\\
    \rule{0pt}{3ex}Helios-r2 & \cite{20KiHeOr}  & \url{https://github.com/exoclime/Helios-r2} \\
\rule{0pt}{3ex}NASA Planetary Spectrum Generator (PSG) & \cite{18ViSmPr}  & \url{https://psg.gsfc.nasa.gov/}\\
    \rule{0pt}{3ex}NEMESIS &  \cite{NEMESIS} & \url{https://nemesiscode.github.io/nemesis.html} \\
    \rule{0pt}{3ex}NemesisPy & \cite{23YaIrBa}  & \url{https://github.com/Jingxuan97/nemesispy/}\\
    \rule{0pt}{3ex}petitRADTRANS & \cite{19MoWaBo.petitRT}  & \url{https://petitradtrans.readthedocs.io/en/latest/} \\
\rule{0pt}{3ex}PICASO & \cite{19BaRoMa}  & \url{https://natashabatalha.github.io/picaso/}\\
\rule{0pt}{3ex}PLATON &\cite{20ZhChKe} & \url{https://github.com/ideasrule/platon} \\
\rule{0pt}{3ex}POSEIDON & \cite{POSEIDON} & \url{https://github.com/MartianColonist/POSEIDON} \\
\rule{0pt}{3ex}Pyrat Bay & \cite{Pyrat-Bay}  & \url{https://pyratbay.readthedocs.io/en/latest/} \\
\rule{0pt}{3ex}TauREx3 &   \cite{21AlChWa} & \url{http://github.com/ucl-exoplanets/TauREx3_public} \\
\hline
    \end{tabular}

\end{table*}

\subsection{Paper structure}

The intention of this community white paper is to ensure the data necessary to interpret observations from exoplanet atmosphere characterisation missions such as Ariel are up to a standard required for successful analyses, and to allow comparability between analyses (for example, line lists which are complete and accurate at the relevant atmospheric temperatures for all expected species, refractive indices for all aerosol types expected in exoplanet atmospheres, etc). We provide a summary of the types of data typically used by atmospheric retrieval and modelling codes, and we identify potential weaknesses in the data and therefore motivate further research. For different data types, as described below, we will provide an overview of what data is currently: 1) available, 2) typically used, 3) being worked on, and 4) needed. Although the focus is on data required for characterising exoplanet atmospheres related to the Ariel space mission, we also include some discussion on more general data usage in the field of astronomical spectroscopy, for example for broader wavelength regions. 
We focus on a number of different data types: molecular and atomic line lists (Section~\ref{sec:linelists}), molecular and atomic line shapes 
(Section~\ref{sec:lineshapes}), computed absorption cross-sections and opacities (Section~\ref{sec:opacities}), collision induced absorption (CIA) and other continuum data (Section~\ref{sec:CIA}), data for aerosols (Section~\ref{sec:aerosols}), data for atmospheric chemistry models (Section~\ref{sec:chemistry}), and data required for UV photodissociation or photoabsorption (Section~\ref{sec:UV}). Needs for data standards, meta data,  existing rules and associated tools are discussed in Section~\ref{sec:data-standards}. 
Current databases providing such data for characterising exoplanet atmospheres are reviewed across the paper in the relevant sections. Our conclusions can be found in Section~\ref{sec:conclusion}. 

The information associated with this paper is made available via the GitHub project\footnote{\url{https://github.com/Ariel-data}}. The goal of this platform is to provide a go-to place both for the data-users and for the data-providers, for the users to inform others about their data needs and make requests, and for the data providers to link to the available or inform about their soon-to-be-available data (see Section~\ref{sec:github}).

\section{Molecular/atomic line lists}\label{sec:linelists}

Molecular and atomic line lists contain information about the frequency and strength of spectroscopic transitions, sometimes referred to as ``lines'', often accompanied by information on the associated initial and final energy states. When it comes to atomic and molecular line lists, the following factors are important: 
accuracy, which refers to the frequency or wavelength location of transition lines, and completeness, which refers to both spectral and temperature coverage; i.e. the number of transitions in the latter case. High-resolution spectra ($\frac{\lambda}{\Delta\lambda}$~$\gtrsim$~25,000) of exoplanets, typically measured from ground-based instruments, require very accurate line positions to be known, usually over a shorter wavelength range than required for observations from low- to mid-resolution space-based telescopes such as Ariel and JWST. The cross-correlation technique used to analyse observed high-resolution spectra is an efficient tool to characterise exoplanet atmospheres, complementary to observations from space telescopes. It has resulted in numerous detections in exoplanetary atmospheres of hot-Jupiters of molecules and atoms, including CO, H$_2$O, TiO, HCN, CH$_4$, NH$_3$, C$_2$H$_2$, OH, VO and CrH, as well Ca, V, Cr, Fe, Ti, Mg, He, K, Na and Li~\citep{22GuFlHe,23GaKeZh}. 

To interpret the lower-resolution ($\frac{\lambda}{\Delta\lambda}$~$\lesssim$~3000) observations, it is particularly important for overall opacity to be provided by a line list, which requires the line list to be considered complete up to the temperature of the atmosphere being modelled. This metric of completeness is related to the number of spectral line transitions between energy levels which are provided by a line list. There will be transitions between so-called hot bands, for example, which would not be present for a line list computed or based on laboratory measurements at room-temperature, but would be important for characterising the atmospheres of exoplanets at high temperatures of around 1000~K~-~2000~K. It is well known that the choice of line list can have a huge impact in interpreting high-resolution spectroscopic observations~\citep[see, for example,][]{21SeNuMo,22ReKeSn}.  
Perhaps less intuitive, but the  accuracy of  line list, especially of the line positions, can be also critical for interpretation of lower-resolution space-based observations, particularly at elevated temperatures $>$~1000~K~\citep[see, for example,][]{2022NatAs...6.1287N}.
Therefore it is important that  molecular opacitites used in the main stream atmospheric retrieval packages are initially built at the highest resolution ($\frac{\lambda}{\Delta\lambda}$ $\sim$ 100~000) before binning them down to the practical resolution ($\frac{\lambda}{\Delta\lambda}$~$\sim$ 3000 -- 15~000) \citep{20ChRoYu}. The aspects of completeness and accuracy for existing molecular line lists will be reviewed below.

\subsection{State of the art - Data  availability}

The main databases which host molecular and atomic line lists are given in Table~\ref{tab:linelists_overview}, with more details for a number of the molecular databases found in the subsections below. 
Of these molecular databases, HITRAN and GEISA provide highly accurate data with the main focus on terrestrial conditions, while the other listed databases typically have more extended coverage, for example covering up to higher temperatures, usually $>$~1000~K. 
Table~\ref{tab:IR:linelists} summarises the wavelength and temperature coverage of molecular line lists from some of the main spectral databases. An approximate temperature up to which the line list is considered complete, in terms of completeness of transition data at a given temperature, is given; either hot ($\geq$~1000~K) or else assumed to be roughly Earth-temperature ($\leq$~500~K). The latter usually applies to species in the HITRAN database, which was originally designed for terrestrial atmosphere applications. We note that most molecules in the ExoMol database are complete up to at least 3000~K, and hotter for some diatomic species. Some stable diatomic species such as H$_2$ and HD$^+$ are even considered complete up to 10,000~K. 
C$_2$H$_4$ 
is considered complete up to 700~K in both the TheoReTs~\citep{16ReDeNi} and ExoMol~\citep{jt729} databases; the same applies to HNO$_3$~\citep{jt614} in the ExoMol database, which is also complete up to 700~K. 
An overview of some molecular and atomic databases (as of September 2022) can also be found in \cite{24Rengel}; additional databases not detailed here include that of JPL Molecular Spectroscopy\footnote{\url{https://spec.jpl.nasa.gov/}} and the Cologne database for molecular spectroscopy (CDMS)\footnote{\url{https://cdms.astro.uni-koeln.de/}}. 

Atomic line list data can be obtained from various sources such as the NIST\footnote{\url{https://physics.nist.gov/PhysRefData/ASD/lines_form.html}}~\citep{NISTWebsite} and Kurucz\footnote{\url{https://lweb.cfa.harvard.edu/amp/ampdata/kurucz23/sekur.html}; \url{http://kurucz.harvard.edu/atoms.html}}~\citep{17Kurucz} databases, and the Vienna atomic line database (VALD3)\footnote{\url{http://vald.astro.uu.se}}~\citep{15RyPiKu}. 
The appendix of \cite{21GrMaKi} compares the opacities available from these three databases. Atomic partition functions are included in the Kurucz and VALD3 databases, and can also be computed by using various sources such as \cite{81Irwin}. We note that while the Kurucz database is well-used and considered accurate for atomic line lists, the majority of the Kurucz molecular line lists are more approximate than the other line list databases described here.

\begin{figure*}
     \centering
     \includegraphics[width=0.47\linewidth]{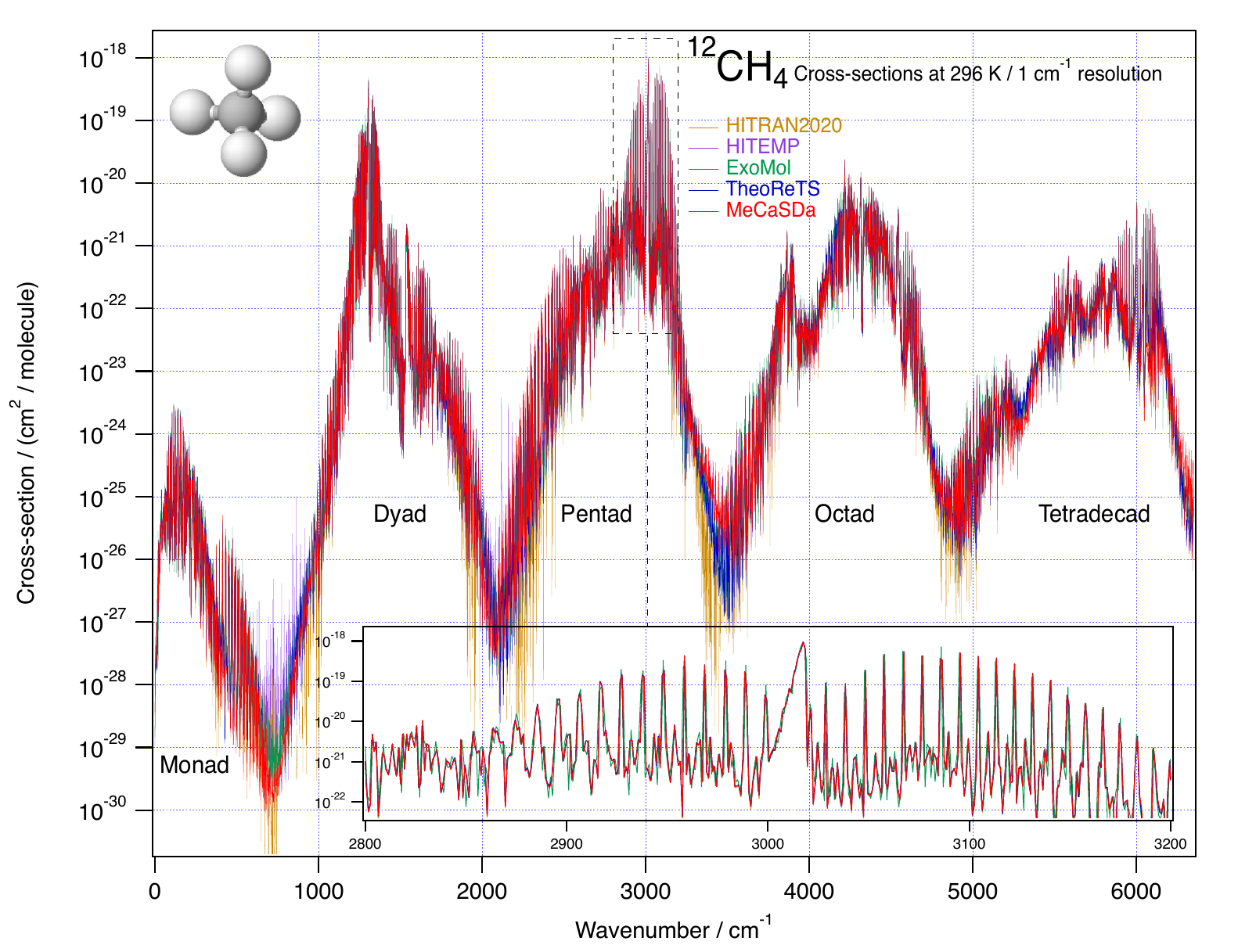}
     \includegraphics[width=0.47\linewidth]{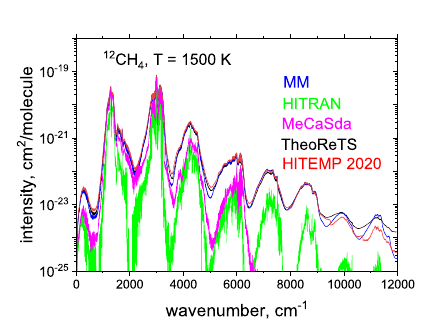}
     \caption{Comparison of $^{12}$CH$_4$ cross-sections at 296~K (left) and 1500~K (right) in the far and mid-infrared regions for different databases:
 MeCaSDa \citep{refcasda}, TheoReTs \citep{TheoReTs}, the ExoMol MM line list \citep{24YuOwKe}, HITRAN2020 \citep{2022JQSRT.27707949G} and HITEMP \citep{HargreavesAJSS2020}. }
     \label{fig:CaSDa}
\end{figure*}

Figure \ref{fig:CaSDa} (left panel) shows a comparison for methane cross-sections computed at 296~K between MeCaSDa \citep{MeCaSDa}, TheoReTs \citep{TheoReTs}, the latest ExoMol MM line list \citep{24YuOwKe}, HITRAN2020 \citep{2022JQSRT.27707949G} and HITEMP \citep{HargreavesAJSS2020}. These databases are detailed in the subsections below. It can be seen that all databases are very similar at this temperature: the differences are mostly in the  ``transparency windows'' between the strongly absorbing polyads. Transitions in these regions are extremely weak and can be purely extrapolated or even absent; thus, those regions should be considered with care. For example, HITRAN2020 lacks many weak lines, but they are present in HITEMP.

The situation is however very different at high temperatures. This is when the line list completeness  or lack of it becomes especially evident and the choice of the line list to be used in retrievals can make a huge difference. In the right display of  Fig.~\ref{fig:CaSDa} we now compare cross-sections of methane generated for $T=1500$~K. The low temperature line lists HITRAN2020 and MeCaSDa lack the essential absorption practically everywhere, with MeCaSDa being more complete at lower wavenumbers. The so-called ``hot line'' lists of TheoReTs, ExoMol MM and HITEMP all demonstrate a high level of completeness and generally agree with each across the entire IR region, but also show discrepancies in the near infrared (NIR). The latter are caused by the differences in the theoretical models and essentially reflect the lack of the underlying laboratory data used to constrain these models. Work on the analysis of the experimental spectra of methane in NIR is currently in progress.

\small
\begin{table*}
    \centering
    \caption{An overview of some line list databases used by the (exo)planetary atmosphere community.}
    \resizebox{\textwidth}{!}{
    \begin{tabular}{llll}
    \hline 
\rule{0pt}{3ex} Database & Type of data & Reference& Link\\
    \hline 
    \rule{0pt}{3ex}CaSDa&  Theoretical line lists up to high temperatures & \cite{refcasda,casda24}& \url{https://vamdc.icb.cnrs.fr}  \\
\rule{0pt}{3ex}ExoMol&  Theoretical line lists up to high temperatures & \cite{jt810}&\url{https://exomol.com}  \\
\rule{0pt}{3ex}GEISA&  Accurate line lists for terrestrial conditions & \cite{GEISA2015,21DeArSc.GEISA}  &\url{https://geisa.aeris-data.fr}\\
\rule{0pt}{3ex}HITEMP&  Line lists up to high temperatures & \cite{2010JQSRT.111.2139R}&\url{https://hitran.org/hitemp}  \\
\rule{0pt}{3ex}HITRAN&  Accurate line lists for terrestrial conditions & \cite{2022JQSRT.27707949G}&  \url{https://hitran.org}\\
\rule{0pt}{3ex}Kurucz& Semi-empirical line lists up to high temperatures  &  \cite{Kurucz_1995} &\url{http://kurucz.harvard.edu/molecules.html}\\
\rule{0pt}{3ex}MoLLIST&  Laboratory line lists up to high temperatures & \cite{MOLLIST}&\url{https://bernath.uwaterloo.ca/molecularlists.php}  \\
\rule{0pt}{3ex}NASA Ames &  Semi-empirical line lists up to high temperatures & e.g. \cite{23HuFrTa}& \url{http://huang.seti.org}\\
\rule{0pt}{3ex}NIST& Atomic line lists up to high temperatures &  \cite{NISTWebsite} &\url{https://physics.nist.gov/PhysRefData/ASD/lines_form.html}\\
\rule{0pt}{3ex}S\&MPO& Spectroscopy and Molecular Properties of Ozone  &  \cite{21TyBaMi}& \url{http://smpo.iao.ru} \\
\rule{0pt}{3ex}TheoReTs  &  Theoretical line lists up to high temperatures & \cite{TheoReTs}& \url{https://theorets.tsu.ru}  \\
\rule{0pt}{3ex}VALD3& Atomic line lists up to high temperatures &  \cite{15RyPiKu} &\url{http://vald.astro.uu.se}\\
\hline
    \end{tabular}
}
   \label{tab:linelists_overview}
\end{table*}
\normalsize

\begin{table*} 
\caption{An overview of the molecules, wavelength coverage, and temperature coverage for some of the major line lists  available from listed databases for the (exo)planetary atmosphere community. See the footnotes to the table for a key of database and wavelength range. Line lists which are considered complete up to $\geq$~1000~K are labelled with (\textit{H}), and all others are generally assumed to be complete up to $\sim$~500~K. A few exceptions are noted in the text. 
}
 \label{tab:IR:linelists}
\begin{minipage}{0.45\textwidth}
\begin{tabular}{lcc}
\toprule
Molecule& $\lambda$ range & Database \\
\midrule
\rule{0pt}{0ex} AlCl		&	 IR, Vis, UV	&	 Ex(\textit{H}),Ku(\textit{H}),Mo(\textit{H})   	\\
\rule{0pt}{0ex} AlF			&	 IR, Vis, UV	&	 Ex(\textit{H}),Ku(\textit{H}),Mo(\textit{H})   	\\
\rule{0pt}{0ex} AlH			&	 IR, Vis, UV	&	 Ex(\textit{H}),Ku(\textit{H}),Mo(\textit{H})   	\\
\rule{0pt}{0ex} AlO			&	 IR, Vis, UV	&	     Ex(\textit{H})     	\\
\rule{0pt}{0ex} AsH$_3$		&	     IR     	&	     Ex     	\\
\rule{0pt}{0ex} BeH			&	 IR, Vis, UV	&	     Ex(\textit{H})     	\\
\rule{0pt}{0ex} C$_2$		&	 IR, Vis, UV	&	     Ex(\textit{H}),Ku(\textit{H})     	\\
\rule{0pt}{0ex} C$_2$H$_2$	&	     IR     	&	   Ex(\textit{H}),HI   	\\
\rule{0pt}{0ex} C$_2$H$_4$	&	     IR     	&	Ex,HI,Th, SDa \\
\rule{0pt}{0ex} C$_2$H$_6$	&	     IR     	&	   HI   	\\
\rule{0pt}{0ex} C$_2$N$_2$	&	     IR     	&	   HI  	\\
\rule{0pt}{0ex} C$_4$H$_2$	&	     IR     	&	   HI   	\\
\rule{0pt}{0ex} C$_3$		&	     IR     	&	         Ex(\textit{H})	\\
\rule{0pt}{0ex} CaF			&	 IR, Vis, UV	&	   Ex(\textit{H}),Mo(\textit{H})   	\\
\rule{0pt}{0ex} CaH			&	 IR, Vis, UV	&	   Ex(\textit{H}),Mo(\textit{H})   	\\
\rule{0pt}{0ex} CaO			&	 IR, Vis, UV	&	     Ex(\textit{H})    	\\
\rule{0pt}{0ex} CaOH		&	   IR,Vis   	&	     Ex(\textit{H})    	\\
\rule{0pt}{0ex} CF$_4$		&	     IR     	&	   HI,SDa,Th    	\\
\rule{0pt}{0ex} CH			&	 IR, Vis, UV	&	   Ex(\textit{H}),Ku(\textit{H}),Mo(\textit{H})    	\\
\rule{0pt}{0ex} CH$^+$			&	 IR, Vis, UV	&	     Ex(\textit{H})     	\\
\rule{0pt}{0ex} CH$_2$		&	     IR     	&	     Th(\textit{H})    			\\
\rule{0pt}{0ex} CH$_3$		&	     IR     	&	   Ex(\textit{H}),Th(\textit{H})   	\\
\rule{0pt}{0ex} CH$_3$Br	&	     IR     	&	    HI  	\\
\rule{0pt}{0ex} CH$_3$Cl	&	     IR     	&	    Ex(\textit{H}),HI,SDa,Th  	\\
\rule{0pt}{0ex} CH$_3$CN	&	     IR     	&	    HI  	\\
\rule{0pt}{0ex} CH$_3$F		&	     IR     	&	   Ex,HI,Th	\\
\rule{0pt}{0ex} CH$_3$I	    &	     IR     	&	    HI  	\\
\rule{0pt}{0ex} CH$_3$OH	&	     IR     	&	    HI  	\\
\rule{0pt}{0ex} CH$_4$		&	   IR, NIR  	&	 Ex(\textit{H}),HI,HT(\textit{H}),Th(\textit{H}),SDa(\textit{H}) \rule{0pt}{0ex} \\
\rule{0pt}{0ex} ClO  	&	     IR     	&	    HI  	\\
\rule{0pt}{0ex} ClONO$_2$  	&	     IR     	&	    HI  	\\
\rule{0pt}{0ex} CN			&	 IR, Vis, UV	&	   Ex(\textit{H}),Ku(\textit{H}),Mo(\textit{H})    	\\
\rule{0pt}{0ex} CO			&	     IR     	&	   Ex(\textit{H}),HI,HT(\textit{H}) 	\\
\rule{0pt}{0ex} CO$_2$		&	     IR     	&	  Am,Ex(\textit{H}),HI,HT(\textit{H})	\\
\rule{0pt}{0ex} COCl$_2$  	&	     IR     	&	    HI  	\\
\rule{0pt}{0ex} COF$_2$  	&	     IR     	&	    HI  	\\
\rule{0pt}{0ex} CP			&	     IR     	&	     Ex(\textit{H})     	\\
\rule{0pt}{0ex} CrH			&	     NIR    	&	   Ex(\textit{H}),Mo(\textit{H})    	\\
\rule{0pt}{0ex} CS			&	     IR     	&	    Ex(\textit{H}),HI  	\\
\rule{0pt}{0ex} CS$_2$		&	     IR     	&	    Am,HI  			\\
\rule{0pt}{0ex} FeH			&	     NIR    	&	   Ex(\textit{H}),Mo(\textit{H})   	\\
\rule{0pt}{0ex} GeH$_4$		&	     IR     	&	    HI,SDa,Th     			\\
\rule{0pt}{0ex} H$_2$		&	 IR, Vis, UV	&	   Ex(\textit{H}),HI(\textit{H})   	\\
\rule{0pt}{0ex} H$_2$CO		&	     IR    		&	  Ex(\textit{H}),HI,Th  	\\
\rule{0pt}{0ex} H$_2$CS		&	     IR    		&	     Ex(\textit{H})     	\\
\rule{0pt}{0ex} H$_2$O		&	     IR    		&	    Ex(\textit{H}),HI,HT(\textit{H})  \\
\rule{0pt}{0ex} H$_2$O$_2$	&	     IR   	 	&	    Ex(\textit{H}),HI   	\\
\rule{0pt}{0ex} H$_2$S		&	     IR   	 	&	    Ex(\textit{H}),HI   	\\
\rule{0pt}{0ex} H$_3^+$		&	     IR    		&	     Ex(\textit{H})    	\\
\rule{0pt}{0ex} H$_3$O$^+$	&	     IR     	&	     Ex(\textit{H})     	\\
\rule{0pt}{0ex} HBO			&	     IR     	&	   Ex(\textit{H})  \\
\rule{0pt}{0ex} HBr			&	     IR     	&	   Ex(\textit{H}),HI(\textit{H})  \\
\rule{0pt}{0ex} HCl			&	     IR    		&	    Ex(\textit{H}),HI(\textit{H})   	\\
\rule{0pt}{0ex} HCN			&	     IR    		&	    Ex(\textit{H}),HI   	\\
\rule{0pt}{0ex} HC$_3$N			&	     IR    		&	    HI   	\\
\rule{0pt}{0ex} HCOOH			&	     IR    		&	    HI   	\\
\rule{0pt}{0ex} HDO			&	     IR    		&	    Ex(\textit{H}),HI   	\\
\rule{0pt}{0ex} HD$^+$		&	 IR	&	   Ex(\textit{H})   	\\
\rule{0pt}{0ex} HeH$+$		&	 IR	&	   Ex(\textit{H})   	\\
\rule{0pt}{0ex} HF			&	     IR    		&	    Ex(\textit{H}),HI(\textit{H})   	\\
\rule{0pt}{0ex} HI			&	     IR    		&	    Ex(\textit{H}),HI(\textit{H})   	\\
\rule{0pt}{0ex} HNO$_3$		&	     IR    		&	    Ex,HI   	\\
  \hline
\end{tabular}

\end{minipage} \hfill
\begin{minipage}{0.45\textwidth}
\begin{tabular}{@{\vline}lcc} 
\toprule
\rule{0pt}{0ex} Molecule& $\lambda$ range & Database \\
\midrule
\rule{0pt}{0ex} HO$_2$		&	     IR    		&	     HI     		\\
\rule{0pt}{0ex} HOBr		&	     IR    		&	     HI     		\\
\rule{0pt}{0ex} HOCl		&	     IR    		&	     HI     		\\
\rule{0pt}{0ex} KCl			&	      IR     		&	    Ex(\textit{H})  	\\
\rule{0pt}{0ex} KF			&	           		&	    Ex(\textit{H}),Mo(\textit{H})   	\\
\rule{0pt}{0ex} KOH			&	     IR    		&	     Ex(\textit{H})     	\\
\rule{0pt}{0ex} LaO		&	          Vis,UV 		&	    Ex(\textit{H}),Mo(\textit{H})   	\\
\rule{0pt}{0ex} LiCl		&	      IR     		&	    Ex(\textit{H}),Mo(\textit{H})   	\\
\rule{0pt}{0ex} LiF		&	      IR     		&	    Ex(\textit{H}),Mo(\textit{H})   	\\
\rule{0pt}{0ex} LiH			&	        IR   		&	    Ex(\textit{H})   	\\
\rule{0pt}{0ex} LiH$^+$		&	        IR   		&	    Ex(\textit{H})   	\\
\rule{0pt}{0ex} LiOH			&	        IR   		&	    Ex(\textit{H})   	\\
\rule{0pt}{0ex} MgF 		&	 IR, Vis, U 	&	 Ex(\textit{H}),Mo(\textit{H}) 	\\
\rule{0pt}{0ex} MgH 		&	 IR, Vis, U 	&	 Ex(\textit{H}),Mo(\textit{H}) 	\\
\rule{0pt}{0ex} MgO 		&	 IR, Vis, U 	&	 Ex(\textit{H}) 		\\
\rule{0pt}{0ex} N$_2$ 		&	 IR,UV 			&	 Ex(\textit{H}),HI 		\\
\rule{0pt}{0ex} N$_2$O 		&	 IR 			&	 Am,Ex(\textit{H}),HI,HT(\textit{H})  	\\
\rule{0pt}{0ex} NaF 		&	 				&	 Ex(\textit{H}),Mo(\textit{H})  		\\
\rule{0pt}{0ex} NaH 		&	 IR, Vis, UV 	&	 Ex(\textit{H})  		\\
\rule{0pt}{0ex} NaCl		&	      IR     		&	    Ex(\textit{H})   	\\
\rule{0pt}{0ex} NaO 		&	 IR, Vis, UV 	&	 Ex(\textit{H})  		\\
\rule{0pt}{0ex} NaOH 		&	 IR 			&	 Ex(\textit{H})  		\\
\rule{0pt}{0ex} NF$_3$ 		&	 IR 			&	 HI,Th 			\\
\rule{0pt}{0ex} NH 			&	 IR, Vis, U 	&	 Ex(\textit{H}),Ku(\textit{H}),Mo(\textit{H}) 		\\
\rule{0pt}{0ex} NH$_3$ 		&	 IR 			&	 Ex(\textit{H}),HI 	\\
\rule{0pt}{0ex} NO 			&	 IR, Vis, U 	&	 Ex(\textit{H}),HI,HT(\textit{H}) \\
\rule{0pt}{0ex} NO$^+$ 			&	 IR 	&	 HI \\
\rule{0pt}{0ex} NO$_2$ 		&	 IR 			&	 HI,HT(\textit{H}) 			\\
\rule{0pt}{0ex} NS 			&	 IR 			&	 Ex(\textit{H}) 		\\
\rule{0pt}{0ex} O$_2$ 		&	 IR, Vis, UV 	&	 HI  			\\
\rule{0pt}{0ex} O$_3$ 		&	 IR 			&	 HI,SM  		\\
\rule{0pt}{0ex} OCS 		&	 IR 			&	 Am,Ex(\textit{H}),HI \\
\rule{0pt}{0ex} OH 			&	 IR, Vis, UV 	&	 Ex(\textit{H}),HI,HT(\textit{H}),Mo(\textit{H})  	\\
\rule{0pt}{0ex} OH$^+$ 		&	 IR, Vis, UV 	&	 Ex(\textit{H}),Mo(\textit{H})  	\\
\rule{0pt}{0ex} PH 			&	 IR, Vis, UV 	&	 Ex(\textit{H})  		\\
\rule{0pt}{0ex} PH$_3$ 		&	 IR 			&	 Ex(\textit{H}),HI,Th(\textit{H})  	\\
\rule{0pt}{0ex} PN 			&	 IR 			&	 Ex(\textit{H})  		\\
\rule{0pt}{0ex} PO 			&	 IR 			&	 Ex(\textit{H})  		\\
\rule{0pt}{0ex} PS 			&	 IR 			&	 Ex(\textit{H})  		\\
\rule{0pt}{0ex} S$_2$ 		&	 UV 			&	 HI  	\\
\rule{0pt}{0ex} ScH 		&	 IR, Vis, UV 	&	 Ex(\textit{H})  		\\
\rule{0pt}{0ex} SF$_6$ 	&	 IR 			&	 HI,SDa,Th   			\\
\rule{0pt}{0ex} SH 			&	 IR, Vis, UV 	&	 Ex(\textit{H}),Mo(\textit{H})  	\\
\rule{0pt}{0ex} SiF$_4$ 	&	 IR 			&	 SDa   			\\
\rule{0pt}{0ex} SiH 		&	 IR, Vis, UV 	&	 Ex(\textit{H}),Ku(\textit{H})  		\\
\rule{0pt}{0ex} SiH$_2$ 	&	 IR 			&	 Ex(\textit{H})  		\\
\rule{0pt}{0ex} SiH$_4$ 	&	 IR 			&	 Ex(\textit{H}),Th,SDa   	\\
\rule{0pt}{0ex} SiO 		&	 IR, Vis, UV 	&	 Ex(\textit{H}),Ku(\textit{H})  		\\
\rule{0pt}{0ex} SiO$_2$ 	&	 IR 			&	 Ex(\textit{H})  		\\
\rule{0pt}{0ex} SiH 		&	 IR, Vis, UV 	&	 Ex(\textit{H})  		\\
\rule{0pt}{0ex} SiS 		&	 IR 			&	 Ex(\textit{H})  		\\
\rule{0pt}{0ex} SO 			&	 IR, Vis, UV 	&	 Ex(\textit{H}),HI  	\\
\rule{0pt}{0ex} SO$_2$ 		&	 IR 			&	 Am,Ex(\textit{H}),HI \\
\rule{0pt}{0ex} SO$_3$ 		&	 IR 			&	 Ex(\textit{H}),HI  	\\
\rule{0pt}{0ex} TiH 		&	 Vis			&	 Ex(\textit{H}),Mo(\textit{H})  				\\
\rule{0pt}{0ex} TiO 		&	 IR, Vis, UV 	&	 Ex(\textit{H}),Mo(\textit{H}),Ku(\textit{H}) \\
\rule{0pt}{0ex} VO 			&	 IR, Vis, UV 	&	 Ex(\textit{H})  		\\
\rule{0pt}{0ex} YO 			&	 IR, Vis, UV 	&	 Ex(\textit{H})  \\
\rule{0pt}{0ex} ZnO 		&	 IR, Vis, UV 	&	 Ex(\textit{H})  	\\
\rule{0pt}{0ex} ZrO 		&	 IR, Vis, UV 	&	 Ex(\textit{H}) 		\\
\\
  \hline
\end{tabular}

\end{minipage}

\textit{Databases: Am=NASA Ames; Ex=ExoMol; HI=HITRAN; HT=HITEMP; Mo=MoLLIST; Ku=Kurucz; SDa=TFSiCaSDa, MeCaSDa, TFMeCaSDa, GeCaSDa, SiCaSDa, ECaSDa; SM=S\&MPO; Th=TheoReTS; see Table~\ref{tab:linelists_overview}.} \\
\textit{$\lambda$ ranges: UV~=~0.1~-~0.4~$\mu$m; Vis~=~0.4~-~0.8~$\mu$m; IR~$\geq$~0.8~$\mu$m.} \\
\end{table*}

\subsubsection{Descriptions of molecular databases}

\paragraph{CaSDa}\label{sec:casda}

The CaSDa (Calculated Spectroscopic Databases) databases\footnote{\url{https://vamdc.icb.cnrs.fr}}~\citep[see][]{refcasda,casda24} contain synthetic line lists resulting from fits of the effective Hamiltonian and dipole moment parameters using assigned experimental laboratory spectra. The models used for the calculations are based on the tensorial formalism developed in the Dijon group, as explained in \cite{reftds}. Currently, there are 10 CaSDa databases for 10 different molecules, which amount to 39 isotopologues in total. A new database, SiCaSDa, has been released very recently for three isotopologues of SiH$_4$ (a molecule already observed in the interstellar medium). All these databases are listed in Table \ref{tab:casda}. For the present topic of exoplanetary spectroscopy, the most relevant ones are, of course, CH$_4$, C$_2$H$_4$ and GeH$_4$, which are present in the atmospheres of the giant planets of the Solar System. CF$_4$ and SiF$_4$ are naturally present in Earth's atmosphere due to volcanic emissions; SiF$_4$ has been suggested as a possible species on Io by \cite{refio} and thus may be relevant in the case of potential volcanic exoplanets; CH$_3$Cl is present in comets \citep{hardy2023high} and thus could be of interest for exo-comets \citep{23JaPaRi}. It was also suggested and explored as a biomarker  in exoplanetary atmospheres \citep{13SeBaHu.CH3Cl}. In the near future, 
the intent is for new data (new bands and new isotopologues, like for instance CH$_3$D in MeCaSDa) to be added to these databases~\citep{casda24}, and 
for further ones to be developed, especially for some molecules of astrophysical interest, like C$_3$H$_6$O$_3$ (trioxane that may be relevant for comets, see \cite{richard2022high}).

These databases contain calculated line lists for all spectral regions of these molecules that could be assigned in experimental laboratory spectra and fitted line-by-line using effective Hamiltonians and dipole moments. The databases contain extra information, such as energy levels with all quantum numbers and eigenvectors. When available, different isotopologues data are included in each database. The given accuracy of line positions and intensities is equal to the standard deviation of the fit; extrapolation is limited to only a few rotational quantum number values outside the existing assignment range. The data output is in the 160-character HITRAN2004 format \citep{rothman2005hitran}. CaSDa  use some specific notations for line assignments in the case of CH$_3$Cl and C$_2$H$_4$ since they are based on use of a tensorial formalism similar to that of CH$_4$. It is also possible to compute absorption cross-sections.

\begin{table*}
    \centering
    \caption{Calculated molecular absorption lines in the CaSDa databases.}
    \label{tab:casda}
    \begin{tabular}{llccrc}
    \hline
    HITRAN ID & Molecule & DB name & Isotopologues & \multicolumn{1}{c}{Total nb. of lines} & Relevant for Ariel \\
    \hline 
        06 & CH$_4$     & MeCaSDa   & 3 & 12,988,898 & * \\
        24 & CH$_3$Cl   & ChMeCaSDa & 2 & 12,152     \\
        30 & SF$_6$     & SHeCaSDa  & 4 & 491,500    \\
        38 & C$_2$H$_4$ & ECaSDa    & 1 & 96,397 & *     \\
        42 & CF$_4$     & TFMeCaSDa & 1 & 258,208 & *    \\
        52 & GeH$_4$    & GeCaSDa   & 5 & 60,878 & *     \\
        -- & RuO$_4$    & RuCaSDa   & 9 & 30,205     \\
        -- & SiF$_4$    & TFSiCaSDa & 3 & 210,401 & *     \\
        -- & UF$_6$     & UHeCaSDa  & 8 & 110,129    \\
        -- & SiH$_4$    & SiCaSDa   & 3 & 20,090 & * \\
    \hline
    \end{tabular}
\end{table*}

\paragraph{ExoMol}
\label{sec:ExoMol}

ExoMol\footnote{\url{www.exomol.com}} is a provider of the molecular and since 2023 also atomic data necessary for modelling spectroscopy of exoplanets~\citep{jt939}. The main data products are line lists, with the emphasis on the molecules important for exoplanetary and stellar atmospheres, but also for spectroscopy of planets, comets  as well as industrial applications~\citep[see][]{21TeYu}. As of today, the ExoMol database contains line lists for 92 molecules and 245 isotopologues, see Table~\ref{tab:exomol} for the current snapshot of the molecular line lists available in ExoMol.  Up to the ExoMol 2020 release \citep{jt810}, the emphasis had been on the completeness of the data in terms of the temperature and wavelength coverage especially aimed at lower to medium resolution exoplanetary studies, usually involving transit spectroscopy, see Table~\ref{tab:IR:linelists}. ExoMol line lists are generally constructed to be accurate for resolving power $R\leq 10000$ \citep{jt939}, i.e. appropriate for the Ariel and JWST missions. Higher temperatures applications with the atmospheres of hot Jupiters studied by HST instruments being a typical application meant that the line lists had an emphasis on completeness~\citep{jt572}. Since 2020, the focus of the ExoMol database has shifted to high resolution (HR) applications, i.e. ground-based HR cross-correlation applications. The ExoMol line lists are currently  being upgraded to the quality of the HITRAN line lists, where the underlying high-resolution data is available. Other conceptual shifts in the ExoMol data structure are the introduction of uncertainties for the line positions to enable the data users to select the lines according with their application needs; provision of UV photoabsorption and photodissociation cross-sections \citep{22PeTeYu,23TePeZh} (see Section~\ref{sec:UV}); broadening due to predissociation~\citep{23TePeZh,jt922}.  

The ExoMol line lists are written as bzip2-compressed ascii files using the two file structure: a States file (.states) and a Transition file (.trans). The States file provides a full description of the (rovibronic) states, including the energy (\cm), total degeneracy, the total angular momentum, energy uncertainty (\cm), lifetimes, Lande-g factors as well as the corresponding quantum numbers, with each entry identified by an integer number (State ID). The Transition files provide the Einstein coefficients in the following compact form: $
i  \;\;  j \;\; A_{ij} $,
where $i$ and $j$ are the upper and lower state IDs, respectively, and $A_{ij}$ is a corresponding Einstein A coefficient (s$^{-1}$). The quantum numbers in the States file, when available, correspond to the standard spectroscopic convention for a given system, but can also include alternative conventions. This is strongly dependent on how the line lists have been produced.

\begin{table*} 
\centering
\caption{An overview of the line lists provided by ExoMol, \url{www.exomol.com}}
 \label{tab:exomol}
\begin{tabular}{llllll}
\toprule
Molecule& Reference & Molecule& Reference &  Molecule& Reference   \\
\midrule
 \ce{AlCl}         &   \cite{jt887}          &  \ce{H2CS}   &   \cite{jt886}         &  NH$_3$         &  \cite{jt771}        \\
AlF                &  \cite{18YoBexx.AlF}    &  \ce{H3+}    &   \cite{jt890}         &  $^{15}$NH$_3$  &  \cite{jt952}        \\
 \ce{AlH}          &   \cite{jt922}          &  \ce{H3O+}   &   \cite{jt805}         &  NO             &  \cite{jt686}        \\
 AlO               &   \cite{jt598}          & HBO          &  \cite{24LiQiLi.HBO}   &  \ce{NO}        &  \cite{jt831}        \\
AsH$_3$            &  \cite{jt751}           & HBr          &  \cite{15CoHaxx.HF}    &  NS             &  \cite{jt725}        \\
BeH                &  \cite{jt722}           &  HCCH        &   \cite{jt780}         &  \ce{OCS}       &  \cite{jt943}        \\
 C$_2$             &   \cite{jt736}          & HCl          &  \cite{13LiGoHa.HCl}   & OH              & \cite{16BrBeWe.OH}   \\
C$_2$              &  \cite{13BrBeSc.C2}     &  HCN/HNC     &   \cite{jt570}         & OH$^+$          & \cite{18HoBiBe.OH+}  \\
 C$_2$H$_4$        &   \cite{jt729}          & HD           &  \cite{19AmDiJo}       & P$_2$H$_2$      & \cite{19OwYuxx.P2H2} \\
 \ce{C3}           &   \cite{jt961}           & HD$^+$       &  \cite{19AmDiJo}       & PF$_3$          & \cite{jt752}         \\
CaF                &  \cite{18HoBexx.CaF}    &  HDO         &   \cite{jtHDO}         &  PH             &  \cite{jt765}        \\
 CaH               &   \cite{jt529}          & HeH$^+$      &  \cite{19AmDiJo}       &  PH$_3$         &  \cite{jt592}        \\
 \ce{CaH}          &   \cite{jt858}          & HF           &  \cite{15CoHaxx.HF}    &  PN             &  \cite{jt954}        \\
 CaO               &   \cite{jt618}          &  HNO$_3$     &   \cite{jt614}         &  PO             &  \cite{jt703}        \\
 \ce{CaOH}         &   \cite{jt858}          &  HS          &   \cite{jt725}         &  PS             &  \cite{jt703}        \\
CH                 &  \cite{14MaPlVa.CH}     &  KCl         &   \cite{jt583}         & ScH             & \cite{jt599}         \\
CH$_3$             &  \cite{19AdYaYu.CH3}    & KF           &  \cite{16FrBeBr.NaF}   &  SH             &  \cite{jt776}        \\
 CH$_3$Cl          &   \cite{jt733}          &  \ce{KOH}    &   \cite{jt820}         &  SiH            &  \cite{jt711}        \\
CH$_3$F            &  \cite{19OwYaKu.CH3F}   & LaO          &  \cite{22BeDoLi.LaO}   & SiH$_2$         & \cite{jt779}         \\
 CH$_4$            &   \cite{jt564}          & LiCl         &  \cite{18BiBexx.LiF}   &  SiH$_4$        &  \cite{jt701}        \\
 \ce{CH+}          &   \cite{jt913}          & LiF          &  \cite{18BiBexx.LiF}   &  \ce{SiN}       &  \cite{jt858}        \\
 \ce{CH4}          &   \cite{jt926}          & LiH          &  \cite{jt506}          &  \ce{SiO}       &  \cite{jt847}        \\
CN                 &  \cite{21SyMcXx.CN}     & LiH$^+$      &  \cite{jt506}          & SiO             & \cite{11Kurucz.db}   \\
CO                 &   \cite{15LiGoRo.CO}    &  \ce{LiOH}   &   \cite{jt905}         &  \ce{SiO2}      &  \cite{jt797}        \\
 \ce{CO2}          &   \cite{jt804}          & MgF          &  \cite{17HoBexx.MgF}   &  SiS            &  \cite{jt724}        \\
CP                 &  \cite{14RaBrWe.CP}     &  \ce{MgH}    &   \cite{jt858}         &  \ce{SO}        &  \cite{jt924}        \\
CrH                &  \cite{06ChMeRi.CrH}    & MgH          &  \cite{13GhShBe}   &  SO$_2$         &  \cite{jt635}        \\
 CS                &   \cite{jt615}          &  MgO         &   \cite{jt759}         &  SO$_3$         &  \cite{jt641}        \\
FeH                &  \cite{10WEReSe.FeH}    & N$_2$        &  \cite{18WeCaCr.N2}    & TiH             & \cite{05BuDuBa.TiH}  \\
H$_2$              &  \cite{19RoAbCz.H2}     &  \ce{N2O}    &   \cite{jt951}         &  TiO            &  \cite{jt760}        \\
 H$_2$$^{16}$O     &   \cite{jt734}          &  NaCl        &   \cite{jt583}         &  VO             &  \cite{jt644}        \\
 H$_2$$^{17,18}$O  &   \cite{jt665}          & NaF          &  \cite{16FrBeBr.NaF}   &  \ce{VO}        &  \cite{jt923}        \\
 H$_2$CO           &   \cite{jt597}          &  NaH         &   \cite{jt605}         &  \ce{YO}        &  \cite{jt921}        \\
 H$_2$O$_2$        &   \cite{jt638}          &  \ce{NaO}    &   \cite{jt854}         & YO              & \cite{jt774}         \\
 H$_2$S            &   \cite{jt640}          &  \ce{NaOH}   &   \cite{jt820}         & ZrO             & \cite{23PeTaMc.ZrO}  \\
 H$_3^+$           &   \cite{jt666}          & NH           &  \cite{24PeMcxx.NH}    &                 &                      \\
\hline
    \end{tabular}
\end{table*}

\begin{figure*}
    \centering
    \includegraphics[width=0.8\linewidth]{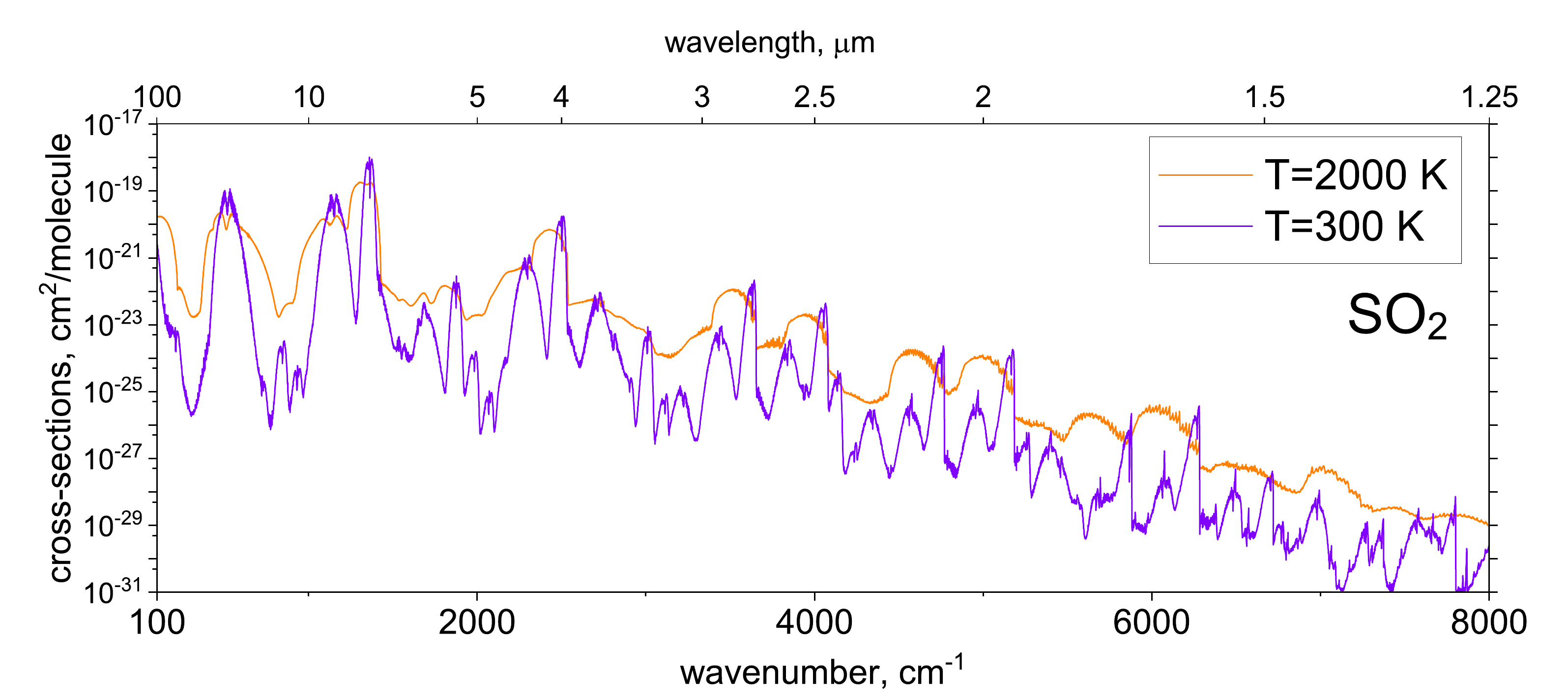}
    \caption{ExoMol Spectral Atlas~\citep{18TeYuxx.exo}: Cross-sections of SO$_2$ computed with the ExoMol line ExoAmes \citep{jt635}.}
    \label{fig:so2}
\end{figure*}

As part of the ExoMol services, a number of useful tools are provided, including an online cross-sections app, some efficient programs ExoCross and PyExoCross  to turn huge line lists into compact temperature and pressure dependent cross-sections \citep{jt708,jt914}, molecular radiative lifetimes \citep{jt904}, molecular specific heats \citep{jt899} and molecular opacities in  four different formats appropriate for a variety of  atmospheric retrieval codes \citep{20ChRoYu}; details of the ExoMolOP opacity database are given in Section~\ref{sec:opacities}. A detailed  Atlas of molecular spectra computed using the ExoMol line lists can be found in \citet{18TeYuxx.exo}, see Fig.~\ref{fig:so2} as an example, where the ExoAmes \citep{jt635} cross-sections of SO$_2$ are shown. 

The majority of the line lists in the ExoMol database have been produced by the ExoMol group using a combination of quantum mechanical calculations based on high-level spectroscopic models (i.e. potential energy surfaces, spin-orbit and other coupling etc.), tuned by fitting to the experimental data and the so-called post-production MARVELisation~\citep{2007JMoSp.245..115F}, where some of the computed energies are replaced by experimental values (see Section~\ref{sec:MARVEL}). Apart from these ExoMol line lists (covering some 70 molecules), the ExoMol database also hosts line lists from other data providers, including a large number of diatomics~\citep{20WaTeYu} from the MoLLIST (Molecular Line Lists, Intensities and Spectra) project~\citep{MOLLIST}, some of the high-temperature species from HITRAN (HF, HCl, HBr)~\citep{13LiGoHa.HCl}, and  CO from \cite{15LiGoRo.CO}. The HITEMP line list for NO~\citep{19HaGoRo.ll} is based on the ExoMol NOname line list~\citep{jt686}, combined with other data.

\paragraph{HITRAN and HITEMP}

The HITRAN molecular spectroscopic database has been a go-to source of molecular data for 50 years \citep{2021NatRP...3..302R}. HITRAN, an acronym of high-resolution transmission, was primarily established to model the radiative transfer through the Earth’s atmosphere and consequently included line-by-line parameters to model the dominant terrestrial absorbers. Over the years, as observations, experiments, and analytical techniques have improved, the HITRAN database has been regularly updated and expanded to accommodate state-of-the-art molecular parameters from the microwave through to the near UV spectral range. Papers describing quadrennial editions of the database typically receive thousands of citations. HITRAN2020 \citep{2022JQSRT.27707949G} – the most recent edition of the database – features a compilation of line-by-line parameters, both experimentally-derived and calculated, for 55 molecules and is freely available from HITRAN\textit{online}\footnote{\url{www.hitran.org}}. Table 1 of \citet{2022JQSRT.27707949G} provides a summary of the spectral coverage and the number of lines included in HITRAN for each isotopologue of every molecule. Each transition in HITRAN includes the line position, intensity, lower-state energy, line shape parameters, and assignment information that are necessary for calculating the absorption of a transition at a given temperature and concentration. HITRAN line-by-line parameters are provided at a reference temperature of 296\,K with intensities that account for the typical natural abundance on Earth. Alongside Voigt line shape parameters for air- and self-broadening (i.e., the broadening due to the terrestrial atmosphere and pure gas, respectively), recent editions of HITRAN have also accommodated parameters for high-level line shapes and broadening parameters that are applicable to planetary atmospheres other than Earth. This includes atmospheres dominated by H$_{2}$/He, CO$_{2}$ \citep{16WiGoKo,2022ApJS..262...40T}, and H$_{2}$O \citep{2019JGRD..12411580T}. A summary of the additional broadening available via the HITRAN database is provided in Table~\ref{tab:add_broadening_HITRAN}.

\begin{table}
    \centering
    \caption{Reference sources for Voigt line broadening parameters available in HITRAN, in addition to air- and self-broadening which are provided for every line. W16 refers to \protect\cite{16WiGoKo}, T19 to \protect\cite{2019JGRD..12411580T}, T22 to \protect\cite{2022ApJS..262...40T}, D18 to \protect\cite{2018JQSRT.219..360D}, L20 to \protect\cite{2020Icar..33613452L},  and self refers to self-broadening parameters~\citep[see][]{2022JQSRT.27707949G}. } 
    \label{tab:add_broadening_HITRAN}
    \begin{tabular}{clcccc}
    \hline
HITRAN        & &   \multicolumn{4}{c}{Broadening Parameters}   \\
 ID      & Molecule    &   H$_2$    &   He    & H$_2$O    &    CO$_2$   \\
\hline
1  & H$_2$O      &   $-$    &   $-$    &    Self   & $-$    \\
2  & CO$_2$      &   T22    &   T22    &    T19    & T22    \\
4  & N$_2$O      &   $-$    &   T22    &    T19    & T22    \\
5  & CO          &   T22    &   T22    &    T19    & T22    \\
6  & CH$_4$      &   $-$    &   $-$    &    T19    & $-$    \\
7  & O$_2$       &   $-$    &   $-$    &    T19    & $-$    \\
9  & SO$_2$      &   W16    &   W16    &    $-$    & D18    \\
11 & NH$_3$      &   W16    &   W16    &    T19    & W16    \\
13 & OH          &   T22    &   T22    &    $-$    & $-$    \\
14 & HF          &   W16    &   W16    &    $-$    & W16    \\
15 & HCl         &   W16    &   W16    &    $-$    & W16    \\
19 & OCS         &   T22    &   T22    &    $-$    & L20    \\
20 & H$_2$CO     &   T22    &   T22    &    $-$    & T22    \\
23 & HCN         &   T22    &   T22    &    $-$    & $-$    \\
26 & C$_2$H$_2$  &   W16    &   W16    &    $-$    & W16    \\
28 & PH$_3$      &   T22    &   T22    &    $-$    & $-$    \\
31 & H$_2$S      &   T22    &   T22    &    T19    & T22    \\
45 & H$_2$       &   Self   &   $-$    &    $-$    & $-$    \\
52 & GeH$_4$     &   T22    &   $-$    &    $-$    & $-$    \\
\hline
    \end{tabular}
\end{table}

In addition to line-by-line parameters, HITRAN also provides experimental absorption cross-sections for over 300 molecules for which no reliable quantum mechanical models exist \citep{2019JQSRT.230..172K} (see Section~\ref{sec:lab_xsc}), collision-induced absorption data for a variety of collisional complexes \citep{19KaGoVa} (see Section~\ref{sec:CIA}), aerosol properties (see Section~\ref{sec:aerosols}), and water vapor continuum data \citep{2012RSPTA.370.2520M} (see Section~\ref{sec:CIA}). HITRAN also provides auxiliary data necessary for carrying out calculations of spectra at different thermodynamic conditions, including Total Internal Partition Sums \citep{jt692,2021JQSRT.27107713G}. The HITRAN Application Programming Interface, HAPI \citep{2016JQSRT.177...15K}, allows one to efficiently download HITRAN data and carry out calculations of spectral absorption, emission, and transmission. HITRAN provides line lists for a small number of species which are considered complete up to high temperatures of around 4000--5000~K:  HF, HCl, HBr, HI, and H$_2$~\citep{13LiGoHa.HCl}.

HITEMP \citep{2010JQSRT.111.2139R} is in principle updated alongside the HITRAN database and is specifically aimed at modelling high-temperature ($\gtrsim$~1000~K) environments where HITRAN data cannot be reliably applied. This includes the atmospheres of stars, brown dwarfs, and planets at elevated temperatures, such as hot Jupiters, hot super-Earths, or Venus. At the moment HITEMP provides line lists for eight molecules (H$_{2}$O, CO$_{2}$, N$_{2}$O, CO, CH$_{4}$, NO, NO$_{2}$, OH) and is consistent with HITRAN where possible, but typically includes a substantial proportion of calculated \textit{ab initio} or semi-empirical parameters. These additional lines account for hot bands, and other transitions that are negligible (i.e., extremely weak) at terrestrial temperatures and therefore not included in the HITRAN line list. While the number of lines required for each molecule can reach billions, one goal of HITEMP has been to remain practical to use in line-by-line radiative transfer calculations. Therefore, the underlying \textit{ab initio} line lists are analyzed and filtered to reduce the number of lines necessary for each molecule, as demonstrated for the addition of CH$_{4}$ \citep{HargreavesAJSS2020} to HITEMP.

In summary, HITRAN and HITEMP both provide essential input parameters for many radiative transfer codes. These data are used for a variety of applications, including the modelling of exoplanet atmospheres at high-resolution.

\paragraph{NASA Ames database}\label{sec:nasa_ames}

The database from NASA Ames 
uses the ``Best Theory + Reliable High-Resolution Experiment (BTRHE)'' strategy~\citep{21HuScLe} to provide accurate and complete line lists for species such as SO$_2$ (5 isotopologues up to 500~K)~\citep{15HuScLe,16HuScLe}, CO$_2$ (up to 3000~K)~\citep{23HuFrTa}, NH$_3$~\citep{22HuSuTo}, N$_2$O \citep{2023HuScLe.N2O}, OCS, CS$_2$ and their isotopologues. The most recent NASA Ames releases include hot line lists are for the four most abundant isotopologues of CO$_2$  ($^{12}$C$^{16}$O$_2$, $^{13}$C$^{16}$O$_2$, $^{16}$O$^{12}$C$^{18}$O, and $^{16}$O$^{12}$C$^{17}$O) \citep{22HuScFr.CO2} and extensive room temperature line lists for N$_2$O with its 12 isotopologues \citep{2023HuScLe.N2O}. The ExoMol SO$_2$ line list \citep{jt635} was also produced jointly with NASA Ames. Some of the earlier NASA Ames line lists were formatted using the HITRAN-like line-by-line ascii format, while the more recent line lists are provided in the ExoMol-like format with two parts, states and transitions. Although these are ascii files, for transitions (Einstein coefficients and states IDs), NASA Ames uses a simple compacting scheme based on representing real numbers as integers. The line lists can be accessed online\footnote{\url{https://huang.seti.org/} and \url{https://data.nas.nasa.gov/ai3000k}}.

\paragraph{TheoReTS}\label{sec:theorets}

Precise knowledge of high-energy molecular states and intensities of rovibrational transitions is essential for the modelling of various planetary atmospheres as well as for the understanding of spectral properties, even in extreme dynamical and temperature 
conditions. This clearly demonstrates the need of having consistent line-by-line spectroscopic databases, with a large coverage in terms of both wavelengths and temperatures. Because of their completeness, variational calculations are well designed for the modelling of planetary atmospheres, unlike traditional spectroscopic effective models for which wavenumber extrapolation beyond the range of observed data turns out very limited. Within that context, the TheoReTS (Theoretical Reims-Tomsk Spectral data) project \citep{TheoReTs} $-$ whose philosophy closely follows that of ExoMol \citep{jt810} $-$ was born in 2016\footnote{\url{https://theorets.tsu.ru/} and \url{https://theorets.univ-reims.fr/}}. For the construction of accurate variationally-computed molecular line lists allowing a 
complete description of the main spectral features, highly-optimised numerical methods and symmetry tools for solving the nuclear-motion equation as well as high-level {\it ab initio} 
calculations are required. 

The TheoReTS line lists are simple ascii files with three columns: $\nu_{ij}$, $S_{ij}$, $E_{low}$.  For ease of use, quantum numbers and symmetry labels can be provided upon request using a HITRAN-like format. In addition, variational energy levels can be also replaced by rotation-vibration levels computed from empirical effective Hamiltonians whose parameters are fitted to experimental data, as in the case of methane \citep{Rey2018Icarus}. This ensures almost the same quality as HITRAN2020 \citep{2022JQSRT.27707949G} for the line positions of cold bands while many new hot bands can be accurately predicted. For  modelling  hot exoplanets and brown dwarfs, a methane line list for the HITEMP \citep{HargreavesAJSS2020} spectroscopic database was constructed using the TheoReTS predictions \citep{Rey2017apj,Wong2019apj}. Moreover, in order to facilitate the manipulation of the so-called quasi-continuum formed by billions of lines (e.g. due to line lists of heavy molecules at high-temperatures), a set of super-lines can be generated at different temperatures to drastically reduce the total number of lines by several orders of magnitude~\citep{TheoReTs}.

In its current form, TheoReTS contains line lists for 9 molecules (CH$_4$, PH$_3$, SF$_6$, C$_2$H$_4$, CF$_4$, GeH$_4$, SiH$_4$, CH$_3$F, NF$_3$) and 28 isotopologues. New line lists will be released in 2024, either by improving the existing ones or by including new ones for both {\it semirigid} and {\it nonrigid molecules} \citep{Rey2023JCP}.   Among the next candidates of primary importance $-$ potentially relevant for the study of exoplanetary atmospheres are CH$_2$, CH$_3$, NH$_3$, H$_2$CCN, H$_2$NCN, CH$_3$Cl, C$_3$H$_4$, H$_2$O$_2$, CCl$_4$, CH$_3$CN or C$_2$H$_6$ that will be gradually uploaded in TheoReTS. The construction of comprehensive line lists is now being revisited using a novel methodology developed in Reims to derive {\it ab initio} effective models \citep{Rey20222JCP}, combining the small dimensionality of the spectroscopic polyad models and the completeness of the variational calculation. Fine-tuning of the {\it ab initio}-based spectroscopic  parameters allows matches to observed data to be made without much computational effort, unlike a refinement procedure of the PES which is generally more demanding.  Effective parameters plan to be shared in the upgraded version of TheoReTS.

\subsubsection{Laboratory spectroscopy measurements}\label{sec:labspecs}

As might be expected, the vast majority of laboratory studies have been undertaken with terrestrial applications in mind, and these studies form the basis for many of the line lists discussed throughout this section. Laboratory studies not only provide empirical measurements of critical parameters such as line positions, intensities, broadening coefficients, and temperature dependencies \citep[e.g., see][which overviews experimental works included for HITRAN2020]{2022JQSRT.27707949G}, but they are also used to refine and develop theoretical models, they are essential for constraining calculated parameters, and they are often used directly as is the case for absorption cross-section measurements (see Section~\ref{sec:lab_xsc}). However, the experimental conditions of manystudies, such as pressure and temperature, are not always directly applicable to non-terrestrial planetary environments, which means spectroscopic parameters derived from them have limitations when applied elsewhere. Within our Solar System there are atmospheric temperatures, pressures, and constituents that are beyond those encountered in the Earth's atmosphere. Moreover, the variety of exoplanet atmospheres that are already accessible to transit measurements, via JWST observations or high-resolution ground-based observations (and those that will be observed by the near-future Ariel mission), mean that the experimental requirements far exceed atmospheric environments previously studied and necessitate a range of planetary-applicable laboratory measurements to better model these observations \citep{2019astro2020T.146F}.

The value of experimental spectroscopy for characterizing planetary atmospheres cannot be overstated. However, this section is not meant to be a comprehensive overview of the history of spectroscopic works, but instead aims to provide a summary of recent experimental measurements that highlight key techniques, spectroscopic parameters, and molecules currently being investigated for planetary applications.

\paragraph{Experimental data for Solar System planets} Experimental data can support both the study of our Solar System bodies and contribute significantly to the understanding of exoplanet spectroscopy.

\begin{figure*}
    \centering
    \includegraphics[width=0.45\linewidth]{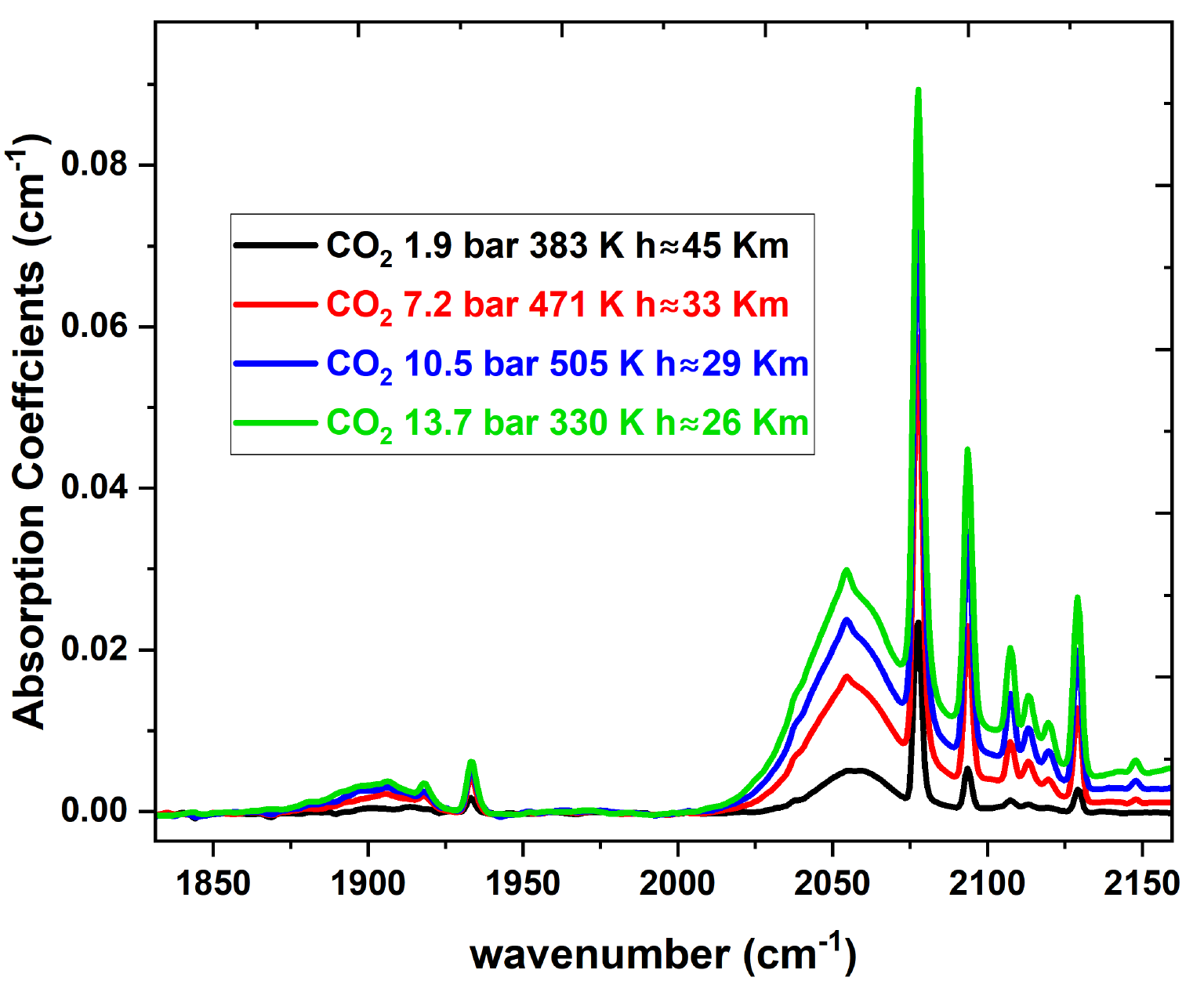}
    \includegraphics[width=0.45\linewidth]{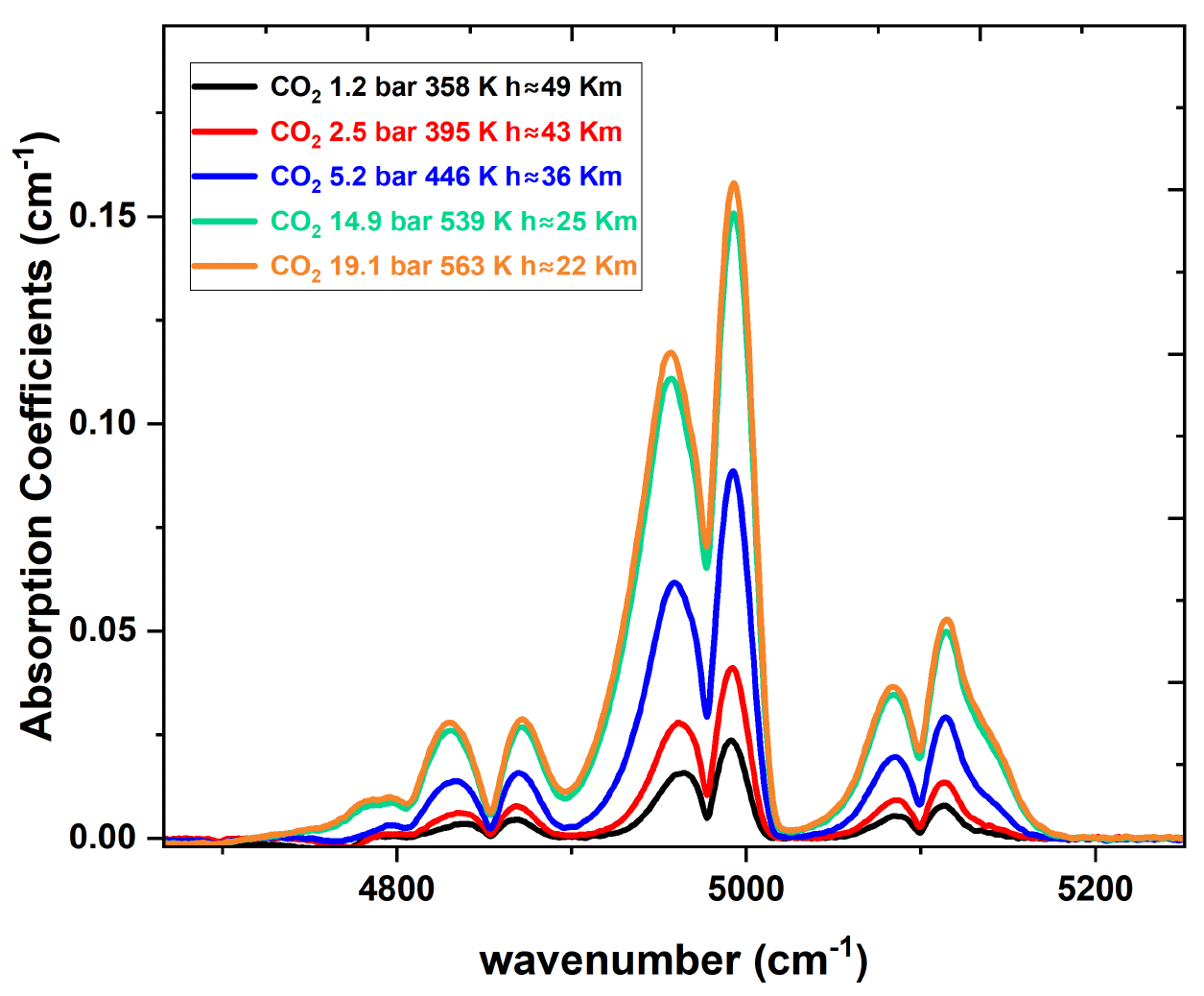}
    \includegraphics[width=0.45\linewidth]{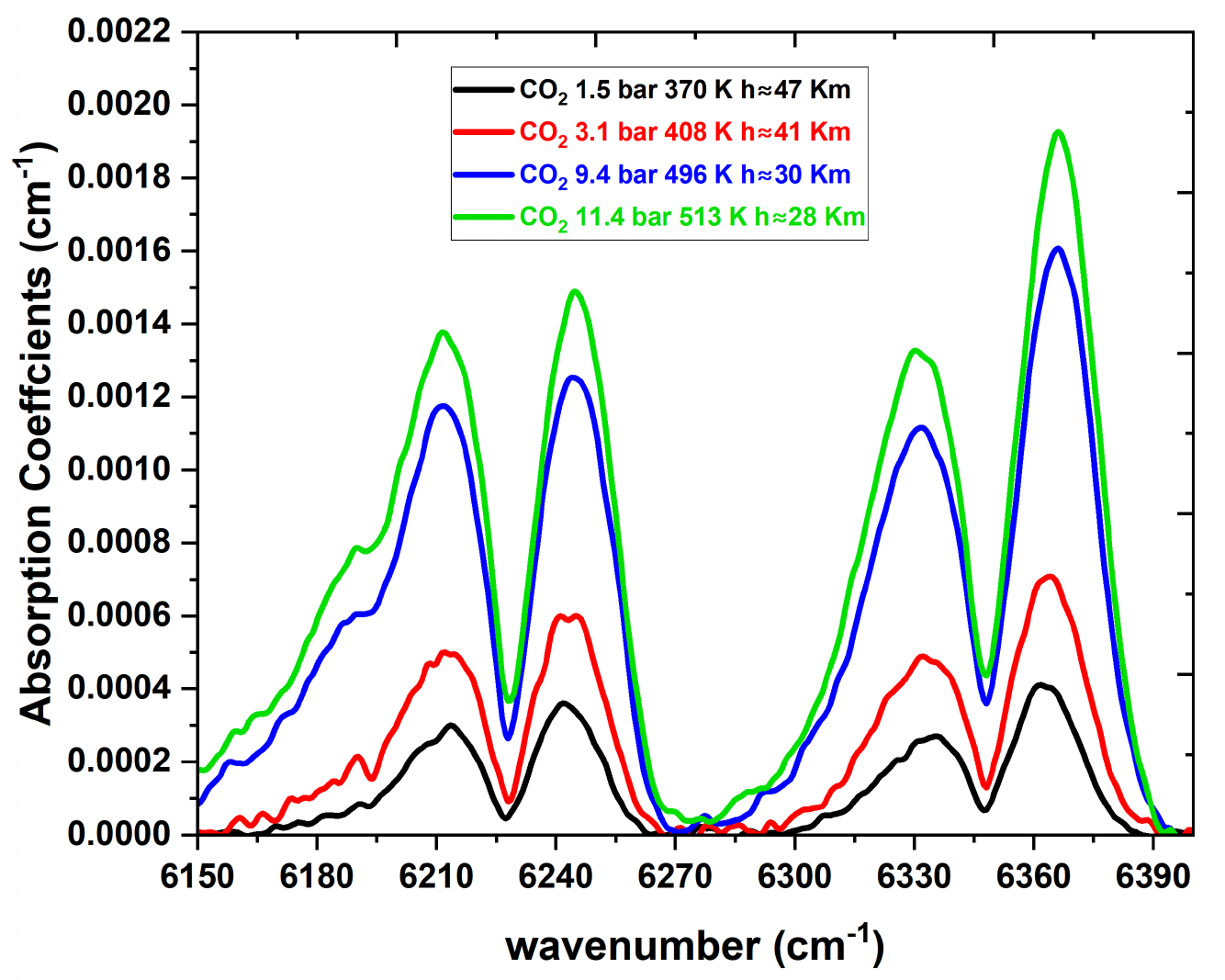}
    \includegraphics[width=0.45\linewidth]{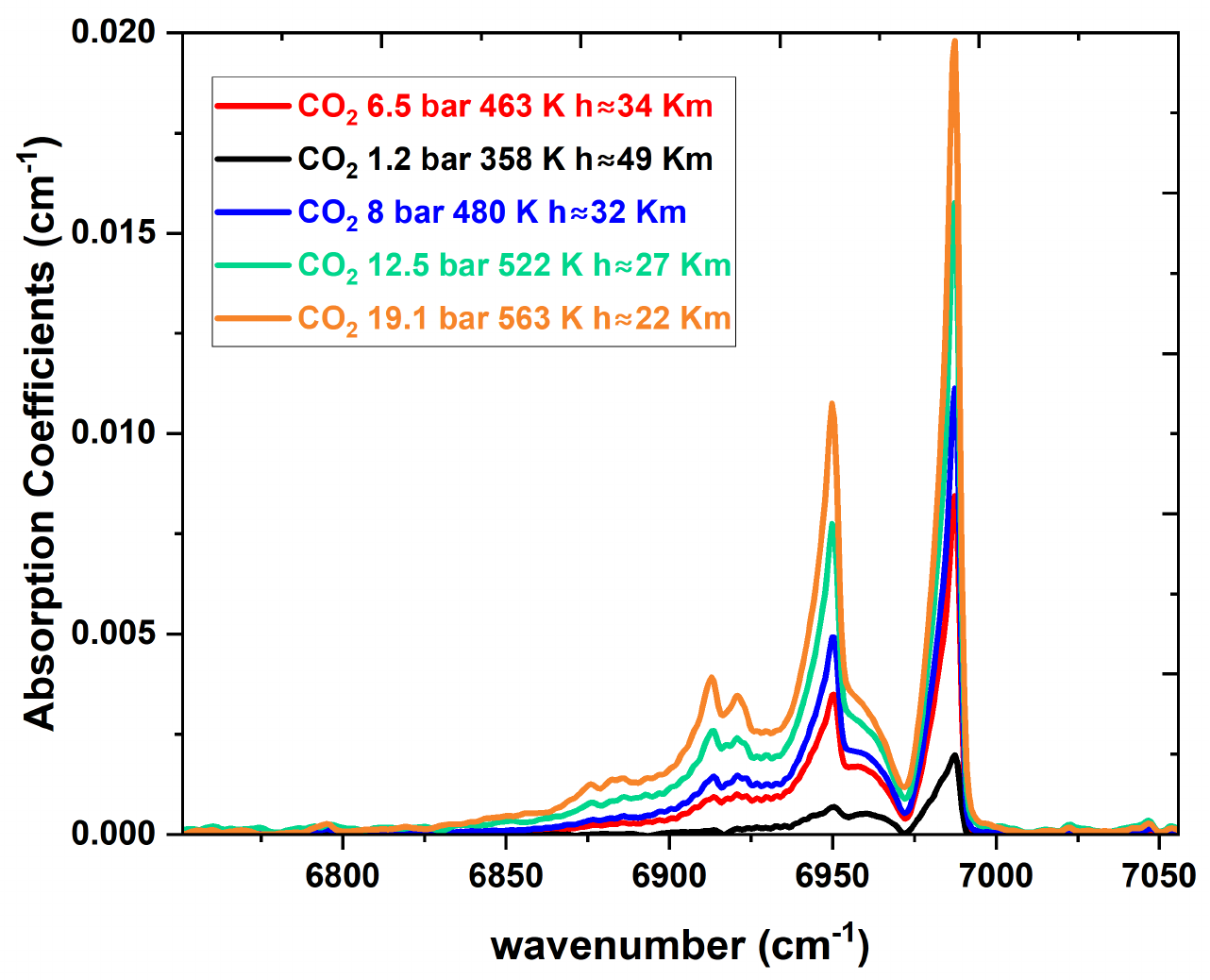}
    \caption{Measured spectra of CO$_2$ at different temperatures and pressures relevant to the atmosphere of Venus, from \protect\cite{Stefani_2013}. }
    \label{fig:CO2_lab}
\end{figure*}

The advent of the ESA Venus Express mission to Venus and the NASA/Juno and ESA/JUICE missions to the Jovian system pushed the interest on new laboratory experiments to study the properties of CO$_2$ at high temperatures and pressures, to simulate the Venusian case, and H$_2$ and H$_2$+He mixtures as representative of collision induced absorption (CIA) in Jupiter's atmosphere. 
For example, two different experimental apparatus have been used to obtain the spectra representative of Venus \citep{Stefani_2013} and Jupiter \citep{21SnStBo}, due to the different pressure and temperature values covered. In the case of Venus, CO$_2$ spectra were obtained in the 1-10~$\mu$m spectral band~\citep{CO2_CIA_database_stefani,CO2_database_Stefani}, exploring temperature and pressure values according to a real Venus vertical profile \citep{Seiff_1985}. An example is shown in Figure \ref{fig:CO2_lab}. For the Jupiter case, collision induced absorption (CIA) coefficients of H$_2$ and H$_2$-He were measured (see Section~\ref{sec:CIA}). Furthermore, empirical absorption cross-sections for key absorption species observed in the atmospheres of Gas Giant planets have traditionally been implemented in retrievals when line lists were incomplete, as done for CH$_{4}$ in the visible \citep{94Karkoschka}. Moreover, numerous experimental measurements of broadening parameters applicable to planetary atmospheres (e.g., H$_{2}$, He, CO$_{2}$, and H$_{2}$O) and their temperature dependencies have been combined into functional forms by \citet{16WiGoKo} and \citet{2022ApJS..262...40T} to populate these parameters in spectroscopic databases. 

Titan, with its complex atmosphere, has motivated data-focused experiments for decades \citep[e.g.][]{1989Icar...80..361M, 1995P&SS...43...25D, 2008Icar..195..792C, 2013Icar..226.1499S, 2020Icar..34413460H}, valuable not just for the analysis of the Saturnine moon but for studies of the early Earth and its exoplanetary analogues. These studies reflect the need for spectroscopic data of hydrocarbons and other organic molecules in nitrogen-rich environments and have already significantly improved our understanding of the opacities of CH$_3$D and C$_2$H$_6$ for instance \citep{22HeSeHo,horst2008origin, vuitton2019simulating}.

\paragraph{High-temperature experimental measurements}\label{sec:lab_high_temp}

Experimental measurements are relatively sparse outside of the terrestrial temperature range, but there have been some efforts to address this for planetary applications in recent years. 
In particular, we focus on the spectral region covered by Ariel at elevated temperatures, due to the atmospheric temperatures found in the  proposed exoplanet target lists \citep{19EdMuTi,2022AJ....164...15E}. Laboratory studies at high temperatures  consistent with hot planetary atmospheres (i.e., above room temperatures and typically up to $\sim$1000--2000~K), using a variety of experimental techniques, have provided crucial data for exoplanetary studies. However, acquiring spectra at high temperatures is challenging, primarily because of the physical properties of laboratory equipment, but also due to decomposition of the sample molecule, which can lead to reaction products contaminating observations.

Many existing experimental line lists have been built upon spectra recorded with Fourier transform spectrometers (FTS) and typically span the infrared region. These measurements provide relatively wide spectral coverage (e.g. multiple vibrational modes) at high spectral resolution ($\lesssim$~0.01~cm$^{-1}$). It is common to couple Fourier transform spectrometers with cells capable of reaching elevated temperatures. Analyses from these measurements have provided new insight for line positions \citep[][for NH$_{3}$]{2012JQSRT.113..670H}, intensities \citep[][for CH$_{4}$]{2019RScI...90i3103G}, line shape parameters \citep[][for H$_{2}$O]{2016JQSRT.184..316D}, and absorption cross-sections \citep[][for C$_3$H$_8$]{2016JQSRT.182..219B} sufficient for accurate retrievals of planetary conditions. 
Complementary to FTS, cavity ring down spectrometers (CRDS) are capable of providing similar frequency accuracy to FTS, but the increased effective path provided by the cavity results in much higher sensitivity for weak transitions. This high sensitivity makes CRDS well-suited to weaker `window' regions of molecular absorption that are difficult to study with other methods. For example, line parameters obtained for weak CH$_{4}$ transitions in the NIR \citep{2023PCCP...2532778C} can be included in spectroscopic databases directly or used to refine line positions in future iterations of \textit{ab initio} line lists. Moreover, CRDS measurements enable constraints to be applied to the water vapor continuum \citep{2023JQSRT.29608432K} , which is essential for accounting for the total water absorption through long path lengths (see Section~\ref{sec:CIA}). Laser-based techniques, for example, have been able to probe limited spectral regions at even higher resolutions comparable with millimeter rotational studies (e.g. $\Delta\nu$~$<$~1.0$\times$10$^{-6}$~cm$^{-1}$). More recently, frequency combs have enabled highly accurate spectroscopic studies to be acquired over moderately large spectral regions, which can now include whole vibrational bands. For example, comb spectroscopy has been used for measurements of high-temperature spectra of H$_2$O~\citep{SchroederJQSRT2017, RutkowskiJQSRT2018}, CH$_4$~\citep{MalarichJQSRT2021}, and CO$_2$~\citep{23CoTrHo} in the 6250–7570 cm$^{-1}$ range. Water spectra measured in a pre-mixed methane/air flat flame~\citep{RutkowskiJQSRT2018} were assigned using the MARVELised POKAZATEL line list, while H$_2$O~\citep{SchroederJQSRT2017} and CH$_4$~\citep{MalarichJQSRT2021} spectra measured in a range of temperatures and sub-atmospheric pressures were used to assess the accuracy of current editions of HITRAN and HITEMP databases. The CO$_2$ spectra were measured at pressures up to 25 bar to test and improve line mixing models and their temperature dependence~\citep{23CoTrHo}. Another development relevant for high-temperature applications is the optical-optical double resonance spectroscopy with continuous-wave pump and comb probe, which allows measurement and assignment of hot-band transitions with sub-Doppler resolution without the need to heat the sample~\citep{FoltynowiczPRL2021, SilvadeOliveiraNatComm2024}. This technique has so far been used to measure and assign hot band transitions in the 3$\nu$$_3$ $\leftarrow$ $\nu$$_3$ range of CH$_4$~\citep{FoltynowiczPRA2021, SilvadeOliveiraNatComm2024}, finding good agreement with the TheoReTS predictions~\citep{14ReNiTy.CH4,HargreavesAJSS2020}.  See Section~\ref{sec:freq_comb} for a detailed spotlight overview of  frequency comb spectroscopy for planetary applications. 

In addition to providing measurements of line positions, intensities, and line shape parameters, broadband FTS measurements of high-temperature gases enclosed in a heated tube cell (up to $\sim$1500~K) can also provide essential testing of the accuracy of current line lists available in the literature, such as for the CO$_{2}$ line list in HITEMP \citep{2007JQSRT.103..146B, 2015JQSRT.157...14A}. Such measurements have also provided valuable absorption cross-sections that can be used for atmospheric characterisation when line lists have been incomplete for planetary studies \citep[e.g.][for CH$_{4}$]{2003JQSRT..82..279N}. In addition, spectra obtained from high-enthalpy sources \citep{2019RScI...90i3103G} and oxyacetylene torches with temperatures up to 3000~K \citep{2005JChPh.122g4307C}, or time-resolved spectroscopy in glow discharges~\citep[][and references therein]{22PaClYuCi.NH}, can be used to investigate combustion environments and non-local thermodynamic equilibrium (non-LTE) conditions \citep{ 2023Icar..39415421D}. Spectroscopic parameters obtained from these studies can then been used to constrain theoretical line lists, improving the accuracy throughout.  While FTS measurements combined with tube furnaces apparatus have formed the bulk of many studies at high temperatures, often laser based analyses are used in high-temperature remote sensing applications of combustion processes \citep{2020OptL...45..583T}. Combining laser spectroscopy techniques with shock-tube apparatus enables sensitive measurements at higher temperatures and pressures \citep{2019JQSRT.224..396D, 2019JQSRT.222..122S, 2021OExpr..2930140T, 2024A&A...681A..39C}, which can be difficult to reach with tube furnaces. In addition, these techniques can also be combined with frequency combs to provide accurate measurements at  high-temperature that cover broad spectral ranges, such as recent investigations for the line-positions of CH$_{4}$\citep{2021MeScT..32c5501P, MalarichJQSRT2021}.

Spectroscopic studies of atomic transitions and diatomic molecules at high temperatures have historically taken advantage of the Sun's photosphere ($\sim$5800~K), providing accurate line positions. For example, telluric-free, high-resolution Solar spectra recorded from orbiting satellites have been used to extend laboratory studies, as done for CH \citep{1989JMoSp.134..305M,2010JMoSp.263..120C}, NH \citep{2010JMoSp.260..115R}, and OH \citep{1995JMoSp.174..490M,2009JMoSp.257...20B,23CiPaFe}; similarly \citet{jt563} used extensive sunspot spectra due to \citet{95CaKlDu.SiO} to help create an extensive SiO line list. This can provide key measurements of spectral lines positions for molecules that are difficult to measure in the laboratory. A recent review of middle-to-near IR emission spectra of four simple astrophysically relevant molecular radicals, OH, NH, CN and CH, was given by \cite{23CiPaFe}. In addition, key diatomic species that characterise spectra of brown dwarfs are also expected to be present in exoplanet atmospheres. These molecules are challenging for theoretical calculations, and measurements of line positions and intensities are essential for improving models.  Semi-empirical line lists have been developed for these molecules by analysing archived high-temperature, high resolution spectra, such as carried out for FeH \citep{2010AJ....140..919H}, and MgH \citep{07ShHeLe,13GhShBe,2013ApJS..207...26H,jt858},
and TiO~\citep{2022ApJ...926...39D}. These semi-empirical works have often been directly implemented into brown dwarf \citep{2012MNRAS.419.1913B} and exoplanet \citep{2023ApJ...942...71M} atmospheric models to improve characterisation.

Water is a key molecule for our atmosphere, and has been prominent in exoplanet observations since early transmission spectra \citep{jt400,2009ApJ...704.1616S}. The water line lists discussed above have relied on numerous experimental studies to accurately determine spectroscopic parameters (e.g. positions, intensities, line shapes, broadening coefficients, pressure shifts, temperature dependencies), and to refine assignments.
Indeed, some of the current theoretical line lists of H$_{2}$O suitable for high temperature analyses have their origin in earlier works that identified new transitions of high-temperature water vapor on the Sun \citep{1995Sci...268.1155W, 1997Sci...277..346P, 1998ASPC..154..718V}.  Some examples of recent high-temperature water vapor studies that use FTS \citep{2008MNRAS.387.1093Z,2015JQSRT.157...14A}, scanned laser spectroscopy \citep{2018JQSRT.214....1M,20MeSaNa.H2O} and frequency comb studies \citep{SchroederJQSRT2017, RutkowskiJQSRT2018} have all helped to refine  spectroscopic parameters for H$_{2}$O line lists and to constrain \textit{ab initio} calculations. Hot (1723 K) laboratory studies have
also been used to test and confirm the accuracy of the available line lists \citep{20MeSaNa.H2O}.

Small polyatomic molecules such as ammonia and methane can have complex infrared spectra when observed at high resolution, and spectroscopic analyses at high temperatures provide opportunities for extending rovibrational assignments \citep{2015JQSRT.167..126B}. Simplified empirical line lists have been produced by analysing multiple measurements at a variety of different temperatures, as done for NH$_{3}$ \citep{2012JQSRT.113..670H} and CH$_{4}$ \citep{2017JQSRT.203..410B, 2018JQSRT.215...59G, MalarichJQSRT2021}.  These simplified line lists lack spectroscopic assignments but constitute empirical line positions, intensities, and lower-state energies, and are capable of reproducing the experimental spectra over the temperature range of the studies and can be used to characterize brown dwarf and exoplanet atmospheres. Analyses of these line lists can then be performed to determine line assignments and constrain calculated parameters. Furthermore, the  measured spectra can be converted into experimental absorption cross-sections \citep{Wong2019apj} that can themselves be used as input data to planetary models in place of line lists, or can be used to compared the accuracy of multiple high temperature line lists as done for CH$_{4}$ \citep{HargreavesAJSS2020}. Measurements of absorption cross-sections up to temperatures of  1600~K \citep{2014JQSRT.149..241E, 2014JMoSp.303....8A, Wong2019apj}, become particularly necessary for molecules when no complete line lists exists (also Section~\ref{sec:lab_xsc}).

Each edition of the HITRAN database incorporates a substantial proportion of experimentally-derived parameters and readers are directed toward the paper describing HITRAN2020 \citep[][and references therein]{2022JQSRT.27707949G} for a summary of relevant spectroscopic experimental studies that apply to molecules given in Table~\ref{tab:IR:linelists}. In addition, it should be noted that the MARVEL (measured active rotation-vibration energy levels) methodology \citep{2007JMoSp.245..115F} (see Section~\ref{sec:MARVEL}), which is used to empirically correct energy levels of \textit{ab initio} line lists \citep[e.g.][]{jt847}, is built upon an extensive survey of experimental measurements, and readers are also directed to each work for a summary of relevant literature.

It is worth stressing that experimental measurements can be particularly challenging for reactive (or toxic) species, especially at high temperatures, which can limit the availability of parameters. In most cases, the absolute intensities are not available experimentally for  species and therefore relying on the theory is the only option. 
It is therefore not uncommon for some spectroscopic parameters, including  intensities (and associated lifetimes) as well as  pressure-broadening \citep{2022ApJS..262...40T}, to be based only on limited laboratory measurements for some molecules (e.g. H$_{2}$ broadening of CO$_{2}$).

\paragraph{Spot-light: Frequency comb-based spectroscopy}\label{sec:freq_comb}

This section presents a deeper dive into different aspects of frequency comb spectroscopy with the emphasis on its potential for providing spectroscopic data for exoplanetary spectroscopic studies. 

Recent advancements in frequency comb sources and comb-based spectroscopic techniques \citep{GohlePRL2007, FoltynowiczPRL2011, MaslowskiPRA2016, KowzanOL2016, CoddingtonOptica2016, ChangalaAPB2016, WeichmanJMS2019, YcasNatPhot2018, KrzempekOE2019, MuravievSciRep2020, FoltynowiczPRL2021} have provided new opportunities for broadband precision molecular spectroscopy. The 
combination of large bandwidth, high spectral resolution, and high frequency accuracy provided by frequency combs make them ideal sources for high-precision measurements of spectroscopic molecular parameters. Such high-resolution measurements are an essential component to improving the accuracy of theoretically derived line lists (see Sections~\ref{sec:lab_high_temp} and~\ref{sec:MARVEL}). Frequency combs can either be used directly as light sources for broadband high-resolution spectroscopy (which is referred to as direct frequency comb spectroscopy) or as references to calibrate the frequency axis of tunable-laser-based spectrometers (referred to as comb-referenced spectroscopy). While the simultaneous spectral coverage of most frequency combs cannot yet compete with incoherent sources, the spectral resolution of comb-based spectrometers is orders of magnitude better than that of an FTS spectrometer based on an incoherent light source, allowing the measurement of undistorted line profiles at low pressures with sub-MHz accuracy on line positions. Moreover, the spectral and spatial coherence of the combs make possible the use of path length enhancement methods, such as multi-pass cells \citep{AdlerOE2010} and cavities \citep{AdlerARAC2010}, to increase the absorption sensitivity and detect weak absorption bands. Comb-referenced cavity-enhanced tunable-laser-based spectrometers provide even better absorption sensitivity and - when combined with saturation spectroscopy - line positions with kHz accuracy. 

Below we give an overview of precision measurements using direct frequency comb spectroscopy and comb-referenced spectroscopy, with a focus on measurements that led to new line lists or substantial improvement of existing ones for different molecules. We note that an exhaustive coverage of all measurements involving frequency combs is beyond the scope of this work.

Table~\ref{tab:combs} summarises a selection of experimental line lists obtained from measurements using different direct frequency comb spectroscopy techniques. Most of these measurements were performed at room temperature using direct absorption in multi-pass cells or enhancement cavities. Many of the works provided also line intensities, and pressure broadening and shift parameters. Three of the room temperature line lists have already been used to update the databases, namely the $\nu_4$ band region (3000–3160 cm$^{-1}$) of CH$_3$I~\citep{SadiekJQSRT2020} and the $\nu_1$+$\nu_3$ band region (2110–2200 cm$^{-1}$) of CS$_2$~\citep{KarlovetsJQSRT2020} were used in updates to the HITRAN2020 database~\citep{2022JQSRT.27707949G}, and the $\nu_6$ and $\nu_4$ band region (1250–1380 cm$^{-1}$) of H$_2$CO measured by \cite{GermannJQSRT2024} was used to improve the accuracy of the ExoMol line list for H$_2$CO~\citep{jt597,21AfTeYu}.

To study complex molecules, whose spectra are unresolved at room temperature, buffer gas cooling has been implemented to reduce the number of populated rotational and vibrational states. Combining buffer gas cooling with cavity-enhanced comb spectroscopy allowed the measurement of rotationally resolved spectra of nitromethane (CH$_3$NO$_2$), naphtalene (C$_{10}$H$_{8}$), adamantane (C$_{10}$H$_{16}$), and hexamethylenetetramine (C$_6$N$_4$H$_{12}$)~\citep{SpaunNature2016}, and vinyl bromide (CH$_2$CHBr)~\citep{ChangalaAPB2016} in the 2850–3070 cm$^{-1}$ range, as well as the buckminster fullerene (C$_{60}$) in the 1180–1190 cm$^{-1}$ range~\citep{ChangalaScience2019}. 

Cavity-enhanced comb spectroscopy has also been combined with velocity-modulation to measure four bands of a molecular ion HfF$^+$ in the 11500–13000 cm$^{-1}$ range~\citep{CosselCPL2012}. Finally, the time-resolved capabilities of comb spectroscopy have been used to measure high-resolution spectra of trans- and cis-DOCO transients (products of the OD + CO reaction) in the OD stretch region 2380–2750 cm$^{-1}$~\citep{BuiMolPhys2018}, as well as the pressure broadening of absorption lines of the short-lived  CH$_2$OO Criegee intermediate~\citep{LuoOL2020} in the 1217–1287 cm$^{-1}$ range. 

Table~\ref{tab:comb_ref} summarises a selection of experimental line lists obtained from measurements using different comb-referenced spectroscopy techniques. 
Many works using comb-referenced cavity-enhanced techniques in the near-infrared provided transition frequencies for overtone and combination bands with kHz level accuracy for, e.g., H$_2$O \citep{ChenJQSRT2018, KassiJCP2018, TobiasNatComm2020,24ToDiCoUb}, CO$_2$ \citep{ReedJQSRT2021, GuoJQSRT2021,FleurbaeyPCCP2023}, CO \citep{MondelainJQSRT2015, WangJQSRT2021} and CH$_4$ \citep{VotavaPCCP2022}. In the mid-infrared, sub-MHz and kHz level accuracy was obtained for the fundamental bands of CH$_4$ \citep{KocherilJQSRT2018}, N$_2$O \citep{KnabeOE2013, TingJOSAB2014, AlSaifJQSRT2018}, CHF$_3$ \citep{VicentiniJQSRT2020}, and CH$_3$OH \citep{SantagataOptica2019} using saturation and Doppler-broadened spectroscopy in absorption cells, and for CH$_4$ using sub-Doppler cavity-enhanced spectroscopy \citep{OkuboOE2011, AbeJOSAB2013}. Three of these works have already been used for updating the databases and are included as part of HITRAN2020~\citep{2022JQSRT.27707949G}. This includes line positions for the 3$\leftarrow$0 and 4$\leftarrow$0 bands of CO \citep{MondelainJQSRT2015} and the $\nu_3$ region of CH$_4$~\citep{AbeJOSAB2013}, with the work of \citet{FleurbaeyJQSRT2021} included in a fit for line positions of the 1.27~$\mu$m band of O$_2$. Line intensities were provided only by a couple of works using CRDS in the Doppler-limited regime, since saturation spectroscopy often does not yield absolute intensities.

\begin{table*}
    \centering
    \caption{Molecular line lists obtained from direct frequency comb spectroscopy measurements using different methods: DCS - dual comb spectroscopy; FTS - Fourier transform spectroscopy; Vernier - Vernier spectroscopy; VIPA - spectrometers based on virtually imaged phased array; PAS - photoacoustic spectroscopy; OODR - optical-optical double resonance spectroscopy; CE - cavity enhanced. The spectral coverage and temperature range are also indicated; RT - room temperature. All works listed here provided line positions, and some, indicated in the 'Parameters' column, provided also intensities, $\textit{S}$, self, air and reaction matrix broadening parameters, $\gamma_{\rm self}$, $\gamma_{\rm air}$, $\gamma_{\rm p}$, and self shift parameters, $\delta_{\rm self}$.} 
    \label{tab:combs}
    \begin{tabular}{lcccccl}
    \hline
    Molecule & Range [cm$^{-1}$] & Range [$\mu$m] & Method & T [K] & Parameters  &  Reference \\
    \hline 
        H$_2$O  &  6800–7200 & 1.39~-~1.47  & DCS   & 296-1300 & $\gamma_{\rm self}$, $\gamma_{\rm air}$ &  \cite{SchroederJQSRT2017} \\
         &  6250–6670 & 1.50~-~1.60  & CE FTS & 1950 & $\textit{S}$ &\cite{RutkowskiJQSRT2018} \\
        CO$_2$  &  5064–5126 & 1.95~-~1.97 & CE Vernier & RT  & $\textit{S}$, $\gamma_{\rm self}$ & \cite{SicilianideCumisJCP2018} \\
         &  6800–7000 & 1.43~-~1.47  & DCS & 495-977 & line mixing & \cite{23CoTrHo} \\
        N$_2$O  &  1250–1310 & 7.63~-~8.00 & FTS & RT &  & \cite{HjaltenJQSRT2021} \\
        CO  &  6310–6365  & 1.57~-~1.58 & CE VIPA & RT & $\gamma_{\rm self}$, $\delta_{\rm self}$ & \cite{KowzanSciRep2019} \\
         &  2040–2230  & 4.48~-~4.90 & FTS & RT &  & \cite{NishiyamaMeas2024} \\
        CH$_4$  &  1250–1380  & 7.25~-~8.00 & FTS   & RT & $\textit{S}$ & \cite{GermannJQSRT2022} \\
          &  2900–3050 & 3.28~-~3.45  & DCS   & RT &  & \cite{BaumannPRA2011} \\
         &  5870–6130 & 1.63~-~1.70  & DCS & RT &  & \cite{ZolotJQSRT2013} \\         
          &  5900-6100 & 1.64~-~1.69  & OODR   & 111 &  & \cite{FoltynowiczPRA2021} \\
          &  5910-5980 & 1.67~-~1.69  & OODR &  RT &  & \cite{SilvadeOliveiraNatComm2024}\\
          &  6770–7570 & 1.32~-~1.48  & DCS   & 296-1000 & $\gamma_{\rm self}$, $\delta_{\rm self}$ & \cite{MalarichJQSRT2021} \\
        $^{13}$CH$_4$  &  1250–1380  & 7.25~-~8.00 & FTS   & RT & $\textit{S}$ & \cite{GermannJQSRT2022} \\
                $^{14}$CH$_4$  &  2910–3110 & 3.22~-~3.44  & PAS   & RT &  & \cite{KarhuOL2019} \\
        H$_2$CO  &  1250–1380 & 7.25~-~8.00  & FTS   & RT & $\textit{S}$ &\cite{GermannJQSRT2024} \\        
        H$^{13}$CN  &  6390–6535 & 1.53~-~1.56  & DCS   & RT & $\textit{S}$ & \cite{GuayOL2018} \\
        C$_2$H$_2$  &  6430–6630 & 1.51~-~1.56  & DCS   & RT &  & \cite{ZolotJQSRT2013} \\
        CH$_3$I  &  3000–3160 & 3.16~-~3.3  & FTS   & RT & $\textit{S}$ & \cite{SadiekJQSRT2020} \\
          &  2930–3160 & 3.16~-~3.57  & FTS   & RT & $\textit{S}$ &  \cite{HjaltenJQSRT2023}\\
        CS$_2$  &  2110–2200  & 4.55~-~4.74 & DCS & RT  & $\textit{S}$ &\cite{KarlovetsJQSRT2020} \\   
        HfF$^+$ &  11500–13000  & 0.77~-~0.87 & CE VIPA  & 823 &   &\cite{CosselCPL2012} \\
        CH$_3$NO$_2$  &  2940–3090 & 3.24~-~3.40  & CE FTS   & 10-20 &  & \cite{SpaunNature2016} \\
        CH$_2$CHBr  &  3023–3033 & 3.30~-~3.31  & CE FTS   & 10-20 &  & \cite{ChangalaAPB2016} \\
        DOCO  &  2380–2750 & 3.64~-~4.20  & CE FTS  & RT &  & \cite{BuiMolPhys2018} \\        
        C$_{60}$  &  1180–1190 & 8.40~-~8.47  & CE FTS   & 150 &  & \cite{ChangalaScience2019} \\
        CH$_2$OO  &  1217–1287 & 7.77~-~8.22  & DCS   & RT & $\gamma_{\rm p}$  & \cite{LuoOL2020} \\
        CH$_2$Br$_2$  &  2960–3120 & 3.21~-~3.38  & FTS   & RT &   &\cite{SadiekPCCP2023} \\
    \hline
    \end{tabular}
\end{table*}

\begin{table*}
    \centering
    \caption{Molecular line lists obtained from comb-referenced spectroscopy measurements using different methods: CRDS - cavity ring down spectroscopy; NICE-OHMS - noise-immune cavity enhanced optical heterodyne molecular spectroscopy; CRSS - comb-referenced saturation spectroscopy, DAS - direct absorption spectroscopy; CEAS - cavity enhanced absorption spectroscopy, NICE-VMS - noise-immune cavity enhanced velocity modulation spectroscopy. The spectral coverage of each is also indicated. }
    \label{tab:comb_ref}
    \begin{tabular}{lccccl}
    \hline
    Molecule & Band & Range [cm$^{-1}$] & Range [$\mu$m] & Method &  Reference \\
    \hline 
    H$_2^{16}$O  & (013 $\leftarrow$ 000) &  12,621~-~12,665  & 0.789~-~0.792 &  CRDS  &  \cite{ChenJQSRT2018} \\
 			     & (101 $\leftarrow$ 000) &  7164~-~7185  & 1.391~-~1.396   &  CRDS  &  \cite{KassiJCP2018} \\
 	 & (200 $\leftarrow$ 000) &  7000~-~7350 & 1.36~-~1.43   &  NICE-OHMS$^a$  &  \citet{TobiasNatComm2020,DioufMolPhys2022} \\
   &&&&& \citet{24ToDiCoUb} \\
 	 H$_2$$^{17}$O & (200 $\leftarrow$ 000)$^{b}$  &  7228~-~7289  & 1.37~-~1.38   &  NICE-OHMS  &  \cite{MelossoJPCA2021} \\
 	H$_2$$^{18}$O & (200 $\leftarrow$ 000) &  7000~-~7350  & 1.36~-~1.43  &  NICE-OHMS$^a$  &  \cite{Diouf2021,24ToDiCoUb} \\
H$_2$O$^c$ & Several$^{d}$ & 8041~–~8633 & 1.15~-~1.24 & CRDS & \cite{KorolevaJQSRT2023} \\

$^{12}$C$^{16}$O$_2$  & (30012 $\leftarrow$ 00001) & 6170~–~6370 & 1.57~-~1.62 & CRDS & \cite{ReedJQSRT2021} \\
					& (30013 $\leftarrow$ 00001) & &  &  &  \\
					 & (30012 $\leftarrow$ 00001)$^{d}$ & 6170~–~6370 & 1.57~-~1.62 & CRDS & \cite{GuoJQSRT2021} \\
      & (20012 $\leftarrow$ 00001) & 4830~–~5010 & 1.99~-~2.07 & CRDS & \cite{FleurbaeyPCCP2023} \\
					& (20013 $\leftarrow$ 00001) & &  &  &  \\

      					 & (01111 $\leftarrow$ 01101) & 2306~–~2312 & 4.33~-~4.34 & CRSS & \cite{GalliMolPhys2013} \\
$^{14}$C$^{16}$O$_2$ & (00011 $\leftarrow$ 00001) & 2190~–~2250 & 4.44~-~4.57 & CRDS & \cite{GalliMolPhys2011} \\

N$_2$O & 3$\nu_1$ & 6518~-~6578  & 1.52~-~1.53 & CRDS & \cite{LiuJQSRT2019} \\
		& 3$\nu_1$ & 6518~-~6578  & 1.52~-~1.53 & DAS & \cite{IwakuniJMolSpec2022} \\
		& $\nu_3$ & 2190~–~2210 & 4.52~-~4.57 & DAS & \cite{KnabeOE2013} \\
		& $\nu_3$ & 2130~-~2270  & 4.41~-~4.69 & CRSS & \cite{TingJOSAB2014} \\
    & $\nu_1$ & 1250~-~1310 & 7.63~-~8.00 & DAS & \cite{AlSaifJQSRT2018} \\
		
N$_2$O$^e$ & -- $^{d}$ & 8320~–~8620  & 1.16~-~1.20 & CRDS & \cite{KarlovetsJQSRT2021,KarlovetsJQSRT2022} \\
CO & (7 $\leftarrow$ 0)$^{d}$ & 14300~-~14500 & 0.689~-~0.699 & CRDS & \cite{BalashovJCP2023} \\
  & (3 $\leftarrow$ 0) & 6170~-~6420 & 1.56~-~1.62 & CRDS & \cite{MondelainJQSRT2015} \\
     & (3 $\leftarrow$ 0) &  6170~-~6420 & 1.56~-~1.62  & CRDS & \cite{KowzanJQSRT2017} \\
     & (3 $\leftarrow$ 0) &  6170~-~6420 & 1.56~-~1.62  & CRDS & \cite{WangJQSRT2021} \\

   & (4 $\leftarrow$ 0)$^{d}$ &  8206~-~8465 & 1.18~-~1.22 & CRDS & \cite{BordetJQSRT2021} \\
  
CH$_4$ & 2$\nu_3$ & 6015~–~6115 & 1.64~-~1.66 & CRDS & \cite{VotavaPCCP2022} \\
& $\nu_3$ & 2890~-~3120 & 3.21~-~3.46 & CEAS & \cite{OkuboOE2011,AbeJOSAB2013} \\
& $\nu_3$ & 2890~-~3120 & 3.21~-~3.46 & CRSS & \cite{KocherilJQSRT2018} \\
       
O$_2$ & B band$^{d}$ & 14539~–~14550 & 0.687~-~0.688 & CRDS  & \cite{DomyslawskaJQSRT2020} \\
 & (0,0)$^{d}$ & 7800~–~7960 & 1.26~-~1.28 & CRDS  & \cite{FleurbaeyJQSRT2021} \\
H$^{13}$CN & 2$\nu_3$ & 6390~–~6536 & 1.53~-~1.56 & CRSS  & \cite{HrabinaOL2022} \\
D$_2$H$^+$ & $\nu_1$ & 2588~–~2930 & 3.41~-~3.86 & NICE-VMS  & \cite{MarkusJMolSpec2019} \\
CHF$_3$ & $\nu_5$ & 1156~-~1160  & 8.62~-~8.65  & CRSS & \cite{VicentiniJQSRT2020} \\
CH$_3$OH & $\nu_5$ & 970~-~973  & 10.29~-~10.30 & CRSS & \cite{SantagataOptica2019} \\
Benzene & $\nu_{11}$ & 675~-~689 & 14.5~-~14.8 & DAS & \cite{LampertiCommPhys2020} \\
    \hline
    \end{tabular}
    \flushleft{$^{a}$: Assisted with spectroscopic networks; $^{b}$: Hyperfine-resolved; $^c$: Six isotopologues; 
     $^{d}$: Intensities provided; $^e$: Five  isotopologues}
\end{table*}

\subsubsection{Alternative parameterisations}

\paragraph{Absorption cross-section measurements}\label{sec:lab_xsc}

Determining an accurate line list for a large proportion of known molecules is not practical, or possible, due to the complexity of assigning congested spectra \citep{2019PCCP...2118970S,22ZaMc,22ZaMcKem,23ZaMc}. In such instances, numerous individual lines typically blend into unresolved spectral bands. However, these molecules can still form strong detectable features in atmospheric spectra, and their spectroscopic signatures are required in retrieval frameworks for detection, in addition to improving the retrieval accuracy of  planetary characteristics or other atmospheric species. To provide this necessary data, it is common practice to carefully measure experimental absorption cross-sections. These measurements are required to cover the expected temperature and pressure conditions of the appropriate planetary atmosphere, as well as including the appropriate broadening gases (e.g. H$_{2}$, He, N$_{2}$, etc) at a resolution that provides sufficient details.

Cross-sections are not just restricted to the infrared or visible wavelength regions; many valuable cross-sections which cover the UV region are regularly employed in Earth observations, for example. There are many cross-sections included in the HITRAN database that cover the UV and visible regions, such as SO$_2$ and O$_3$. While this section will mainly discuss infrared and visible absorption cross-sections due to the relevance for Ariel, UV absorption cross-sections are also important for exoplanet atmosphere models used to inform our understanding of observed planets. UV cross-sections also have special continuum features related to photodissociation and photoabsorption effects; see Section~\ref{sec:UV} where we discuss these separately.

The Pacific Northwest Nation Laboratory (PNNL) produced a large catalogue \citep{2004ApSpe..58.1452S} of infrared absorption cross-sections for many chemical compounds recorded at 1.0 atm with a Fourier transform spectrometer at temperatures of 5, 25 and 50$^{\circ}$C. Species included compounds that were difficult to measure in the laboratory. These cross-sections are no longer available from PNNL directly, but the majority of these measurements are now included as part of the HITRAN database \citep{2022JQSRT.27707949G}. The HITRAN database provides experimental absorption cross-sections of over 300 molecules \citep{2019JQSRT.230..172K}, in addition to the line-by-line molecules described above. HITRAN also regularly expands the available data with each new edition \citep{2022JQSRT.27707949G}.  

It should be noted that many chemical compounds have been measured due to signatures appearing in Earth observation measurements. For example, chlorine-containing compounds (e.g. chlorofluorocarbons) are intensely monitored because their reactions in the atmosphere lead to the depletion of stratospheric ozone. Therefore, the temperature and pressure coverage of absorption cross-sections is regularly expanded for improved accuracy of retrievals, for example, using Fourier transform measurements of the molecule contained in a cold cell \citep{2018AMT....11.5827H}. 

Other planets of the Solar System (and their moons) contain many trace gases, which are typically retrieved using experimental absorption cross-sections. For example, Titan, with its rich hydrocarbon atmosphere, requires cross-section measurements of nitrogen broadened species at low temperatures \citep[e.g.][]{2013Icar..226.1499S, 18SuToDr,vuitton2019simulating}. Similarly, H$_{2}$ and He broadened measurements are required for the Jovian planets \citep[e.g.][]{2016Icar..271..438S, 2020JQSRT.25307131D}. It is expected that some of these species will be present at elevated temperatures in warm Gas Giant exoplanetary atmospheres \citep{2023ApJ...956..134F}, therefore recent studies have also provided experimental absorption cross-sections at elevated temperatures \citep[e.g.][]{2014JMoSp.303....8A, 2015MolAs...1...20H, 2016JQSRT.182..219B, 2021JQSRT.26907644B, 2023JQSRT.30008522A}.

The limited number of absorption cross-section measurements for specific temperatures and pressures makes them less flexible than line lists. However, techniques have been developed to interpolate between measurements with polynomial coefficients \citep{2022JAMES..1403239B} as well as producing pseudo-line lists \citep{2023JQSRT.31008730S}. Each technique does not identify quantum mechanical transitions, as is the case for spectroscopic line lists (see earlier in Section~\ref{sec:linelists}), but instead empirical parameters are determined that can reproduce the absorption cross-section intensity over the range of measurements. However, care should be taken when extrapolating these techniques to conditions beyond those used from the initial studies.

Ariel will cover the $\sim$2500--5000 cm$^{-1}$ ($\sim$2~-~4~$\mu$m) region at a resolving power of 100 (see Table~\ref{tab:instruments}), which spans the spectral range where key molecular functional groups (e.g. C-H stretch at $\sim$~3000-3300 cm$^{-1}$) exhibit strong absorption. The majority of the absorption cross-sections described above and included in HITRAN have spectral features within this region. Some of these molecules have already been proposed as biosignatures \citep{2021AsBio..21..765Z} or technosignatures \citep{2022PSJ.....3...60H}, due to their distinct spectral features and limited pathways for synthesis.  Indeed, it was in this spectral range that the recent tentative detection of dimethyl sulphide (DMS or (CH$_{3}$)$_{2}$S) was made from transit spectra of K2-18b recorded using JWST \citep{2023ApJ...956L..13M} and absorption cross-sections in HITRAN. 

\textbf{Polycyclic aromatic hydrocarbons (PAHs)} and tholins are known to contribute significantly to the optical properties of Solar System planetary spectra~\citep{16AbMa,18ZhKaXu}. They have an optical slope similar to cloud/hazes, potentially causing a degeneracy in the interpretation of spectral data caused by aerosols (see Section~\ref{sec:aerosols}, and, for example, Figure~1 of \citet{22ErRaMo}). \cite{22ErRaMo} explore the detectability of PAHs observed with Ariel~\citep{18TiDrEc.exo} and Twinkle~\citep{19EdRiZi}, finding there is a possibility of detection. \citet{23DuGrAr} investigate the formation processes of various PAHs on the thermalised atmospheres of hot Jupiter exoplanets, and found that planets with an effective temperature of around 1300~K are the most promising targets for investigating PAHs, particularly for atmospheres with a high carbon-to-oxygen ratio (C/O) and high metallicity. \red{A follow-up study by Gr{\"u}bel et al. (MNRAS, submitted) focuses on the inter-comparison of different cloud/haze, PAHs and tholin models while retrieving the best case scenario for a particular planet, Wasp-6~b.} The cross-sections presented by \citet{22ErRaMo}, based on calculations of \citet{07DrLi}, represent a combined effect observed from a collection of crystalline PAHs. These cross-sections are consistent with observations of the interstellar medium (ISM) and thus presumably originate from a diverse PAH mixture. Given the sheer amount of different molecules, they currently represent the most convenient way to implement PAHs in atmospheric retrievals. The NASA Ames PAH IR Spectral Database is a source of cross-section data for PAHs~\citep{14BoBaRi,18BaRiBo,20MaHuBo}\footnote{\url{https://www.astrochem.org/pahdb/}}, offering extensive data on thousands of PAHs, although primarily for the identification within interstellar emission spectra rather than for their detection in exoplanetary atmospheres.

\paragraph{Vibrational frequency calculations}\label{sec:vib_freq}

The computation of a full ro-vibrational line list for one species can be very time consuming, and often takes years to produce. There has thus been work on computing the vibrational spectra for a large number of species, and then adding rotational sub-structure (for example using software such as PGOPHER~\citep{PGOPHER}), in order to indicate where prominent spectral features are expected. Root mean square errors of such computations are typically on the order of 15~-~25cm$^{-1}$~\citep{22ZaMcKem}, with higher order computational chemistry or laboratory-derived data required to be incorporated for higher accuracies. 
There are some notable absorption features in the observed spectra of some exoplanet atmospheres, for example in WASP-39b's atmosphere at 4.56~$\mu$m as observed by JWST~\citep{23AlWaAl}, which have yet to be identified among the species with available ro-vibrational spectra. There is benefit in identifying potential species or types of species which could fit such features, which could then give incentive for more detailed work into the full ro-vibrational spectra. The computation of such vibrational frequencies for large numbers of molecules is facilitated by studies into the best practice for appropriate model chemistry choices for harmonic frequency calculations, such as \cite{22ZaMcKem,23ZaMc}. VIBFREQ1295~\citep{22ZaMc} is a database of vibrational frequency calculations. RASCALL (Rapid Approximate Spectral Calculations for ALL)~\citep{2019PCCP...2118970S} is an approach which can simulate spectra consisting of approximate band centres and qualitative intensities for a huge number of species. The approximate spectra for a large number of phosphorus-bearing species was computed by \cite{21ZaSyRo}, building upon the RASCALL database by using quantum chemistry to improve on the accuracy of the approximate spectra for these species. 

In addition, computations are available which focus on the vibrational spectra of nanoclusters, such as (TiO$_2$)$_n$, which can be important for exoplanet atmosphere models or observations in the IR~\citep{23SiHeGo}. Vibrational frequencies of other nanoclusters have been computed, such as (VO)$_n$ up to $n$~=~10~\citep{24LeGoSi}, (VO$_2$)$_n$ up to $n$~=~10~\citep{24LeGoSi}, (V$_2$O$_5$)$_n$ up to $n$~=~4~\citep{24LeGoSi}, Ti$_x$O$_y$~\citep{JeoChaSed2000} and (MgO)$_x$~\citep{ChenFelm2014}.

\subsubsection{How high-resolution laboratory spectroscopy can help improve the accuracy of theoretically computed line lists}\label{sec:MARVEL}

There are several ways in which laboratory spectroscopy can contribute to the data needs of exoplanetary atmospheric studies, some of which are already in active use and some require more investment. The most common and efficient methodology of the production of spectroscopic data 
is semi-empirical, in which molecular spectra are first measured at given laboratory conditions (temperature, pressures, wavelength range etc), analysed and then inter- or extrapolated to conditions dictated by the retrieval frameworks. 
A typical example illustrating the need for such extrapolations is that most of the existing experimental molecular spectra have been recorded
at, or close to, the ambient atmospheric conditions on Earth, i.e. room temperature and pressure, and for specific spectroscopic ranges dictated by the experimental setup. A typical retrieval procedure, on the other hand, requires a broad range of temperatures, pressures and wavelengths, on dense grids of the corresponding values (for example, pressures between at least 1$\times$10$^{-5}$~-~100~bar, and temperatures $\sim$100~K~-~3000~K). This is achieved by representing the experimental data in some functional forms, so-called spectroscopic models, of different complexity and extrapolation power. 

This is a powerful, well established procedure, which is however very hungry for experimental data. Some molecules have almost no spectra measured, especially those corresponding to extreme conditions such as  suggested for the atmospheres of the so-called Lava planets, including SiO$_2$, KOH, NaOH and MgOH~\citep{12ScLoFe.exo,17TeYuxxi}. 
Apart from the mainstream atmospheric molecules,
such as those described in Section~\ref{sec:lab_high_temp},
with relatively accessible spectra, there is a serious lack of analysed laboratory spectra for a large number of molecules at higher temperatures and lower wavelengths. 

The so-called effective Hamiltonians (such as used by CaSDa) based on Taylor-type expansions with respect to the angular momenta with a large set of empirical spectroscopic constants provide very accurate description of the experimental spectra (usually recorded at room temperature) but with rather limited extrapolation power (e.g. to higher temperatures and especially to higher energies). These methods usually benefit from an extensive rotational coverage of individual vibrational bands in a given spectroscopic regions and thus are ideally suitable for the analysis of the FTS-like broad band techniques, as mentioned above. 
The so-called empirically adjusted \textit{ab initio} techniques (employed by ExoMol, ThoReTS and NASA Ames) are based on the empirical refinement of the underlying inter-nuclear \textit{ab initio} spectroscopic properties, such e.g. potential energy surfaces (PES). These methods hugely benefit from analysed laboratory data (line positions and energies), ideally with an extensive coverage of the vibrational excitations at the lowest rotational excitations to minimise the coupling between the rotational and vibrational degrees of freedom. Put simply, experimental information on vibrations is especially important to increase the quality of the extrapolation of the empirical model, while the rotational information is more useful for interpolation and benchmarking of the models. These type of empirical  techniques are less accurate than effective Hamiltonians when it comes to matching the experiment, but are very powerful when extrapolating to higher temperatures and even higher energies. 

The MARVEL (measured active rotation-vibration energy levels)~\citep{2007JMoSp.245..115F} process involves analysing high-accuracy laboratory measured spectra to get out experimentally-determined energy levels. These energy levels are then used in theoretically computed line lists to improve the accuracy of all transitions which involve those levels. In the case of the ExoMol database, the MARVEL process has been applied to molecular line lists for species such as C$_2$H$_2$~\citep{jt705}, TiO~\citep{mckemmish2017marvel}, H$_2$CS~\citep{jt886}, C$_2$~\citep{16FuSzCs.C2,20McSyBo}, CaOH~\citep{20YiOwTe}, H$_2$S~\citep{18ChNaKe}, NH$_3$~\citep{15AlFuTe.NH3,jt784}, AlO~\citep{21BoShYu}, H$_2$CO~\citep{21AfTeYu,GermannJQSRT2024}, H$_3^+$ \citep{jt869}, OCS \citep{jt916} as well as ongoing efforts to refine the line list of PH$_3$.
Due to new laboratory data becoming available, the MARVEL process for improving the accuracy of theoretial line lists need to be assessed periodically. The opacities computed from line lists therefore also ideally need recomputing periodically, as the accuracy and coverage of line lists improve (see Section~\ref{sec:opacities}). 

Regardless of the method used to improve the accuracy of theoretically computed line lists, their key common feature is the strong dependence on experimental laboratory data. Even the best modern \textit{ab initio} methods, except perhaps of very limited systems, do not have the required quality to be used in atmospheric applications and need to be either completely constructed from scratch, such as for the Effective Hamiltonians, by fitting the corresponding expansion, so-called spectroscopic, parameters or refined, such as for the \textit{ab initio} methods, by fitting the corresponding analytic descriptions of the spectroscopic models to the experimental data. 

Our general appeal to the experimental spectroscopists who would like to engage with exoplanetary atmospheric studies is to talk to the atmospheric modelers about their direct needs in terms of the species, temperature and wavelength coverage and also talk to the line list providers about their needs in terms of the vibrational and rotational coverage. 

\subsection{Data used by retrieval codes}

Owing to the importance of the line lists for atmospheric retrievals, it would not be an exaggeration to say that all, or close to all, the line lists 
discussed throughout this section, have already been or are in the process of being included in the main (exo-)planetary retrieval frameworks. 
It is the usual practice of exoplanetary atmospheric modelers to implement new line lists or important updates into their retrieval programs periodically; this usually involves building molecular opacities in the form of cross-sections or k-tables. Details on opacities used in various retrieval codes can be found in Section~\ref{sec:opacities}.

\subsection{What's being worked on?}

\begin{enumerate}
  \item \textbf{ExoMol:} at the time of writing, the following line lists are in  progress by the ExoMol project (see Section~\ref{sec:ExoMol}), either as new species or improvements/updates to existing line lists; diatomics:  OH, PN, CO, CH, CS,  OH$^+$, NO$^+$, NiH, CrH, AlF, BH, ZrS; triatomics in IR: HCN, HOD; triatomics (electronic): HCN; larger polyatomic (IR): C$_2$H$_2$, C$_2$H$_6$. MARVEL studies for N$_2$O isotopologues (see \citet{jt908} for $^{14}$N$_2^{16}$O), CO$_2$ (see \citet{jt925,jt932}, CO, HCN/HNC, HCO$^+$, BH, CS$_2$, C$_2$H$_4$  and $^{15}$NH$_3$ are being performed as well as updates to  $^{14}$NH$_3$, HCCH, and inevitably, water. 
  \item \textbf{TheoReTS:} line lists for several species including CH$_2$, CH$_3$, NH$_3$, H$_2$CCN, H$_2$NCN, CH$_3$Cl, C$_3$H$_4$, H$_2$O$_2$, CCl$_4$, CH$_3$CN or C$_2$H$_6$ are being worked on, and will be gradually uploaded to the TheoReTS database (see Section~\ref{sec:theorets}). 
\item \textbf{CaSDa:} the CaSDa database is working on producing new IR data for the existing species listed in Table~\ref{tab:casda}, including new bands and new isotopologues (e.g. for CH$_3$D). They also plan to develop new line lists for species such as SiH$_4$ and C$_3$H$_6$O$_3$ (see Section~\ref{sec:casda}). 
\item \textbf{NASA Ames:} line lists for sulphur-bearing species OCS and CS$_2$ (see Section~\ref{sec:nasa_ames}). 
\item \textbf{HITRAN and HITEMP:} Work toward the next edition of the HITRAN database is underway, which will include works highlighted in the HITRAN2020 paper \citep{2022JQSRT.27707949G} and additional recent literature. Examples include expanding the 3800-4800 cm$^{-1}$ region of NH$_{3}$, the addition of new molecules such as S$_2$ \citep{24GoHaGo.S2}.
\item \textbf{Non-LTE conditions:} although most studies of exoplanet atmospheres assume species to be in local thermodynamic equilibrium (LTE), effects arising from non-LTE conditions are known to be important in Earth’s upper atmosphere, as well as in the atmospheres of other Solar System planets. In the most accurate treatment, non-LTE radiative transfer calculations determine the populations of all the atomic and molecular states involved by considering the relevant species' rates with respect to radiative decay and collisional excitation, de-excitation as well as potentially ionisation and other processes~\citep{22WrWaYu}. However, complete data sets are rarely available for this purpose; even for the Non-LTE of H atoms, some of the collisional data are not so well known. In particular, data for collisions of H atoms with heavy particles (atoms/molecules) are lacking~\citep{19GaSc}. Studies such as \cite{22WrWaYu} are working towards improving data and models for non-LTE effects in exoplanet atmospheres.  In a recent work, \citet{24GaRaFa} also considered non-LTE effects and emphasised the importance of H$_2$O-H$_2$O excitation collisions, especially for the vibrational modes of the molecule.
\item \textbf{Vibrational modes of molecular clusters:} vibrational spectra from DFT calculations are in progress for species such as (TiO)$_n$ clusters up to $n$~=~10 and (SiO)$_n$ clusters up to $n$~=~20 (Lecoq-Molinos, in prep.).
\item \textbf{Quantum computing approaches:} \textit{ab initio} line list parameters are currently calculated using modern high-performance computers, particularly crucial for line lists containing millions to billions of transitions. 
However, the algorithms used can become computationally intensive and time consuming, scaling with the complexity of the molecule and temperature coverage.
Quantum computers could be an alternative solution, even though fault-tolerant quantum computers are unlikely to be available in the immediate future. Variational Quantum Algorithms (VQAs) and, in particular, Variational Quantum Eigensolvers (QVEs) are promising class of algorithms to estimate the eigenstates and the corresponding eigenvalues of a given Hamiltonian~\citep{2020arXiv201209265C,jt860}. This kind of technology demonstrates the potential, from a theoretical point of view, to describe the electronic structure of complex molecules faster. In the near future, a quantum speedup would be ideal to describe precise line lists~\citep{2023arXiv230507902I}.

\end{enumerate}

\subsection{Data needs: What's missing and urgent?}\label{sec:linelists_missing}

The main molecules expected to be observed by Ariel and JWST are the usual atmospheric suspects, H$_2$O, CO$_2$ and CH$_4$, as well as the recently detected SO$_2$. The line lists for these species are generally in a good shape both in terms of temperature completeness and in terms of accuracy, at least as far as space-based observations in the IR are concerned. This applies to other potentially detectable, common atmospheric absorbers such as NH$_3$, HCN, N$_2$O, NO, C$_2$H$_2$ etc. While the existing data for these species should be at a high level of readiness for the majority of low-to-medium resolution studies including JWST or Ariel, not all line lists extend to low enough wavelengths to cover the entire range observed by these telescopes - see, for example, the wavelength ranges in Tables 7~-~15 of \cite{20ChRoYu} for a number of high-temperature line lists and should be considered when interpreting observed spectra which includes a wavelength region not covered by a given line list. Not taking this into account could cause some biases in analyses which extend beyond the wavelength range of current line lists. Extending current line lists into the optical is therefore very important. Less mainstream or more exotic molecules are not at such a good readiness level, some of which we review in the following. The line lists for high resolution (HR) applications require even more work. With some exceptions, such as H$_2$O or CO$_2$ in the mid-IR, HR applications demand further improvement of the quality and quantity of the laboratory data at high temperatures for most of the molecules. The data needs discussed below are mostly for the IR and optical regions as directly imposed by to the observational and associated retrieval coverage of Ariel and JWST. However, the data needs in shorter wavelengths, UV and even far-UV, are dictated by the atmospheric chemistry models (see Section~\ref{sec:chemistry}).

\begin{enumerate}
  \item \textbf{Species occurring in JWST studies:} 
The recent potential detection of CS$_2$ in the exoplanetary atmosphere of TOI-270d \citep{24HoMa.exo,24BeRoCo} is an example of a molecule requiring improvement of the spectral coverage of associated molecular data, highlighting the need for line lists of sulphur species such as CS$_2$ up to high temperatures. The recent observation of SO$_2$ in the atmosphere of hot gas giant exoplanet WASP-39~b~\citep{23TsLePo.wasp39b,24PoFeLe} and in the atmosphere of warm Neptune WASP-107~b~\citep{24DyMiDe} prompted an increase in interest in sulphur species, with theoretical studies such as \cite{23JaWoHe} predicting the importance of them also in hot rocky exoplanet atmospheres. The SO$_2$ line list of \cite{jt635} is considered complete up to high temperatures of $\sim$2000~K in the wavelength range it covers, but only covers wavelengths down to 1.25~$\mu$m. This will lead to an underestimation of opacity in the wavelength region below 1.25~$\mu$m for model atmospheres including SO$_2$. The main reason for the lack of wavelength coverage of the line list is that the quantum chemistry computations used for computing a high-temperature line list can become extremely computationally extensive with increasing energy (decreasing wavelength), especially for molecules composed of more and heavier atoms. Considering the recent SO$_2$ detections in exoplanet atmospheres, having a high temperature line list extending to lower wavelengths is a high priority.

\item \textbf{Ti- and V-bearing species:} \cite{20HoSePi.hearts} predict significant concentrations of gas-phase TiO$_2$, VO$_2$, and TiS in the atmosphere of hot Jupiter exoplanet WASP-121b. These could be important absorbers, potentially accessible both by the high and low resolution observational methods, which currently do not have any available ro-vibrational (IR) or rovibronic (Vis and UV) line lists.
\item \textbf{Species relevant to lava planets:} recent studies suggested that modelling the atmospheric properties of hot-Super Earth (lava) exoplanets require different types species to be considered, including KOH, NaOH, SiO$_2$, CaOH, FeO, MgO, MgOH~\citep{12ScLoFe.exo,17TeYuxxi,17MaHeMi}. There have therefore been calls for line lists for these species spanning the IR and visible spectral region. We note these calls have been somewhat addressed and there are now available line lists from ExoMol for KOH and NaOH (IR) \citep{jt820}, SiO$_2$ (IR) \citep{jt797}, CaOH (IR, Vis) \citep{22OwMiYu}, MgO (IR, UV) \citep{jt759}. Theoretically, FeO is particularly challenging; there are more than fifty low-lying electronic states \citep{11SaMiMa,17TeYuxxi}. 
\item \textbf{Carbon and hydrocarbon species:} for example, C$_3$ which can be important for high-temperature carbon atmospheres~\citep{09ArGiNo}. An ExoMol IR line list for \ce{C3} has just been completed \citep{jt961}.
\cite{23MaZhDo} report recent laboratory work on the infrared spectra of C$_3$, with a combined fit of available infrared and optical laboratory data performed using the PGOPHER~\citep{PGOPHER} package; these and other data were used
as input for a MARVEL study \citep{jt915}.
Molecular line lists are required for the following hydrocarbon species, in IR for direct retrievals and at shorter wavelengths for atmospheric chemistry modellings (see more in Section \ref{sec:chemistry}). C$_4$H  is considered to be important in the upper atmosphere of low-metallicity objects~\citep{13BiRiHe.exo}. 
Allene (CH$_2$CHCH) and propyne (CH$_3$CCH)
are isomers of one another which have both been detected on Titan~\citep{19LoNiGr,13NiJeBe}. 
Propene has also been observed on Earth~\citep{96BlChSm}, and is expected on Saturn~\citep{00MoBeLe}.
There are some known cross-sections for propene~\citep{23BeDoZh}, including some hosted by the HITRAN database~\citep{18SuToDr}, but not a line list. 

\item \textbf{PAHs:} can be important for Earth and Titan-like atmospheres~\citep{13LoDiAd}. As previously mentioned, one of the main sources of PAH cross-sections is the NASA Ames PAH database~\citep{14BoBaRi,18BaRiBo,20MaHuBo}\footnote{\url{https://www.astrochem.org/pahdb/}}, which is tailored for characterising species in the cold temperatures of the ISM. To be able to fully comprehend the abundance and role of PAHs on exoplanets, there is a need for new PAH opacities to be calculated or measured over a broad parameter space of environmental conditions and easily implementable for retrieval applications. This necessity extends beyond the infrared spectrum, as the availability of temperature-dependent UV cross-sections is quite limited (see Section~\ref{sec:UV} for the availability of UV PAH photionisation and photoabsorption cross-sections).
\item \textbf{Species relevant to Archean atmospheres:} another special type of exoplanetary atmosphere with special data requirements are archean atmospheres, necessitating line lists for species including HC$_3$N \citep{19RiFeWa.exo,21RiMaPe}.

\item \textbf{Molecular ions:} \citet{jt793} performed laboratory experiments to help identify ions observable by Ariel in the upper atmospheres
of sub-Neptunes. Their results pointed towards H$_3$O$^+$ and H$_3^+$ being the most promising. Partly as a consequence of this study ExoMol provides IR
line lists for H$_3$O$^+$ \citep{jt805} and H$_3^+$ \citep{jt666,jt890}. The ExoMol database also provides
line lists for a number of other ions including CH$^+$ \citep{jt913}, OH$^+$ \cite{17HoBexx.OH+,18HoBiBe.OH+} and HeH$^+$ \citep{19AmDiJo} covering a very broad wavelength range. Other potentially important ions for which there are no suitable line lists at present include HCO$^+$, N$_2$H$^+$ and  H$_2$O$^+$, as well as NO$^+$ which is the dominant molecular ion in the Earth's ionosphere. 
\item \textbf{Radicals:} such as OH, HO$_2$, C$_2$H, CH$_2$, C$_3$H$_3$, NH$_2$, N$_2$H$_3$ are known to be important on Earth~\citep{16SePa}, Mars~\citep{72McDo}, Titan~\citep{04WiAt}, and other planetary atmospheres~\citep{17CaKa}. Out of these species, the required molecular data are only in a good shape for OH. There are high quality (accurate and complete) rovibronic data for OH applicable for IR and UV studies \citep{MOLLIST,jt933}. Construction of an IR line list for \ce{CH2} is in progress as part of  TheoReTs.
\item \textbf{Line lists are needed} extending into the UV and optical. Extension down to the UV is needed for all major species such as, H$_2$O, CO$_2$, CO, OH, CH$_4$, HCN, N$_2$, NH$_3$, C$_2$H$_2$, C$_2$H$_4$, C$_2$H$_6$, H$_3^+$, H$_2^{+}$, H$_3$O$^+$, OH$^+$, SO, SH, H$_2$S, SO$_2$, OCS, S$_2$, CS. See Section~\ref{sec:UV} for further discussion on UV photoabsorption and photodissociation data. In some of these cases (such as HCN, NH$_3$, C$_2$H$_2$, C$_2$H$_4$, C$_2$H$_6$, H$_3$O$^+$, H$_2$S, SO$_2$, OCS), data in the optical as well as the UV is also lacking or only partially covered for line lists which are appropriate up to high temperatures. Extending even just down to the optical region is even more pressing for direct use of these line lists in interpreting and modelling Ariel and JWST observations. We note that spectroscopic measurements of excited states of some of these species in the UV region are available in the literature, for example for NH$_3$ from \cite{98LaOrMo}. See Section~\ref{sec:MARVEL} for a discussion on how measured data such as these can be used to compute an accurate and complete line list up to high temperatures.
\item \textbf{Isotopologues:} Very recently \citet{lew2024highprecision} used JWST to detect isotopologues of CO. Similar studies have been performed on cool stars
from the ground \citep{jt777,jt799}; such isotope-resolved studies have the potential to provide a lot of extra information about atmospheres. Some atoms have more than
one isotope with relatively large terrestrial  abundances, e.g. Cl is 75\% $^{35}$Cl and 25\% $^{37}$Cl. In other cases, isotopic abundances can be  very different from Solar abundances, e.g. Mg and C. The importance of H means that even if deuterium (D) is a minor fraction, D containing molecules such  HD, HOD and CH$_3$D can be highly abundant.
We note that all these species have different symmetry to the parent (unsubstituted) molecule which means that the spectra of the two species differ
significantly. The ExoMol data base provides line lists for HOD \citep{jt469} and the weak IR spectrum of HD \citep{19AmDiJo}, while  CH$_3$D has been considered by TheoReTs \citep{2014Rey}. Lines for isotopically substituted species which do not break their symmetry are straightforward to compute given a spectroscopic model for the parent molecule, and this done routinely by ExoMol; methods to improve the accuracy of these line lists have been developed \citep{jt665,jt948}.
 A list of natural (terrestrial) abundances of various species included in the HITRAN database can be found online\footnote{\url{https://hitran.org/docs/iso-meta/}}, as well as elemental isotopic compositions from NIST\footnote{\url{http://physics.nist.gov/Comp}}. 
  \item \textbf{Vibrational modes of molecular clusters:} \cite{21KoHeBo} point out that the onset of cloud formation may be observable through vibrational bands of (TiO$_2$)$_4$, (TiO$_2$)$_5$, and (TiO$_2$)$_6$, potentially via the Mid-Infrared Instrument on JWST or the Extremely Large Telescope’s mid-IR imager, but more complete line-list data are required first. We note that the full quantum chemical computation of line lists for such large clusters is challenging, and so may require computations of vibrational frequencies only, at least as a starting point. There is some discussion on this in Section~\ref{sec:vib_freq}. We note some vibrational data for TiO$_2$ nanoclusters have recently been computed~\citep{23SiHeGo}.
\end{enumerate}

\section{Line shapes: molecular and atomic}\label{sec:lineshapes}

 \subsection{Theoretical approaches to line-shape parameters: Molecules}\label{4-1-2}

\textbf{The Voigt profile.}  In order to simulate a spectrum of a single molecule at given pressure and temperature, three main components are required; line position, line intensity and line profile. In this section, the latter is reviewed. The most popular line profile used in the majority of atmospheric applications (and beyond) is the Voigt profile. It refers to the so-called ``intermediate'' pressure regime and accounts for collision (pressure) and Doppler (thermal motion) broadenings considered as statistically independent processes and has been the standard for high-resolution line-by-line modelling of infrared molecular absorption \citep{VOIGT}. The Voigt profile represents a convolution of the associated Lorentzian and Gaussian profiles, representing the collisional and Doppler effects, and tends to these two shapes in the limiting cases of high and low gas densities, respectively. The Voigt profile has no closed analytical form and is often related to the complex probability function. Several special numerical algorithms exist in the literature for efficient evaluation of the Voigt profile~\citep{Humblicek82, Kuntz97}, see the recent computational and numerical
comparison by \citet{jt914}; some numerical approaches allow one to take advantage of the convolution of the entire spectrum at once \citep{21VaPa}.
While for the Doppler contribution, the corresponding (half) line width $\gamma_{\rm D}$ is a simple function of the molecular mass, gas temperature and the line positions (frequency):
\begin{equation} 
\label{eq:alpha}
    \gamma_{\rm D}=\sqrt{\frac{2N_Ak_BT\ln{2}}{M}}\frac{\tilde{\nu}_{fi}}{c},
\end{equation}
there is no simple analytic description of the line (half) width $\gamma_{\rm L}$, defining the Lorentzian contribution, at least in the general case. The molecular parameters $\gamma_{\rm L}$ strongly depend on the system (the molecule and the perturbing partner), as well as on the gas conditions, i.e. pressure and temperature. Together with the line positions $\tilde\nu$, transition probabilities (e.g. oscillator strength $gf$ or Einstein $A$ coefficients), the Lorentzian width (usually half-width-at-maximum, or HWHM) $\gamma_{\rm L}$ are an integral part of spectroscopic databases. They are typically obtained empirically either through direct measurements with different level of parametric modelling or semi-empirically, where a more sophisticated theory tuned to available experimental data is involved such as, e.g., the semi-empirical approaches \citep{Bykov2004} representing a simplification of the Robert-Bonamy formulae to Anderson-type expressions with fitted correction-factor parameters or modified complex Robert-Bonamy calculations \citep{Ma2007} with adjusted atom-atom interaction parameters. Application of fully first-principles methods are currently limited to simple system or low temperatures \citep{21WcThSt}.

The pressure-broadening of spectral lines corresponding to pure {\it rotational} (microwave, MW) and {\it rovibrational} (infrared, IR) transitions is mainly governed by inelastic collisions \citep{DeLucia88} and exhibit (except for light perturbers such as hydrogen and helium) strongly pronounced dependence on the rotational quantum number $J$.

Apart from the line broadening effect, collisions also affect the line positions by introducing the so-called line shift $\delta$. These are also provided by the spectroscopic databases, especially those focusing on high resolution applications, such as terrestrial, planetary \citep{05SuVa}, Solar and in some cases stellar, but so far have not featured very much 
in the low-mid resolution spectroscopy of exoplanets, as would be applicable to the Ariel mission. It is also noteworthy that ro-vibrational line shifts are typically smaller than the associated line widths by some orders of magnitude.

In the context of the Ariel mission, the temperature and pressure dependencies of line widths are crucial. For Voigt line widths $\gamma$ (in cm$^{-1}$atm$^{-1}$), the following power law is almost always assumed:
\begin{equation}
\label{gammaPL}
\gamma(T)=\gamma_0(T_{\rm ref}, P_{\rm ref}) \left(\frac{T_{\rm ref}}{T}\right)^n \frac{P}{P_{\rm ref}},
\end{equation}
where $T_{\rm ref}$ and $P_{\rm ref}$ are some reference temperature and pressure (e.g. 1 atm 
and $296$~K, respectively). Then the parameters $\gamma_0$ (reference value of HWHM) and $n$ (temperature exponents) are tabulated by the databases for each perturber. 
This model works quite well for many molecular systems but for a given set of parameters it can deviate when a description of a large range of temperatures is required.

The Lorentzian (and therefore Voigt)  as an isolated line profile model is known to fail in the far wings, always leading to an overestimation of the opacity at high pressure, especially for the systems with closely spaced lines \citep[Chapter IV]{08HaBoRo_book.broad}, e.g. in heavier molecules.
As a work-around, a cut-off the far wings at some distance from the centre of the line is applied. The common standards are to use a 25~\cm\ cut-off, employed by HITRAN, ExoMol and now also by \textsc{MAESTRO} (Molecules and Atoms in Exoplanet Science: Tools and Resources
for Opacities)~\citep{24GhBaCh}. A detailed analysis of line-wing profiles and cutoffs in the context of exoplanet atmospheres, including the Voigt profile, can be found in \citet{24GhBaCh}.

\textbf{Beyond the Voigt line profile.} Recent experimental studies powered by the development of experimental techniques in the last few decades have demonstrated the importance of the departure of the line shape description from the simple Voigt profile model, at least for high resolution applications. This motivated the spectroscopic databases also to provide parameters for more accurate profile models. One of such departures is due to the effect of collisional (Dicke) line narrowing. 

The line narrowing can be described either as a reduction of the Doppler component (velocity-changing collisions resulting in molecular confinement) or the dependence of the collisional relaxation rates on the relative molecular speed. In the first case, velocity-changing collisions are often considered as strong (no memory effects, Maxwell-Boltzmann re-distribution of velocities after each collision) and the most frequently used associated profile is that of \cite{Rautian66,Rautian67}, although the soft-collision model (light perturbers, many collisions required to modify significantly the active-molecule velocity) developed by \cite{Galatry61} is used by some researchers too. It is now understood that both mechanisms can be important across any pressure regimes, from very low to very high \citep{02DELeRo}.

In the second case, the speed-dependence (SD) of the collisional relaxation rates dominates the line narrowing and the corresponding models forming the so-called Speed-Dependent Voigt (SDV) profiles. The most accurate description of the dependence on the absolute absorber's speed \citep{Berman72,Pickett80} is by the hypergeometric confluent function (hgSDV model), which however, leads to high computational costs. As a more practical approach, a modification of hgSDV based on the mean quadratic speed \citep{Rohart94,Kohler95,Rohart2007} is used (qSDV model, often referred to as simply SDV), for which an efficient numerical-calculation algorithm is available \citep{Boone2007}.

Both mechanisms, with the velocity-changing collisions and with the speed dependence viewed as independent, are accounted for in ``combined'' line-shape models, such as Speed-Dependent Rautian (SDR) or Speed-Dependent Galatry (SDG) profiles. The partial correlation of these effects is considered by the partially Correlated quadratic-Speed-Dependent Hard-Collision profile (pCqSDHCP), developed by \citet{13TrNgHa} (called the Hartmann-Tran profile or HTP profile) and recommended for high resolution spectroscopy in a IUPAC Technical Report \citep{14TeBeCa}. Parameters for this advanced profile have been incorporated into the HITRAN database \citep{2022JQSRT.27707949G} since 2016 for the hydrogen molecule as well as some transitions of water vapor and molecular oxygen. The line narrowing effects, being routinely used in high-resolution terrestrial and planetary applications, have not been considered in the mainstream exoplanetary spectroscopic studies yet, and likely first requires studies to demonstrate the impact of such line profiles on different atmospheric models. 

Another important deficiency of the common line profile methodology used in spectral modelling is due to the breakdown of the isolated lines approximation. This phenomenon, also caused by molecular collisions, is associated with the so-called line-mixing or line-interference effect. While the line narrowing effects are mostly noticeable at lower pressure (well below 1~atm for most gases), line-mixing usually becomes relevant at higher pressure when individual lines start to broaden and overlap. Line-mixing effects can be especailly significant in the line wings regions where intensity is never truly zero, and there is always some degree of overlapping with others lines. Because of this effect, the intensity in the wings of a band falls off faster than predicted by the sum of Lorentzians~\citep[see figures in][]{2010HartmannBoTr}. To a lesser extent, the troughs between individual lines or branches can also be affected. In practice, as stated above, cross-sections calculations employ sums of Voigt profiles with certain wing cut-offs, say $\pm 25$~\cm\ from line centres. Such treatment bypasses a proper account of line-mixing and prevents overestimation of absorption away from band centres which would come from addition of billions of line wings (recall that the wings of the Voigt profile fall off as Lorentzians $\propto \omega^{-2}$ and not exponentially). However, such cut-offs may lead to another non-physical situation, where the calculated cross-sections between strong bands end up being exactly zero. In reality, line-mixing would cause a sub-Lorentzian shape that far enough from strong absorption bands would contribute to continuum absorption (see Section~\ref{sec:CIA} for a discussion on continuum absorption). An in-depth discussion of wing cut-offs is presented in \cite{24GhBaCh}. An empirical method of far-wing corrections involving $\chi$-factors is discussed in the following sections.

At higher pressure where even line cores significantly overlap, line-mixing can lead to a noticeable intensity redistribution within a branch or an entire band. This is known to happen for closely spaced lines in Q-branches and band heads even at sub-atmospheric pressures, and these features may appear stronger and more narrow than one would expect from the sum of isolated lines. This presents potential interest for more reliable confirmation of molecular species' detection, where features like these are often the most noticeable. The case of very high pressure has been considered before in order to describe the deep venusian atmosphere~\citep{2013FilippovAsSi,2013BuldyrevaGeFi}. In such conditions, the rotational structure is entirely collapsed and the band shape differs significantly from the sum of individual spectral lines. This is relevant to modelling of radiative transfer in hot dense atmospheres.

Proper analysis of line-mixing is complicated, and usually a perturbative approach is taken on a band-by-band basis for the most intense bands. 
At present, line-mixing is not typically implemented in suites for producing opacities for exoplanetary atmospheric retrievals, although these effects may become more important with newer instruments that can probe high-pressure regions of around 1~-~10~bar.

\textbf{Calculations of line widths and shifts.} 

The needs of MW/IR pressure-broadened line widths and pressure-induced line shifts for atmospheric applications motivated an extensive development of theoretical methods. The most developed and commonly accepted hypothesis of the impact approximation completed by various choices for the translational dynamics gave birth to semi-classical \citep{Anderson49,Tsao62}, classical \citep{Gordon66}, and quantum-mechanical \citep{Baranger58,Shafer73} approaches. Besides, there are statistical approaches, e.g. \cite{Margenau35}, whose two-particle static limit is the opposite case to the impact approximation. Static limit works well for low molecular velocities (negligible rate of change of intermolecular interactions) and is applicable at large frequency detunings, i.e. in the far spectral wings which are hardly observable at IR wavelegths. Therefore, approaches based on the impact approximation remain major tools for predicting collisional line-shape parameters. High computational cost and absence of refined potential-energy surfaces (PES) set limits to the use of quantum-mechanical methods as soon as polyatomic/``exotic'' molecular pairs and/or high temperatures are concerned.

\textbf{Quantum-mechanical approaches.} The most rigorous method, close coupling (CC), is usually applied to predict parameters of purely rotational transitions at room temperature, and only some studies address hot environments. Such data, provided the vibrational contribution to line-broadening is small, can technically be reused for other lines in the infrared. Certain simplifications of this formalism, such as the coupled-states (CS) approximation~\citep[used, for example, in][]{11ThIvBu}, the infinite-order sudden (IOS) approximation~\citep{77GoGrKo}, and the nearest neighbour Coriolis coupling (NNCC)~\citep{18YaHuZh,22SeVaGr} can make calculations more manageable while preserving the accuracy in the high-temperature limit for most lines in the band. The main issue of quantum-mechanical approaches at higher temperatures is that the number of rotational functions needed to construct a complete basis grows too fast, making calculations of the scattering matrix impractical. A useful feature of quantum-mechanical approaches is that they allow estimating parameters of beyond-Voigt profiles mentioned above. Semiclassical approaches can provide the Lorentzian component of the profile, and with some extra work, the speed-dependence parameters. If all the parameters of the extended Hartmann-Tran profile are required, one could employ the generalised Hess approach along with CC for the scattering matrix computation. This has been done recently for H$_2$ and HD with helium as the perturber~\citep{21WcThSt,21StStJo} up to 1000~K.

\textbf{Semi-classical approaches.} 
The need for accurate potential energy surfaces (PES) and long-time statistics accumulation also disadvantage purely classical computations. 
As a result, advanced semi-classical approaches are typically employed, such as the formalism of Robert and Bonamy \citep{rb79} and its fully complex implementation \citep{Lynch96} further supplied by exact trajectories \citep{Buldyreva99}, a corrected average over the perturber's states \citep{Antony2006,Ma2007} and an account for line coupling \citep{Ma2013}. In cases where large volumes of line-shape data are to be produced with a minimal CPU cost, the semi-empirical method \citep{Bykov2004} is used which represents a simplification of the Robert-Bonamy expressions to the Anderson-Tsao-Curnutte forms with a correction factor with empirically fitted parameters. Although some experimental values are necessary, the adjusted model parameters are practically temperature-independent and can be used for calculations at other required temperatures (not too low to keep the validity of the trajectory notion).

\textbf{Machine Learning.} Given the high volume of line broadening data that is required to model the effects of various different exoplanet compositions, machine learning appears to provide a practical approach to the problem. Recently \citet{jt919} have completed an initial machine learning study of air-broadening based on data in the HITRAN database. Studies which look at the effect of changing the broadener are underway.

\textbf{UV and visible frequency ranges.} Spectra which involve transitions between electronic states (i.e. rotation-vibration-electronic or rovibronic transitions), typically between the ground and an excited electronic state, occur, with some exceptions, in the UV and/or visible region. In these cases, the line broadening due to predissociative effects can dominate the collisional (pressure) broadening by some orders of magnitude. These broadening are caused by the associated lifetimes of \textit{quasi-bound} transitions decreased. A new data format has recently been proposed to capture~\citep{23TePeZh} and use \citep{23YuSzHa} the predissociative line broadening in spectroscopic databases.

In the case of rovibronic transitions between \textit{bound} low-lying vibrational levels of electronic states, available measurements show \textit{no} well pronounced $J$-dependencies in the collisional broadening of rovibronic transitions. The absence of rotational dependence of the collisional (pressure) broadening means that rotational energy transfer is of minor importance and the broadening parameters are mainly influenced by elastic collisions. For such transitions line widths and line shifts have the same orders of magnitude. As mentioned by \cite{Margenau36}, lines of rovibronic (bound-bound) transitions are broadened and shifted similarly to atomic lines of a similar type and a similar energy jump. Indeed, the first theoretical approaches to pressure-induced line widths/shifts used the model of a classical oscillating radiator whose phase is modified during collision by a random amount, leading thus, via Fourier-integral analysis, to a ``dispersion'' (Lorentz) profile. For example, \cite{Lindholm45} and \cite{Foley46} successfully developed such theories. \cite{Foley46}, in particular, argued that non-adiabatic effects are small for atoms due to large gaps between their electronic levels but can occur for molecular IR transitions because of similar energy differences between the rotational levels of the radiator and perturber. Detailed analyses of some representative one-term interactions (dipole-dipole, quadrupole-quadrupole and dispersive) in the framework of phase-shift theory were made by \cite{Mizushima51} who also considered velocity averaging and derived line-width and line-shift expressions (intended for microwave transitions) with explicit temperature dependencies. These expressions served as a basis for the power law of Eq.~(\ref{gammaPL}). Later, \citet{Hindmarsh67} accounted additionally for short-range repulsive forces via a Lennard-Jones 12-6 potential model but for simplicity conducted their derivation with the mean-thermal-velocity approximation. Temperature dependence of collisional broadening and shifts of atomic and molecular rovibronic lines has been also addressed by \citet{Cybulski2013}. Quantum-mechanical adiabatic approaches were initiated by \citet{Jablonski37,Jablonski45} who assumed the gas to be a very large ``molecule'' and calculated the stationary states of internal motion. A unified model relying on the Franck-Condon principle and containing both impact and quasistatic limits was proposed by \citet{Szudy1975}. The authors retrieved the expressions of \citet{Lindholm45} and \cite{Foley46} in the classical limit and suggested also a first-order correcting term leading to line shapes with an asymmetry effect.

When there is an interaction with one or more dissociative states, i.e., predissociations, the broadening can vary significantly  due to a variation in the lifetimes of the excited states. The broadening in this case is a manifestation of such interaction with the continuum. The extent of the line broadening (or lifetime of the state in question) derives from the strength of the predissociation, i.e., the strength of the coupling between the bound excited states and the dissociative states. Typically non-radiative lifetimes of electronically excited states are ca. 10$^{-8}$s which result in widths of $5\times10^{-4}$~\cm. For lifetimes of $10^{-13}$ s the linewidth will be 50 \cm\ which will result in the rotational structure disappearing. For nonradiative lifetimes shorter than $10^{-15}$ s, the vibrational structure disappears and will resemble a continuum. These types of predissociation-induced line broadenings have been well documented experimentally for over half a century (see chapter 7.4.2 in \citet{04LeFe}).

\subsection{Theoretical approaches to line-shape parameters: Atoms}\label{sec:theory_atoms}

The broadening of alkali lines by rare gases is a problem which has been extensively investigated in experimental and theoretical work \citep{allard1982,03BuVo.broad,16AlSpKi.broad,19AlSpLe.broad,20PeYuCh,23AlMyBl,24AlKiMy}.
A theory of spectral line broadening has been developed to calculate neutral atom spectra given the interaction and transition moments for relevant states of the radiating atom with other atoms in its environment. Within this framework it is possible to compute the complete spectrum with a unified approach. Unlike impact theories of
line broadening which predict a Lorentzian line or the
approximation methods of \cite{Szudy1975,Szudy1996} 
the unified theory of \cite{allard1999} provides an accurate spectrum from the line centre to
the extreme wing. Complete details and the derivation of the theory
are given by \cite{allard1999}. 
The approach is based on quantum theory of spectral line shapes by \cite{baranger1958a,baranger1958b} with an adiabatic representation to include the degeneracy of atomic levels.

 The spectrum $I(\Delta \omega)$ can be written as the Fourier transform (FT) of the dipole autocorrelation function $\Phi$(s),
 \begin{equation}
I(\Delta \omega)=\frac{1}{\pi} \, Re \, 
\int^{+\infty}_0\Phi(s)e^{-i\Delta  \omega s} ds \; ,
\label{eq:int}
 \end{equation}
where $\Delta \omega$ is the angular frequency difference from the unperturbed centre of the spectral line. The autocorrelation function $\Phi$(s) is calculated with the assumptions that the radiator is stationary in space, the perturbers are mutually independent, and in the adiabatic approach the interaction potentials give contributions that are scalar additive.
This last simplifying assumption allows 
the total profile $I(\Delta \omega)$ to be calculated when all the perturbers interact, as the FT of the $N^{\rm th}$ power of the autocorrelation function $\phi$(s) of a unique atom-perturber pair. Therefore,
\begin{equation}
\Phi(s)=(\phi(s))^{N}\; ,
\end{equation}
that is to say, 
the interperturber correlations are neglected.
For a perturber density $n_p$,
\begin{equation}
  \Phi(s) = e^{-n_{p}g(s)},
\label{eq:intg}  
\end{equation}
 where decay of the autocorrelation function with time leads to atomic line broadening.
When $n_p$ is high, the spectrum is evaluated by computing the
FT of Eq.~(\ref{eq:intg}).
The real part of $n_pg(s)$ damps $\Phi(s)$ for large $s$ but this calculation is not feasible when extended wings have to be computed at low density because of the very slow decrease of the autocorrelation function.
An alternative is to use the expansion of the spectrum $I(\Delta \omega)$ in powers of the density described in \cite{royer1971}.
For the implementation of alkali lines perturbed by helium and molecular hydrogen in atmosphere codes, the line opacity is calculated by splitting the profile into a core component described with a Lorentzian profile, and the line wings computed using an expansion of the autocorrelation function in powers of density.
The impact approximation determines the asymptotic 
behavior of the unified line shape correlation function. In this way the results are applicable to a more general line profile and opacity evaluation for the same perturbers at any given layer in the photosphere or planetary atmosphere.

In \cite{19AlSpLe.broad} the use of a density expansion in the opacity tables is reviewed.
When the expansion is stopped at the first order it is equivalent to the one-perturber approximation. Previous opacity tables were constructed to third order allowing us to obtain line profiles up to $N_{\rm He}$/$N_{{\rm H}_2}$=$10^{19}$ cm$^{-3}$. The new tables are constructed to a higher order allowing line profiles to $N_{\rm He}$/$N_{{\rm H}_2}$=$10^{21}$ cm$^{-3}$.

\subsection{State of the art - Data availability}

\subsubsection{Molecular line profiles}

\textbf{Voigt parameters:} it is now recognised in the (exo-)atmospheric community that Earth's conditions with the oxygen and nitrogen as the main collisional broadeners are not default ones, especially for hot Jupiter exoplanets, where hydrogen and helium dominate molecular collisions. Although, as noted below, broadening parameters are becoming available for a small number of systems which deviate from the Voigt profile, the vast majority of available data are for Voigt broadening parameters. Even so, collisional broadening parameters which are required for computing Voigt profiles are missing or incomplete for a larger number of the relevant molecules (see Section~\ref{sec:linelists}) and are therefore in practice substituted by crude ``guestimates'' when generating molecular opacities for the subsequent use in the radiative transfer calculations. Indeed, computing opacities without line profiles is meaningless, i.e. ``something'' is always better than ``nothing'', but the errors involved are usually difficult to quantify. The impact of using incorrect broadening parameters was shown by \cite{2022NatAs...6.1287N, 2023ApJ...950L..17N} to impact retrieved exoplanet properties.

HITRAN now includes different broadening species applicable to planetary atmospheres, as shown in Table~\ref{tab:add_broadening_HITRAN}. The addition of these parameters to the database is carried out by searching the literature for accurate experimental measurements, which can be fit to a rotationally-dependent function \citep[see][]{16WiGoKo,2019JGRD..12411580T,2022ApJS..262...40T}. This enables the functions to be applied to the high-temperature line lists of databases such as ExoMol and HITEMP. These broadeners are representative of the typical species in the (exo-)atmosphere under study: a mix of H$_2$ and He for gas giants, CO$_2$ for rocky planets and H$_2$O for tropical atmospheric scenarios (or water-dominated atmospheres). The temperature dependence of line broadening parameters has also recently been under investigation. 
Indeed line shapes are described by different parameters such as the pressure broadening and pressure shift coefficients and their respective dependence on temperature; this applies to both molecules and atoms. 
The species and broadeners for which Voigt broadening parameters are currently available in the ExoMol database can be found online\footnote{\url{https://exomol.com/data/data-types/broadening_coefficients/}}. 
A complete set of air-broadening parameters have been generated using machine learning \citep{jt919} and a large number of broadening parameters based on the theory of \cite{Buldyreva2022RASTI} have
also been added recently.

\textbf{Empirical $\chi$-factor corrections:} Solar System applications have shown that the Voigt profile, tending to a Lorentz profile with increasing pressure, is not applicable for high-pressure conditions where a sub-Lorentzian profile needs to be used~\citep{11BeFeJe}. 
The pressure is so high in the lower layers of the atmosphere that the line-interference (intensity-transfer, also called line mixing) effects, in particular line (and, therefore, band) wings, need to be accurately modeled. For now, the band shapes obtained in the impact-approximation (Markov) limit of collisional-broadening theory (which is invalid for large detunings from the band centre) are empirically corrected via so-called $\chi$-factors, as, e.g., \citet{96ToFiBe,11BeFeJe} for CO$_2$ on Venus and the  recent confirmation by laboratory measurements~\citep{22MoCaFl,23MoCaFl}. Empirical $\chi$-factor corrections were also used to model far-wing absorption
for CO$_2$ at high pressures and temperatures \citep{11TrBoSt,23CoTrHo}, CH$_4$~\citep{11TrBoSt} and H$_2$O~\citep{11TrBoSt}. The artificial $\chi$-factors as well as the forced re-normalisation procedure  in the impact-approximation models become unnecessary if a non-Markovian approach operating with a frequency-dependent relaxation matrix and a symmetric metric in the line space is employed \citep{ECS-IR,2013FilippovAsSi}, but the collision-induced absorption (while relevant) should be correctly modelled; see Section~\ref{sec:CIA}. Even for correctly modelling methane absorption in the cold ice giant atmospheres, for example, having more precise line profiles can be important.
 
 \textbf{Double-power law:}  \cite{18GaVixx} suggested to correct Eq.~(\ref{gammaPL}) by adding a supplementary power term (so-called double-power law), with an analogous two-term expression for the shifts, in order to model line widths and shifts over large temperature intervals with the same parameters' sets. So far, this has been applied to the H$_2$O-CO$_2$ collisional system, with respective  parameters published in \citet{19ReCoGa}, to CO$_2$-H$_2$O reported in~\citep{24ViGa}, and to He-broadening of H$_2$ and HD~\citep{21WcThSt,21StStJo}.

\textbf{The importance of laboratory measured broadening parameters:} as mentioned earlier in this section, accurate laboratory broadening measurements are critical for the production of rotationally-dependent broadening parameters, many of which have been compiled in the HITRAN database (see Table~\ref{tab:add_broadening_HITRAN}). 
A few examples of laboratory measurements of broadening coefficients to study Earth, Mars, Jupiter and other planetary compositions for key absorbers include \cite{smith2014air,sung2020Ch4-H2,devi2016spectral,2020Icar..33613452L}. Available measurements that are relevant to high-temperature atmospheres, are not all necessarily relevant to high-metallicity atmospheres \citep[e.g.][]{19GhHeBe, Yousefi2021CH4,baldacchini2001temperature, goldenstein2013diode}, or are only limited to a small spectral region. Further details can be found in \cite{Hartmann2018review}. Nevertheless, such measurements in the laboratory are an important component of compiling broadening parameters for a large range of species, spectral regions, broadening species, temperatures, and pressures. 

\subsubsection{Atomic line profiles}

\textbf{Na and K:} the resonance line doublets of atoms such as Na and K can have a large impact on the spectra of exoplanets and Brown Dwarfs, as demonstrated by, for example, \cite{19AlSpLe.broad} and \cite{23WhGlCh}. Such strong lines as these require detailed computations to model the line profile and wings far from the line centre. Typically works such as \cite{16AlSpKi.broad} for K perturbed by H$_2$\footnote{\url{https://cdsarc.u-strasbg.fr/viz-bin/qcat?J/A+A/589/A21}}, \cite{24AlKiMy} for K perturbed by He\footnote{\url{https://cdsarc.cds.unistra.fr/viz-bin/cat/J/A+A/683/A188}}, \cite{19AlSpLe.broad} for Na perturbed by H$_2$\footnote{\url{http://cdsarc.u-strasbg.fr/viz-bin/qcat?J/A+A/628/A120}}, \cite{23AlMyBl} for Na perturbed by He\footnote{\url{https://cdsarc.cds.unistra.fr/viz-bin/cat/J/A+A/674/A171}}, or \cite{03BuVo.broad} for Na and K perturbed by a mixture of H$_2$ and He, are used in the computation of opacities for exoplanet and Brown Dwarf atmospheres. 
See also \citet{07ShBuxx.dwarfs} for empirical estimates used for other atomic (and molecular) line broadening parameters. 

For the strongest atomic lines (K, Na, He) individual line absorption can dominate the entire observed spectroscopic region and very accurate shapes are mandatory, as described above. For others (Fe, Ca, Mg etc.), the atomic features are less intense and can be very dense, comparable to the molecular lines. In this case, the spectra are less sensitive to the individual lines. We do not go in detail here on these other species, and instead focus only on Na and K which are most relevant to high-temperature exoplanet atmospheres due to their extensive line wings.

\subsection{Data used by retrieval codes}

As previously highlighted in this section, the vast majority of atmospheric modelling and retrieval codes use the Voigt profile as the current standard for temperature- and pressure-broadened opacities. 

The broadening parameters used for computing opacities (either on-the-fly or pre-computed; see Section~\ref{sec:opacities})
will have an impact on the resulting radiative transfer models~\citep[e.g.][]{22AnChCh,19GhLi}. 
It is important to compute (or choose, if pre-computed) opacities which are relevant for the atmosphere which is being modelled, both in terms of temperature and pressure, and also what the main broadening species are (e.g. H$_2$/He-dominated, H$_2$O-dominated, CO$_2$-dominated atmospheres). Often broadening data are supplied alongside line lists, which may then be typically used in the computation of opacities. One of the main differences between opacity computations is the extent out to which the line wings are calculated from the line centre~\citep[see, for example,][]{24GhBaCh}.
The ExoMolOP database~\citep{20ChRoYu} employs the line profiles of \cite{16AlSpKi.broad} and \cite{19AlSpLe.broad} in computing the resonance line opacities of K and Na, respectively, broadened by H$_2$. For the molecular opacity Voigt line profiles in the ExoMolOP database, averaged-$J$ broadening parameters for H$_2$ and He broadening were used for those species with available data at the time, as outlined in Table~2 of \cite{20ChRoYu}. The Voigt line wings were computed up to 500 Voigt widths from the line centres, up to a maximum cutoff of 25~cm$^{-1}$. The EXOPLINES database~\citep{21GhIyLi} uses a Voigt profile with a variable line-wing cutoff depending on the pressure (30~cm$^{-1}$ for P~$\leq$~200~bar and 150~cm$^{-1}$ for P~$>$~200~bar), but note the lack of available broadening coefficients for the species they compute opacities for (including a number of metal hydrides). They therefore estimated parameters to use in a similar way to \cite{20ChRoYu}, based on the similarity between molecular symmetry or dipole moments that these absorbers have with the available Lorentz coefficients. 
 Some works, such as \cite{19GhLi} and \cite{22AnChCh} use opacities which were computed assuming H$_2$O-broadening parameters, extracted from the HITRAN2020 database~\citep{2022JQSRT.27707949G}. The opacities computed for the Helios-r2 retrieval code~\citep{20KiHeOr,21GrMaKi} are computed using a Voigt profile. The impact of the spectral line wing cut-off has been recently reviewed by \cite{24GhBaCh}, with the recommendation for the following line wing cut-offs to be used for current computations using the Voigt profile: 25~cm$^{-1}$ for P~$\leq$~200~bar and 100~cm$^{-1}$ for P~$>$~200~bar.

\subsection{What's being worked on?}

\textbf{Machine learning approaches:} so far, the theoretical methods used for line broadening are very limited in what they can realistically provide, in terms of the molecular species, spectral ranges and temperature coverage. 
 Machine learning (ML) represents an attractive alternative to fill the existing data gaps in broadening data due to its relative simplicity, which has been recently explored to produce air broadening data in the ExoMol data base \citep{jt919}. Considering the success of ML in a wide range of applications and the fast progress in the development of ML techniques, the work in this direction will continue.

\textbf{(Vib)rotational line-shape parameters:} 
the extreme conditions experienced by most observable exoplanets (high temperatures and huge levels of ionisation) lead to the formation of ``exotic'' species whose spectroscopic parameters are often impossible to measure in standard laboratory experiments. A significant number of required perturbers and non-existing empirical PESs to practice ``standard'' theoretical approaches described in Section~\ref{4-1-2}. Therefore, for a rough $J$-independent theoretical estimate for IR/MW line width (in cm$^{-1}$atm$^{-1}$), the following expression for the Lorentzian HWHM has been recently suggested~\citep{Buldyreva2022RASTI}:
\begin{equation}
\label{gammaRASTI}
\gamma_{\rm L} = 1.7796\cdot 10^{-5}\frac{m}{(m-2)}\frac{1}{\sqrt{T}}\sqrt{\frac{m_{a}+m_{p}}{m_{a}m_{p}}}d^{2},
\end{equation}
where only the index $m=2(q-1)$ of the leading long-range interaction $V \sim R^{-q}$, the active- and perturbing-molecule masses $m_a$, $m_p$ and the kinetic diameter of the molecular pair $d$~=~$(d_a+d_b)/ 2$ are required at the input temperature $T$. Besides the neutral active molecules, Eq.~(\ref{gammaRASTI}) holds also for ionic absorbers. It has been employed to produce 
line-width data \citep{jt957} for more than fifty active species (and their isotopologues) self-perturbed and perturbed by He, Ar, O$_2$, H$_2$, N$_2$, CO, CO$_2$, NO, H$_2$O, NH$_3$ and CH$_4$ at the reference temperature of 296~K. These new data are progressively integrated into ExoMol as new files of ExoMol Diet \citep{17BaHiCz} and into the definition .def files for all line lists of a considered molecule.

\textbf{Rovibronic line-shape parameters:} the same goal of providing at least approximate values of line-shape parameters for electronic transitions in hot-temperature active molecules concerned by Ariel has initiated recently a study \citep{Buldyreva2324} revisiting classical phase-shift theory. Besides a general analysis for arbitrary molecular pairs (including the points on the temperature dependence and Maxwell-Boltzmann averaging over relative velocities), 
improvements of the potential and trajectory descriptions were attempted. Detailed calculations, based on specially computed state-of-the-art PESs, were done for some representative molecular pairs (NO and OH perturbed by N$_2$ and Ar, selected because of available experimental data). As intended, for the considered molecular partners the calculated line-shape parameters were found to be consistent with measurements within 30--40\% in wide temperature ranges (300--3000~K). Although this phase-theory approach requires knowledge of the isotropic interaction PESs associated with the excited and ground radiator's states, it has a much lower computational cost with respect to quantum-mechanical methods which, in addition, become unpractical at high temperatures.

\subsection{Data needs: What's missing and urgent?}

In general, we would advocate for a consistent set of parameters which cover an adequate spectral range for each species of importance. This includes species which require new or improved line lists in Section~\ref{sec:linelists_missing}. The need for the accurate treatment of line broadening effects in atmospheric retrievals is known to play an important role in atmospheric models, and more so with emerging observational facilities, regardless of if they are space-based (lower resolution) such as JWST and Ariel or ground-based (high resolution) such as the VLT or ELT \citep{16HeMaxx.exo}. Studies such as \cite{19GhLi}, \cite{2022NatAs...6.1287N}, and \cite{22AnChCh} highlight the impact of using different broadening species for computing spectral line shapes on model atmospheric spectra, indicating that incorrect line shapes will likely lead to biases when analysing observed spectra. \cite{19GhLi} find that emission spectra are particularly affected by incorrect broadening parameters, with the differences observed highly dependent on the pressure-temperature structure of the atmosphere. There will be an impact on the radiative balance and thermal structure of the atmosphere.
In the frame of the Ariel mission, the requirement of a consistent set of parameters which cover an adequate spectral range for each species of importance is a challenging request as Ariel aims to observe about 1000 exoplanets ranging from Jupiters and Neptunes down to super-Earth size in the visible and the infrared with its meter-class telescope \citep{19EdMuTi}. This implies very different broadeners and pressure and temperature conditions. Line shapes accurate at high temperature, up to 2500~K, are high up on the list. This includes temperature exponents of the line parameters as temperature dependent line mixing effects for the main molecular species at least. The most important broadeners to be considered remain the usual suspects; H$_2$, He, CO$_2$ and H$_2$O. 

\begin{enumerate}
\item \textbf{Species occuring in JWST studies:} see the first point in Section~\ref{sec:linelists_missing}, with line broadening parameters also required up to high temperatures for the listed species.
  \item \textbf{Species relevant to lava planets:} the broadening for molecules important in lava planet environments require very different types of line-broadening to other commonly studied planets, which are generally not currently available~\citep[see, for example,][]{17TeYuxxi}. This includes species such as KOH, NaOH, SiO$_2$, CaOH, FeO, MgO, MgOH~\citep{12ScLoFe.exo,17TeYuxxi,17MaHeMi}, whose spectra will be broadened by exotic moleules.
  
   \item \textbf{H$_2$ and He broadeners at high temperatures:} there has been significant progress recently in the provision of line broadening parameters for H$_2$/He dominated atmospheres such as of hot Jupiters or hot Neptunes. Even so, it should be acknowledged that while many species in HITRAN have broadening parameters for terrestrial-relevant molecules, H$_{2}$ and He broadening parameters are not available for some important species such as CH$_4$.
   
   \item \textbf{N$_2$ broadening:} it is known from studies of the Solar System that broadening by N$_{2}$ is required for the remote sensing of Titan~\citep{22HeSeHo}. Currently, N$_{2}$ is not separated from ``air'' in HITRAN, but is included as a broadening gas for many absorption cross-sections included in HITRAN that have been observed in Titan's atmosphere.
  \item \textbf{Species relevant to heavy atmospheres:} the availability of broadening data with species such as CO$_2$ and H$_2$O as broadeners is also relatively limited (see e.g. Table~\ref{tab:add_broadening_HITRAN}). 
  \item \textbf{Rotational- and vibrational-dependence:} while the rotational ($J$)-dependence is often not known for many species~\citep[see, e.g.,][]{21GhIyLi}, the vibrational dependence is even less studied but generally considered 
  to be less important, see \citet{jt919} for example. 
  \item \textbf{Beyond Voigt profiles:} introducing a new line profile, such as Hartmann-Tran (see Section~\ref{4-1-2}), implies introducing new line parameters. This complicates the release of parameters and recipes to the community. Efforts are made, and for instance, HAPI\footnote{\url{https://hitran.org/hapi/}}~\citep{2016JQSRT.177...15K} enables any user to choose a set of adequate line parameters depending on the atmospheric conditions of their spectra. Nevertheless, issues arise when the available parameters do not entirely cover the required spectral range and different datasets need to be combined. This results in a lack of traceability and poor reliability on the retrievals' results. We would therefore advocate for a consistent set of parameters which cover an adequate spectral range. Note that the various representations of the profiles
  are not orthogonal fits so the parameters cannot be used reliably in other functional forms.
  \item \textbf{Line mixing:} A recent overview of line mixing, which causes potentially large deviations from a Voigt profile at high pressures, can be found in \cite{24GhBaCh}. We are not aware of any exoplanetary models which currently allow for the
effects of line mixing, but these effects may become more important with newer instruments that can probe high-pressure regions of around 1~-~10~bar. 
\item \textbf{Deuterated species:} deuterated molecular species are targeted in the (exo-)planetary community to infer planetary evolution. Typical targets for these kinds of investigation are hydrocarbons and water: C$_2$HD, CH$_3$D, and HDO~\citep{22ViLiAo} all have the property that they have different symmetry to their parent isotopologue. Broadening parameters for these species are important but the change of symmetry means the usual practice of transferring broadening parameters between isotopologues will not work for these cases.
\item \textbf{Atomic species:} line parameters for atomic species are generally not readily available (with the previously noted exceptions of the strong Na and K doublets broadened by H$_2$ and He), which can lead to uncertainties in the calculated cross-sections. 
As shown in \citet{21GrMaKi}, the use of a Voigt profile for atomic lines is not always suitable. For instance the wings of autoionisation lines 
are more appropriately described by a so-called Fano profile \citep{61Fano}, which decreases much faster than a Voigt profile. Using a Voigt profile for these lines will result in an overestimate of the absorption, creating an artificial continuum. Other problematic lines for atomic species are resonance lines, where using a normal Voigt profile can also lead to over- or underestimated absorption.
\item \textbf{Limitations of experimental studies on some species:} as noted in Section~\ref{sec:lab_high_temp}, experimental measurements can be particularly challenging for reactive (or toxic) species, especially at high temperatures, which can limit the availability of parameters. It is therefore not uncommon for some spectroscopic parameters, such as pressure-broadening \citep{2022ApJS..262...40T}, to be based only on limited laboratory measurements for some molecules. For example, inorganic salts are very difficult to sublimate and keep in the gas phase.  These molecules, where possible, could be targets of theoretical studies instead. 

The lack of the line shape data also means the lack of understanding of their importance on the atmospheric characterisation. More work, both experimental and theoretical is required.

\end{enumerate}

\section{Computed cross-sections and opacities}\label{sec:opacities}

Line lists (see Section~\ref{sec:linelists}) give all the information required (such as upper and lower energy levels, transitions, and typically either Einstein-A coefficients or intensities at a reference temperature) to generate a cross-section at a given temperature and pressure. Line lists will have a maximum temperature up to which they are considered complete; cross-sections computed above these temperatures will have opacity missing. This is because hot band transitions between energy levels of higher energy than are populated for the maximum line list temperature will not be taken into account. When computing cross-sections various choices need to be made, such as what line list, line profile, line-wing cutoff, and broadening parameters to use (see Sections~\ref{sec:linelists} and~\ref{sec:lineshapes}). These will partly be determined by the requirements of the cross-sections being computed: will they be used for modelling Earth-like planets, in which case a line list such as HITRAN might be the most appropriate, or hotter exoplanets, in which case a line list computed or measured up to high temperatures is more appropriate, such as those provided by databases such as HITEMP~\citep{2010JQSRT.111.2139R}, ExoMol~\citep{jt810}, or TheoReTs~\citep{TheoReTs}. If the cross-sections are for modelling or retrieving atmospheric properties of a hot gas giant exoplanet, then the atmospheric composition will be typically dominated by H$_2$ and He. The choice of broadening parameters in Section~\ref{sec:lineshapes} will thus be H$_2$ and He for such atmospheres, whereas for a planet with a heavier atmosphere, heavier broadening species such as H$_2$O or N$_2$ will be more appropriate. Partition functions are also required for computing cross-sections; these are typically included as part of a line list (see Section~\ref{sec:linelists}), with tools such as TIPS~\citep{2021JQSRT.27107713G} available for computing partition functions for the HITRAN2020 database. Studies such as \cite{2022NatAs...6.1287N} and \cite{2023ApJ...950L..17N} highlight how different choices, including line lists, broadening coefficients, and far-wing behaviours, can impact interpretations of observed spectra. \cite{2023ApJ...950L..17N} explore the impact of opacity choices on the atmospheric properties of the JWST NIRSpec/G395H transmission spectra of WASP-39~b~\citep{23AlWaAl}. Both \cite{2022NatAs...6.1287N} and \cite{2023ApJ...950L..17N} conclude that the incompleteness and inaccuracy of line lists is the biggest issue in the accurate retrieval of atmospheric properties from observations. \cite{2022NatAs...6.1287N} find strong biases due to incorrect line broadening assumptions, with the effects due to incomplete knowledge of far line wings greater for super-Earths than warm-Jupiters. They find the strongest biases, however, are due to using a line list which is incomplete at the atmospheric temperature being probed (i.e. using HITRAN instead of HITEMP or ExoMol for a hot Jupiter exoplanet atmosphere).\cite{24ViFaKo} present comparison tests designed for the exoplanetary community, in order to identify the effects of different input data choices, such as opacity format and line list source.

Opacity files are typically provided as either: 1) high-resolution cross-sections, 2) sampled cross-sections, or 3) correlated-k coefficients, also known as k-tables. The general process is to compute high-resolution cross-sections (typically with a grid spacing of around 0.1~cm$^{-1}$ or 0.01~cm$^{-1}$, or R~=~$\frac{\lambda}{\Delta\lambda}$~$\sim$~1,000,000) for a grid of temperatures and pressures (typically spanning pressures of at least $\sim$1$\times$10$^{-5}$~-~100~bar, and temperatures $\sim$100~K~-~3000~K), and then, usually for reasons of computational feasibility, to either sample these cross-sections at a lower resolution ($\frac{\lambda}{\Delta\lambda}$~$\sim$~10,000~-~20,000), or compute k-tables at $\frac{\lambda}{\Delta\lambda}$~$\sim$~300~-~5000, if the application is for lower-resolution applications such as characterising Ariel or lower-resolution JWST observations. These resolutions are approximately equivalent to one another for achieving the same accuracy when used in an atmospheric modelling or retrieval code. The general principle of k-tables is to order spectral lines within a given spectral bin, producing a smooth cumulative distribution function to represent opacity, which can then be more efficiently sampled than cross-sections being sampled without any ordering of the spectral lines, with less loss of opacity for the same resolving power~\citep[see, for example,][for further details]{17Min,20ChRoYu,21Leconte}. 

\textbf{Super-lines:} very large line lists can contain billions of lines between millions of energy level states, for example the latest ExoMol CH$_4$ line list consists of over 50~billion transitions between 9~million states~\citep{24YuOwKe}. The computation of opacities from such large line lists can benefit from techniques such as the super-lines method~\citep{TheoReTs,17YuAmTe,jt810}, which yields a vast improvement in computational speed when computing high-resolution cross-sections. Either line intensities can be computed at a given temperature, and then the intensities of all lines within a specified spectral bin summed together to create a so-called super-line. If a high enough resolving power, such as R~=~$\frac{\lambda}{\Delta\lambda}$~$\sim$~1,000,000, is used for the creation of the super-lines then the overall opacity will be conserved~\citep{17YuAmTe,jt810,20ChRoYu}. Alternatively, some line lists, such as the CH$_4$ line lists of ExoMol~\citep{17YuAmTe} and TheoReTs~\citep{14ReNiTy.CH4,TheoReTs}, are already partitioned into a large set of relatively weak lines and a small set of important, stronger lines. When computing opacities, the weaker lines can be combined into a set of super-lines (or effective lines as done for CH$_{4}$ in HITEMP \citep{HargreavesAJSS2020}), with broadening of the stronger lines treated individually. \cite{19IrBoBr}
employed a similar ``pseudo continuum'' approach when employing the NH$_3$ line list of \cite{jt771} to analyse the visible spectrum of Jupiter, finding a great increase in efficiency with very little loss of accuracy.

\subsection{State of the art - Data availability}

There are some programs which have been developed by line list producers to generate cross-section data from line lists, such as ExoCross\footnote{\url{https://github.com/Trovemaster/exocross}}~\citep{jt708} and pyExoCross\footnote{\url{https://github.com/ExoMol/PyExoCross}}~\citep{jt914} (ExoMol), and HAPI (HITRAN)\footnote{\url{https://hitran.org/hapi/}}~\citep{2016JQSRT.177...15K}. Other tools have been developed by line list users, such as HELIOS-K\footnote{\url{https://github.com/exoclime/HELIOS-K}}~\citep{15GrHexx}, Exo-k\footnote{\url{https://perso.astrophy.u-bordeaux.fr/~jleconte/exo_k-doc/index.html}}~\citep{21Leconte}, RADIS\footnote{\url{https://github.com/radis/}}~\citep{RADIS,21VaPa} and others~\citep{17Min}. Various databases are available which have made use of tools such as these to compute opacities. Each will have their own set of assumptions and choices used when computing the opacities, depending on the intended applications: 
some databases provide high resolution cross-sections and some sampled cross-sections or k-tables (see the introduction to this section), often formatted for direct use in a particular atmospheric retrieval or modelling code. Some codes, which employ a so-called line-by-line method, work directly from line lists, and so do not require these opacity files. In such an approach assumptions about line broadening are made by the user at the time of running the code and so it can be considered to be more flexible, however it can become computationally unfeasible for atmospheric retrievals which cover a large range of pressures, temperatures, and wavelengths. Table~\ref{tab:opacity_overview} gives a non-exhaustive list of opacity databases employed by typical codes used to compute models or run retrievals relevant for the Ariel space mission, along with other applications. Some additional notes on these databases are given in Section~\ref{sec:opacity_codes}. It is important, although not always the case, for modelling and retrieval studies to report on the opacities they use and either cite or explain the assumptions that went into computing those opacities. This allows for traceability and for analyses to be reproducible.

\subsection{Data used by retrieval codes}\label{sec:opacity_codes}
Molecular opacities (cross-sections and k-tables) are now included in many radiative transfer codes and associated libraries, such as those listed in Table~\ref{tab:retrieval_codes}. The opacity databases listed in Table~\ref{tab:opacity_overview} are available online and although some are formatted with specific retrieval or modelling codes in mind, they can generally be adapted for more general use or for different codes. 
ExoMolOP~\citep{20ChRoYu} has k-tables and cross-sections for currently around 80 species in the ExoMol database formatted for direct input into TauREx3~\citep{21AlChWa}, ARCiS~\citep{18OrMi.arcis,20MiOrCh.arcis}, NEMESIS~\citep{NEMESIS}, petitRADTRANS~\citep{19MoWaBo.petitRT} (see Table~\ref{tab:retrieval_codes}).
The EXOPLINES~\citep{21GhIyLi}, Gandhi2020~\citep{20GaBrYu}, MAESTRO~\citep{MAESTRO} and DACE~\citep{21GrMaKi} databases/compilations all contain high-resolution cross-sections for various pressures and temperatures. 
These can be sampled down to a smaller desired resolution for practical use, or converted to k-tables. Various tools currently exist for this, including Exo-k\footnote{\url{https://perso.astrophy.u-bordeaux.fr/~jleconte/exo_k-doc/index.html}}~\citep{21Leconte}, with more tools expected to be become openly available in the near future. We note there are various file formats used for storing opacity data which are often different for different retrieval codes; popular choices include HDF5~\citep{19MoWaBo.petitRT,21AlChWa}, pickle~\citep{21AlChWa}, or FITS~\citep{20MiOrCh.arcis}. 

The need to interpret observational spectra covering a wide range of atmospheric conditions robustly and benchmark them across various studies inspired the creation of the MAESTRO (Molecules and Atoms in Exoplanet Science: Tools and Resources for Opacities) database. MAESTRO is a NASA-supported, open opacity service accessible to the community through a web interface\footnote{\url{https://science.data.nasa.gov/opacities/}}. Primary aims of the MAESTRO project are to incorporate the most up-to-date line lists, establish community standards for computing opacity data~\citep[see, e.g.,][]{24GhBaCh}, and offer insights and guidance on assumptions and limitations regarding data relevancy and completeness, particularly concerning different atmospheric compositions. The overall goal is to create one uniform and standard database for opacities. MAESTRO currently hosts high-resolution  data from various sources such as EXOPLINES~\citep{21GhIyLi} and ExoMolOP~\citep{20ChRoYu}, which span a large range of pressures, temperatures, and wavelengths. The ExoMolOP data hosted by the MAESTRO database are the high-resolution cross-sections which were used to compute the k-table or sampled cross-section files mentioned above. There is an option for opacity-producers to submit data to MAESTRO via GitHub\footnote{\url{https://github.com/maestro-opacities/submit-data}}.

We note the importance of checking opacity files for their original line list source to check their accuracy, spectral coverage, and see what temperature they are suitable to be used up to; see Section~\ref{sec:linelists} for a discussion of completeness of the line lists for different species, which depends on the temperature up to which a line list is considered complete.

\begin{table*}
    \centering
     \caption{An overview of some typical opacity databases designed for the (exo)planetary atmosphere community. }
     \resizebox{\textwidth}{!}{
    \begin{tabular}{llll}
    \hline 
\rule{0pt}{3ex} Database & Types of data & Reference& Link\\
    \hline 
    \rule{0pt}{3ex}DACE & Many molecular and atomic cross-sections, 0.01~cm$^{-1}$ spacing & \cite{21GrMaKi}& \url{https://dace.unige.ch/opacityDatabase/}\\
\rule{0pt}{3ex}ExoMolOP & \rule{0pt}{3ex}Sampled cross-sections and k-tables for many molecules & \cite{20ChRoYu}& \url{https://exomol.com/data/data-types/opacity/}\\
\rule{0pt}{3ex}EXOPLINES &  High-resolution cross-sections for selected species & \cite{21GhIyLi}&  \url{https://doi.org/10.5281/zenodo.4458189}\\
\rule{0pt}{3ex}Gandhi2020& High-resolution cross-sections for 6 molecules,  0.01~cm$^{-1}$ spacing &  \cite{20GaBrYu}& \url{https://osf.io/mgnw5/?view_only=5d58b814328e4600862ccfae4720acc3}\\
\rule{0pt}{3ex}MAESTRO & High-resolution cross-sections from various sources & \cite{MAESTRO} & \url{https://science.data.nasa.gov/opacities/}\\
\hline
    \end{tabular}
   }
    \label{tab:opacity_overview}
\end{table*}

\subsection{What's being worked on?}

Opacities need to be updated periodically as new line lists become available or are updated; work is in progress to keep these databases up-to-date, including adding important isotopologues to ExoMolOP. The MAESTRO database is also expanding, with new species and isotopologues to be added in due course. 
  With existing instruments such as HST/UVIS and potential future instruments such as the Habitable Worlds Observatory (HWO), comes the requirement for opacity data which spans into the UV region, starting around 0.1~-~0.2$\mu$m. Work is ongoing to compute opacities down to these wavelengths for the species which already have some UV data available, and for future line lists which are in progress. Such opacities will be incorporated into databases such as ExoMolOP and MAESTRO.

\subsection{Data needs: What's missing and urgent?}

\begin{enumerate}
  \item \textbf{External updates and advances:} as mentioned at the start of this section, there are many variables and assumptions that are made when computing opacities. As new and improved data become available (line lists, broadening parameters, theoretical knowledge), then opacities will ideally need to be recomputed to keep to date with the latest advances. The requirements of new instruments will also have an impact here; opacities for use in characterising atmospheres observed with JWST, for example, need to be at higher resolution than those for use in characterising observations of the same planets using HST or Spitzer. Theoretical advances in line shape theory could also impact new opacities. As mentioned above, opacities which extend into the UV are also needed.
  \item \textbf{Heavy and other atmosphere applications:} a focus on heavier atmospheres will require opacities tailored differently (\emph{i.e.} different broadening parameters) to those for lighter gas giant atmospheres. In addition, stellar atmospheres also have their own requirements~\citep[e.g.][]{07CrStLe}.
  \item \textbf{Isotopologues:} although there are exceptions and updates are being made, opacities and line lists can be focused on the main isotopologues of a species only. For some species, however, isotopologues are more important than in others; for example natural elemental abundances of $^{79}$Br/$^{81}$Br and $^{35}$Cl/$^{37}$Cl require molecules containing these elements to be formed from cross-sections combined from a combination of the two main isotopologues. There is also some interest in observing different isotopologues in exoplanet (and brown dwarf) atmospheres~\citep{19MoSn.exo,23BaMoPa}, which require accurate opacities to be computed for a variety of individual isotopologues for species of interest. 
\end{enumerate}

\section{Collision induced absorption (CIA) and continuum data}\label{sec:CIA}

\subsection{Collision induced absorption (CIA)}\label{sec:CIA_CIA}

Collision-Induced Absorption (CIA) arises as broad features in atmospheric spectra due to additional spectral absorption occurring during collisions between pairs of species, with the interactions inducing a dipole moment. Symmetric molecules, such as O$_2$, N$_2$, H$_2$ do not possess any permanent or vibrationally induced electric dipole and are thus microwave and IR inactive, although even if extremely weak, magnetic dipole and electric quadrupole moments can contribute to the absorption.  Nevertheless,  significant absorption is induced through the creation of a transient dipole via intermolecular interactions during their collisions. This process enables one to reveal the presence of nonpolar species in the atmosphere by detecting collision induced supramolecules signatures.
CIA in dense gases was discovered in 1949 by Crawford and Welsh in the infrared range \citep{49CrWeLo}. Almost immediately after this discovery, it was realized that CIA of electromagnetic radiation should be important in just about any cool and dense environment, including atmospheres of solar and extrasolar planets \citep{64Trafton,95MaQu},
cool white dwarfs \citep{77Shipman,78MoLi,95BeSaWe,99SaJa},
brown dwarfs \citep{06BuBuKi,09GeSaGo},
cool main sequence stars \citep{69Linsky,98HaSt}, 
and so-called first stars \citep{76Rees,04BrLa,09BrYoHe,04RiAb,10Loeb}.

The term ``collision-induced absorption'' is usually used to refer to any bimolecular absorption that is absent from the isolated molecules but reveals itself as a result of intermolecular interaction \citep{17ViMo}.
To understand this process theoretically, a detailed statistical analysis of the wide range of thermally exciting interacting molecules that make up molecular pairs is needed. Three categories of molecular pairs, all related to a range of the potential energy surface of the associated van der Waals complex, can be distinguished: true bound, metastable and free pairs. The simultaneous presence of bound–bound (dimer), bound–free, free–bound and free–free (collisional) transitions in collision-induced processes makes the spectral analysis and the theoretical treatment very complex.

For decades, CIA has been known to contribute significantly to the transmission of radiation in the Earth’s atmosphere, primarily via N$_2$-N$_2$, N$_2$-O$_2$, and O$_2$-O$_2$~\citep{66FaHo,81RiSmRu,82RiSmSe},
as well as to play a crucial role in the radiation budget of other planets \citep{85TiPo}. Different molecular pairs have been identified as major contributors to the opacity of planetary atmospheres, such as H$_2$-H$_2$, H$_2$-He
in gas giant planets~\citep{12RiGoRo}, O$_2$-CO$_2$ and CO$_2$-CO$_2$ in Venus’ atmosphere~\citep{97GrBo} and H$_2$-N$_2$, N$_2$-CH$_4$ in Titan’s~\citep{1993DPS....25.2512B}. In the Solar System, the H$_2$ dimer has been detected on Jupiter and Saturn from the comparison between satellite measurements and laboratory absorption spectra recorded by \cite{88McKellar}. Several ionic complexes were also identified for example in the E-ring of Saturn, where the successive solvation of sodium cations by water was proposed to assign the observed mass spectrum \citep{09PoKeSc}
H$_2$-H$_2$ and H$_2$-He collision-induced absorption (CIA) are considered as a thermometer to measure the thermal structure of the upper tropospheres of the gas giants~\citep{Conrath_98}. The opacity of Jupiter in the far infrared is dominated by the CIA of H$_2$–H$_2$ and H$_2$–He~\citep{64Trafton,84BoMoFr,12RiGoRo}, making them vital components of not only the solar system gas giant planets but also Jupiter-like exoplanetary atmospheres. The importance of CIA warming in exoplanets was highlighted in~\citep{Ramirez_17} in the context of rocky planets with H$_2$-dominated atmospheres. 
Furthermore, the impact of CIA on the radiative budget has been demonstrated in the atmospheres of Earth and Super-Earth mass exoplanets with with H$_2$-enriched atmospheres~\citep{11PiGa}.
Other molecular pairs, including CO$_2$ have been studied to assess the impact of CIA on the surface temperature and to draw conclusion on the climate and habitability of of early Mars~\citep{Turbet_19,turbet2020measurements,20GoRaCa}. These CIA signatures are expected in spectra of hot Jupiters \citep{24BeCrCu} too.  During their campaign of measurements, \cite{Turbet_19} showed the disagreement between experimental data and the theoretical \textit{ab initio} calculations of \cite{17WoKaLo}. Sufficient wavenumber coverage and appropriate temperature ranges for the data are necessary for the CIA effect to be included. When it comes to symmetric molecules, such as H$_2$ or O$_2$, the CIA contribution is the only way to quantify the density of these molecules, bearing in mind that their absorption coefficients depend on the square density~\citep[see, e.g.,][]{12RiGoRo}.

\subsection{Continuum}

Continuum absorption is a process, and therefore dataset, which is typically presented as distinct from CIA. There are two main contributions which make up continuum absorption data: 1) excess opacity due to far line wings (see Section~\ref{sec:lineshapes}), and 2) dimer absorption, i.e. the absorption due to molecules which collide and become loosely bonded; to date this is most commonly associated with water~\citep{11PtShVi,16ShCaMo}. 

There is potentially some community consensus lacking on how the continuum absorption and CIA processes are distinguished in experiment and calculations, and how they are treated numerically in radiative transfer codes. In some cases (see Section~\ref{sec:continuum_available}) continuum data contains some CIA absorption also, so care needs to be taken to not double count these contributions. We note the temperature coverage of continuum data is limited. 

Probably the most well used source of continuum data, particularly for terrestrial applications and Earth-like planets, is the water vapour continuum. The continuum is a source of opacity which should be included in addition to line-by-line molecular absorption of H$_2$O (and other species, where necessary and available). It is typically tailored to be used alongside a particular opacity source; e.g. the MT\_CKD continuum data~\citep{2012RSPTA.370.2520M,23MlCaMa} is designed to work alongside opacities computed from the HITRAN line list with a 25~cm$^{-1}$ Voigt line wing cut-off. This is due to the continuum data making up for the excess opacity due to far line wings.

\subsection{State of the art - Data availability}

\subsubsection{Collision induced absorption (CIA)}

The primary source for CIA data is the HITRANonline website\footnote{\url{https://hitran.org/cia}}, where coverage for the number of pairs of species is greatest. Another compilation of CIA data can be found via the Laboratoire de M{\'e}t{\'e}orologie Dynamique Generic GCM (LMDG) online repository\footnote{\url{https://web.lmd.jussieu.fr/~lmdz/planets/LMDZ.GENERIC/datagcm/continuum_data}}~\citep[see, e.g.,][]{21ChBlBe}, which can be found via the Exo-k website\footnote{\url{https://perso.astrophy.u-bordeaux.fr/~jleconte/exo_k-doc/where_to_find_data.html}}. For hot Jupiters, with a H$_2$- and He-dominated atmosphere, H$_2$-H$_2$, H$_2$-He, and He-He are the most important sources of CIA to be included. However for Earth-like or heavier atmospheres, a wider range of species will need to be considered. 

Table~\ref{tab:cia_HITRAN} summarises the current status of CIA data available in the HITRAN database. An overview of the transitions and pairs of species included at the time of the 2019 update of the HITRAN CIA parameters can be found in \cite{19KaGoVa}, with updates for the 2020 release of the HITRAN database outlined in \cite{2022JQSRT.27707949G}.
Table~\ref{tab:cia_HITRAN} shows the coverage and limitations of the current state of CIA data for the species where data exists and also the wavenumber regions and temperatures for which this data is applicable.
It can be seen that the coverage of CIA data for heavy molecules is seriously lacking, which could have an impact on the modelling efforts of super-Earth atmospheres. CIA data for H$_2$-H$_2$ and H$_2$-He were originally motivated by the Solar System gas giants~\citep{18FlGuOr}, but the wide increase in temperature coverage was due to the interest in higher temperature exoplanet atmospheres such as hot Jupiters~\citep{02Borysow}. 
Light molecular systems 
are generally less complicated to model and measure, which could also explain the more complete wavelength and temperature coverage of data for lighter species. Higher mass molecular systems with many degrees of freedom, on the other hand, are more challenging to understand. Classical molecular dynamics simulations were used for CH$_4$-CO$_2$~\citep{22FaBoTr} and compared to measurements~\citep{Turbet_19}. The agreement was good but the limitations of the model were highlighted.

\begin{table*}
   \caption{\label{tab:cia_HITRAN} Collision induced absorption (CIA) data sets available from HITRANonline with the wavenumber and temperature ranges for which they provide coverage, and the citations for the original sources. 
   }
    \centering
    \begin{tabular}{llllll}
    \hline
        System & $\nu_{\mathrm{range}}$/cm$^{-1 }$ & $\lambda_{\mathrm{range}}$/$\mu$m & $T_{\mathrm{min}}$/K & $T_{\mathrm{max}}$/K & Reference \\ \hline
        H$_2$–CH$_4$ (equilibrium) & 0~-~1,946 & $>$~5.14 & 40 & 400 & \cite{86BoFr}\\
        H$_2$–CH$_4$ (normal) & 0~-~1,946 & $>$~5.14  & 40 & 400 & \cite{86BoFr}\\ 
        H$_2$–H$_2$ & 20~-~10,000 & 1~-~500 & 200 & 3,000 & \cite{11AbFrLi,18FlGuOr}\\ 
        H$_2$–H & 100~-~11,000 & 0.9~-~100 & 1,000 & 2,500 & \cite{03GuFr}\\ 
        H$_2$–He & 20~-~20,000 & 0.5~-~500 & 200 & 9,900 & \cite{12AbFrLi} \\ 
        He–H & 50~-~11,000 & 0.9~-~200 & 1,500 & 10,000  & \cite{01GuFr}\\ 
        N$_2$–H$_2$ & 0~-~1,886 & $>$~5.3 & 40 & 400 & \cite{86BoFrb}\\ 
        N$_2$–He & 1~-~1,000 & 10~-~10,000 & 300 & 300 & \cite{72BaWe}\\ 
        N$_2$–N$_2$ & 0~-~5,000 & $>$~2 & 70 & 330 & \cite{15KaMiHu,05BaLaFr}\\ 
         &  &  &  &  & \cite{96LaSoWe,17HaBoTo}\\
        &  &  &  &  &  \cite{16SuWiVe,19ChFiLo}\\
        N$_2$–air & 1,850~-~5,000 & 2~-~5.4 & 228 & 330 & \cite{05BaLaFr,96LaSoWe} \\ 
        &  &  &  &  & \cite{17HaBoTo,93MeDoBo} \\
        N$_2$–H$_2$O & 1,930~-~2,830 & 3.5~-~5.2 & 250 & 350 & \cite{18HaBoTr} \\ 
        O$_2$–CO$_2$ & 12,600~-~13,839 & 0.7~-~0.8 & 296 & 296  & \cite{09VaTrHa}\\
        O$_2$–N$_2$ & 1,300~-~13,840 & 0.7~-~7.7 & 193 & 296 & \cite{05BaLaFr,96LaSoWe}\\ 
        &  &  &  &  & \cite{97ThMeDo,91OrTyNi}\\
        &  &  &  &  & \cite{93MeDoBo,99MaLuFr} \\
        &  &  &  &  & \cite{06TrBoHa,KarmanNatChem2018}\\
        O$_2$–O$_2$ & 1,150~-~29,800 & 0.3~-~8.7 & 193 & 293 & \cite{04BaLaFr,99MaLuFr}\\ 
        &  &  &  &  & \cite{97ThMeDo,91OrTyNi}\\
        &  &  &  &  & \cite{06TrBoHa,KarmanNatChem2018}\\
        &  &  &  &  & \cite{06TrBoHa,KarmanNatChem2018}\\
        &  &  &  &  & \cite{12SpZa,11SpKiFi}\\
        &  &  &  &  & \cite{13ThVo,KassiJGPRA2021}\\
        &  &  &  &  & \cite{23AdKaCa}\\
        O$_2$–air & 1,300~-~13,839 & 0.7~-~7.7 & 193 & 300 & \cite{99MaLuFr,12SpZa}\\ 
        &  &  &  &  & \cite{97ThMeDo,91OrTyNi}\\
        &  &  &  &  & \cite{06TrBoHa,KarmanNatChem2018}\\
        &  &  &  &  & \cite{17DrBeBr}\\
        CO$_2$–CO$_2$ & 1~-~3,250 & 3.1~-~10,000 & 200 & 298 & \cite{97GrBo,99BaVi}\\ 
        &  &  &  &  & \cite{18Baranov}\\
        CO$_2$–H$_2$ & 0~-~2,000 & $>$~5 & 200 & 350 & \cite{17WoKaLo}\\ 
        CO$_2$–He & 0~-~1,000 & $>$~10 & 300 & 300 & \cite{72BaWe}\\ 
        CO$_2$–CH$_4$ & 1~-~2,000 & 5~-~10,000 & 200 & 350 & \cite{17WoKaLo}\\
        CO$_2$–Ar & 0~-~300  & $>$~33 & 200 & 400 & \cite{turbet2020measurements}\\ 
        &  &  &  &  & \cite{21OdSeBa}\\
        CH$_4$–He & 1~-~1,000 & $>$~10 & 40 & 350 & \cite{88TaBoFr}\\ 
        CH$_4$–Ar & 1~-~697  & 14.3~-~10,000 & 70 & 296 &
        \cite{97SaNaBo}\\
        CH$_4$–N$_2$ & 0~-~1379 & $>$~7.3 & 40 &  400 & \cite{finenko2021fitting,finenko2022trajectory}\\ 
        CH$_4$–CH$_4$ & 0~-~990 & $>$~10.1 & 200 &  800 & \cite{87BoFr}\\
        \hline
    \end{tabular}
\end{table*}

The accurate spectroscopic parameters obtained from the comb-referenced spectroscopy measurements of O$_2$ by \cite{FleurbaeyJQSRT2021} in Table~\ref{tab:comb_ref}, combined with other CRDS measurements~\citep{MondelainJGRA2019,KassiJGPRA2021}, were recently utilised~\citep{23AdKaCa} for fitting a parameterised representation of the theoretical calculations by \cite{KarmanNatChem2018} for O$_2$-O$_2$ and O$_2$-N$_2$ collision-induced absorption.

\subsubsection{Continuum}\label{sec:continuum_available}

There are various databases and literature which provide continuum data, such as the MT\_CKD (Mlawer-Tobin-Clough-Kneizys-Davies) database\footnote{\url{https://github.com/AER-RC/MT_CKD}}~\citep{2012RSPTA.370.2520M,19MlTuPa}. Both self (due to the interaction of the species with itself) and foreign (due to interaction with the species with another species in air) continuum data are available. The MT\_CKD water vapor continuum was recently made available through HITRANonline~\citep{23MlCaMa}. H$_2$O continuum data is also available from the CAVIAR project~\citep{11PtMcSh,16ShCaMo}. \cite{22AnChEl} performed a recent analyses of the impact of continuum data in exoplanet atmospheres and showed that its inclusion led to potential observable shifts in the transit depth in the region of 60 ppm for planets with water-rich atmospheres. Original sources of continuum data measured in the laboratory include works such as \cite{08BaLaMa,20OdTrSi}. As well as H$_2$O, the MT\_CKD continuum model also includes contributions from the so-called foreign continuum involving carbon dioxide, ozone, nitrogen and oxygen~\citep{2012RSPTA.370.2520M}. It includes a number of collision-induced continuum absorption bands, so care needs to be taken not to double count when including both continuum and collision induced absorption (CIA) data~\citep{2012RSPTA.370.2520M}. 

Although we do not go into detail here, we note that other continuum opacities, such as H$^-$, can be important for characterising hot gaseous exoplanet atmospheres. For example, opacity data was computed and implemented into retrieval codes used in the observational study of \cite{20LeWaMa.UV} using data from \cite{88John}. Retrieval analysis found evidence for significant amounts of H$^-$ in the atmosphere of hot Jupiter HAT-P-41b, thought to be produced through a combination of photochemical and collisional processes~\citep{20LeWaMa.UV}. Evidence for H$^-$ was also found by \cite{22MiSiBa} in the atmosphere of ultrahot gas giant exoplanet WASP-121b and by \cite{18ArDeLi} for very hot gas giant exoplanet WASP-18b.

\subsection{Data used by retrieval codes}

\subsubsection{Collision induced absorption (CIA)}

Collision induced absorption data is readily usable in retrieval codes such as ARCiS, TauREx3, petitRADTRANS, and others mentioned in Table~\ref{tab:retrieval_codes}. The HITRAN formatted data (.cia) provides a basis for including CIA data\footnote{\url{https://hitran.org/data/CIA/CIA_Readme.pdf}}~\citep{19KaGoVa}. As noted above, CIA including H$_{2}$ and He is commonly included in the atmospheric modelling of hydrogen and helium rich atmospheres and the corresponding data is distributed with the default data sets for retrieval codes such as TauREx3 and ARCiS. Additional species can be included in the modelling and retrieval codes as they become available or are updated, following the HITRAN format to maximise interoperability. Though data format does not pose a major limitation, restricted wavelength range coverage for the CIA data poses constraints on modelling where CIA between additional species is considered.

The HITRAN CIA format supports different sets of absorption data for the same pair of species but for varying temperatures. While this is provided in some data sets, it relies on implementation in the retrieval code or downstream processing library in order to query based on temperature. This is not always supported. 

\subsubsection{Continuum}

\cite{22AnChEl} modelled the effects of including vs not including continuum data for H$_2$O on examples of low-mass temperate exoplanet transmission spectra using TauREx3~\citep{21AlChWa}. The effect of continuum cross-sections exhibit an inverse relationship with temperature, so are most important for use in modelling atmospheres with temperatures at or close to Earth's. 

As previously mentioned, one aspect of using continuum models such as MT\_CKD is that they rely on specific line wings being used for computations of corresponding molecular cross-sections. For example, the MT\_CKD model assumes that a line wing cutoff of 25~cm$^{-1}$ is used, otherwise there can be double counting of the wing contribution~\cite[see, for example,][]{24GhBaCh}.

 \subsection{What's being worked on?}

\begin{figure*}
    \centering
      \includegraphics[width=0.49\linewidth]{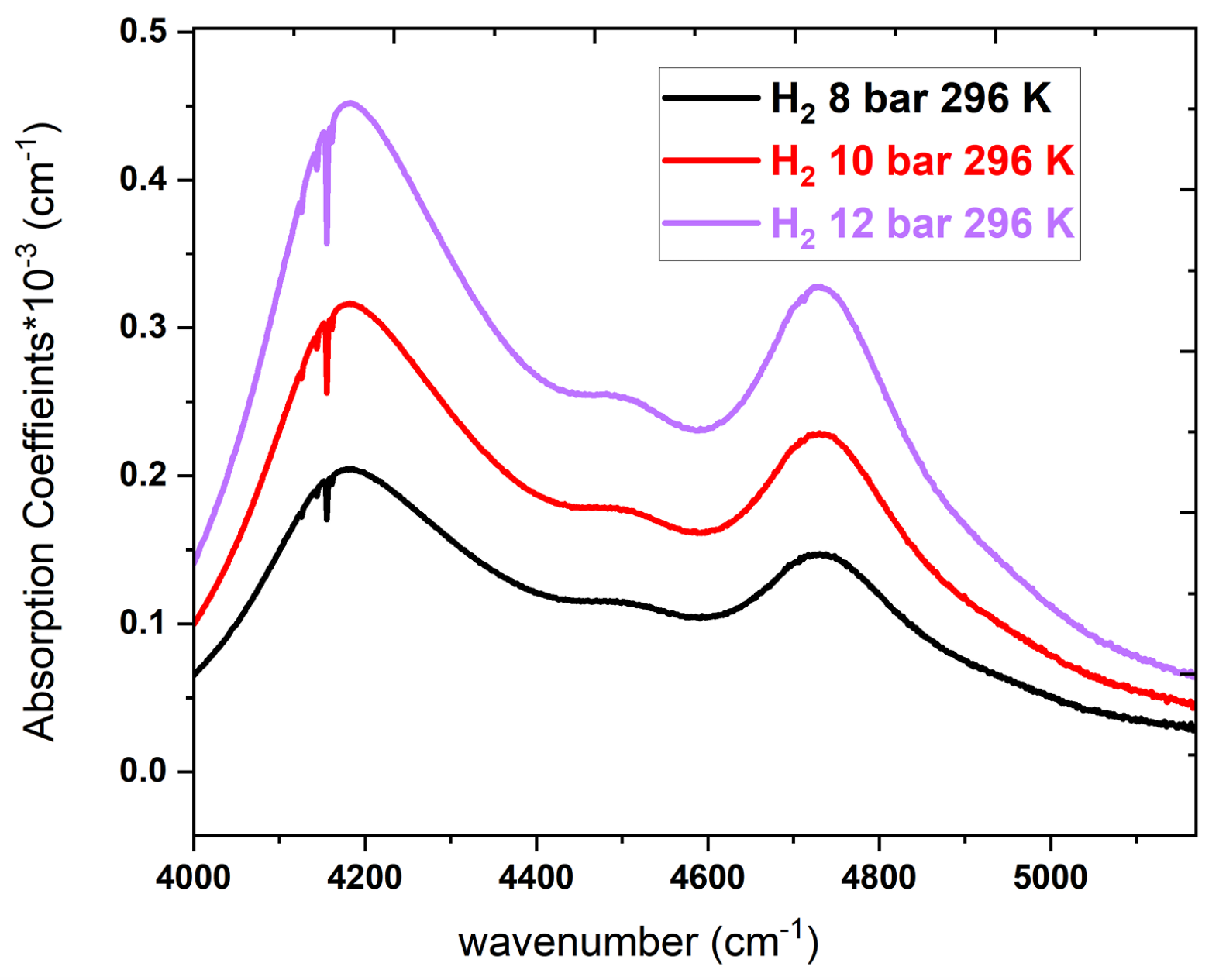}
\includegraphics[width=0.49\linewidth]{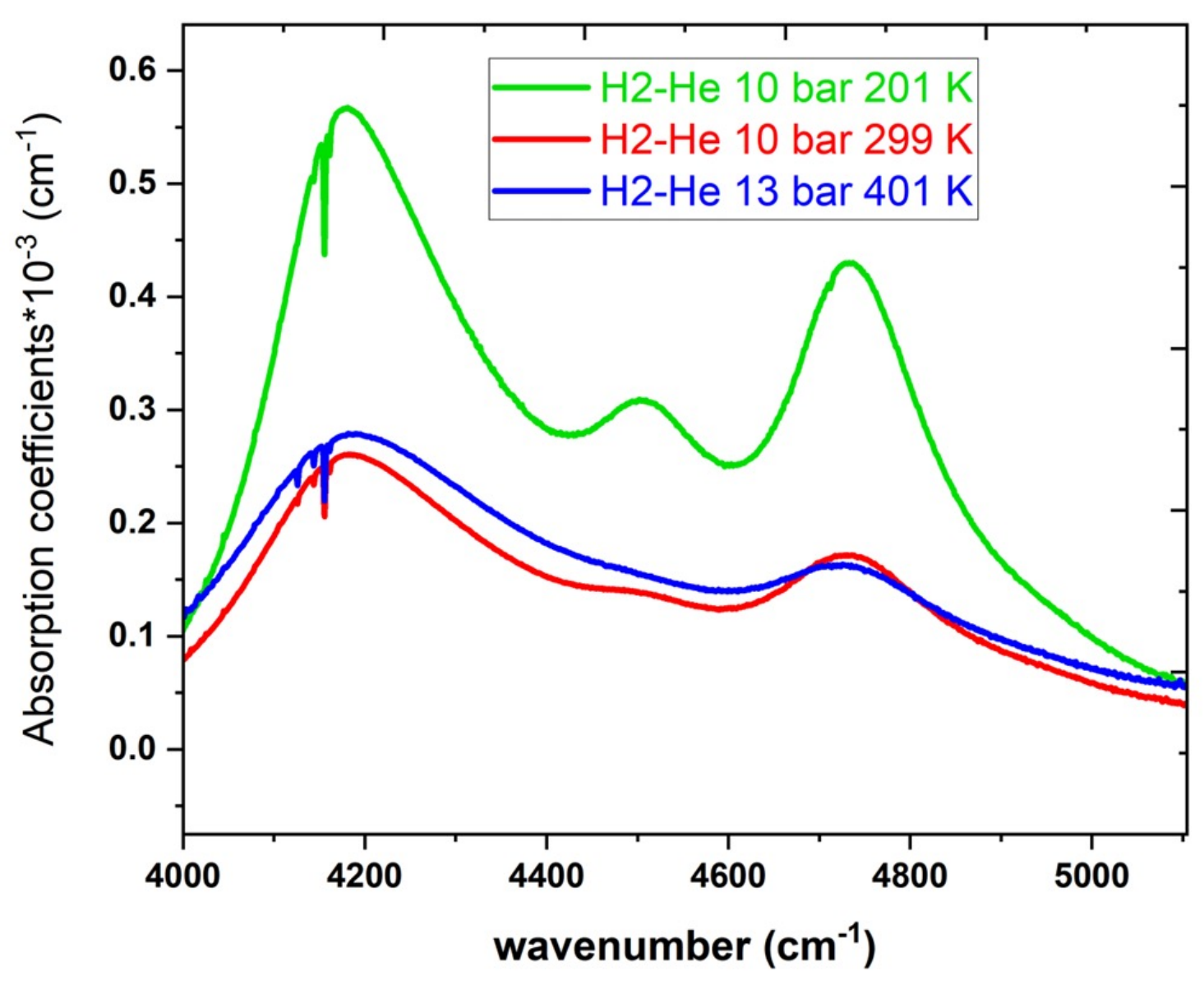}
    \caption{Left: H$_2$ CIA coefficients at different pressures (8, 10, 12~bars) and room temperature (296~K). Right: H$_2$-He CIA coefficients at different pressures (10, 13~bars)) and temperatures (201, 299, 401~K). \citep{24Vitali}}
    \label{fig:H2_lab}
\end{figure*}

Collision induced absorption (CIA) coefficients of H$_2$ and H$_2$-He have been recently measured in the 
3600-5500 cm$^{-1}$ ($\sim$1.8-2.8$\mu$m) spectral range  in order to characterise Jupiter's atmosphere \citep{24Vitali}. These are shown in Figure \ref{fig:H2_lab}, covering pressure values from 10~mbar to 70~bar and temperature values of 120~-~500~K. It is important to have up-to-date CIA measurements, even for H$_2$-H$_2$ and H$_2$-He where there is already data available (see Table~\ref{tab:cia_HITRAN}), either to expand the wavelength region or to compare the newer measurements to previous measurements or theoretical models present in the literature. As was highlighted in Section~\ref{sec:CIA_CIA}, the CIA of H$_2$–H$_2$ and H$_2$–He contribute significantly to the opacity of Jupiter, and therefore Jupiter-like exoplanets~\citep{64Trafton,84BoMoFr,12RiGoRo}.

\begin{figure}
    \centering
      \includegraphics[width=1.0\linewidth]{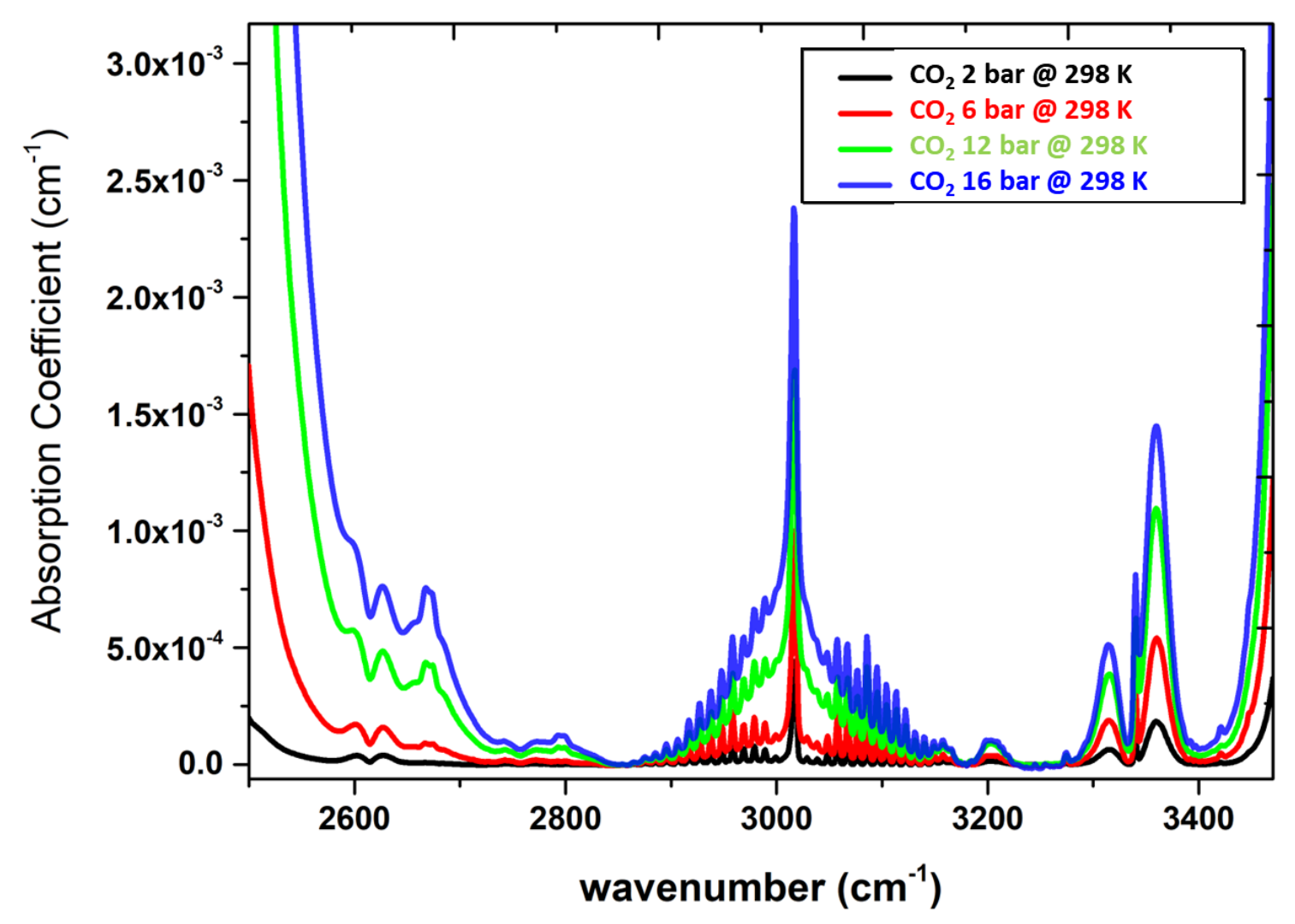}
    \caption{The absorption spectra of 16.03 amagat of carbon dioxide at 243 and 323 K, demonstrating a number of CIA bands, recorded with PASSxS, after subtraction of the strong contributions of the $\nu_3$ fundamental band at 2350 \cm and a weaker Fermi doublet ($\nu_3+\nu_1, \nu_3+2\nu_2$) around 3700 \cm, as well as contributions of the $^{16}$O$^{12}$C$^{18}$O isotopomer \citep{21SnStBo}.
    }
    \label{fig:INAF}
\end{figure}

CIA laboratory measurements have also been published recently from groups in Grenoble \citep{22FlMoFa}. There is a simulation chamber for absorption spectroscopy in planetary atmospheres in INAF \citep{21SnStBo}; see Fig.~\ref{fig:INAF}. In terms of species, molecular systems related to Venus and Early Mars are mostly being targeted, i.e. CO$_2$-H$_2$O and CO$_2$-CO$_2$.

Theoretical work is planned for molecular systems related to Jupiter-like planets.

 \subsection{Data needs: What's missing and urgent?}

 \subsubsection{Collision induced absorption (CIA)}

\begin{enumerate}
 \item \textbf{Data for additional CIA species:} CIA data is currently unavailable for many pairs of species that are thought to be important in exoplanetary atmospheres, for example N$_2$ is currently the only available species for collisions with H$_2$O (see Table~\ref{tab:cia_HITRAN}). H$_2$O, and other species such as CO$_2$ and CH$_4$ have been detected multiple times in exoplanet atmosphere spectra with JWST. CIA data for these molecules or combinations including them is crucial for a correct interpretation of the spectra especially in the search for biosignatures~\citep{23MaTeBa}.

  \item \textbf{Wavelength coverage:} much of the data that is available is limited to bands at wavelength ranges outside the window observed by Ariel (see Table~\ref{tab:cia_HITRAN} for an overview of current coverage of CIA data). The broad CIA features will, however, impact the shape of the Ariel observed spectra. In Jupiter's spectra, some CIA contributions are visible between 2 and 2.5 microns~\citep{18LoMoPa} which is lower in wavelength than the data available for many other pairs. Without taking CIA into account, it is possible that major uncertainties on the retrieved values will remain.
  \item \textbf{Temperature dependence:} data is often available only at room temperature, and the extension to higher temperatures is desirable, especially for modelling hot Jupiters and lava planets. We note some work has been done towards this, e.g. \cite{18StSnPi} for CO$_2$, along with data for H$_2$-H$_2$, H$_2$-H, H$_2$-He, He-H, detailed in Table~\ref{tab:cia_HITRAN}. 
  \item \textbf{Isotopologues:} CIA data coverage does not currently extend far in the consideration of isotopologues; generally the main isotopologue is taken for theoretically derived data, while the terrestrial ratios are used for experimentally derived data. The potential impact of a variation in isotopologues for CIA is small, meaning isotopologue data is less urgent than for the above mentioned pairs, though it is potentially relevant for the HD-H$_2$ pair \citep{19KaGoVa}.
\end{enumerate}

\subsubsection{Continuum}

\begin{enumerate}
    \item \textbf{Species other than H$_2$O:} although the MT\_CKD database contains some other species such as CO$_2$, O$_3$, N$_2$, and O$_2$, the main focus of continuum absorption has been on H$_2$O. (Exo)planets with atmospheres dominated by species other than H$_2$O have a particular need for continuum data.
     \item \textbf{Temperature dependence:} continuum absorption data for H$_2$O is currently only available below temperatures of 500~K, so there is not available data to use for modelling exoplanets with higher temperature atmospheres. We do note, however, that the impact of the H$_2$O continuum absorption on spectra decreases as the temperature increases~\citep{22AnChEl}, so this is not thought to be of urgent concern for the community.
\end{enumerate}

\section{Aerosols and Surfaces}\label{sec:aerosols}

It is well known that the types of cloud or haze particles, collectively known as aerosols, in the atmospheres of exoplanets will have an effect on their observed transmission or emission spectra~\citep{15WaSi,17PiMa,20MiOrCh.arcis,23MaItAl,23TaPa}, and their albedo~\citep{15GaIs,21FrMaSt,23GoMaLe,23ChStHe}, through the cloud particle's absorption and scattering properties. Generally, haze particles are smaller than cloud particles and are formed via photochemistry, which is a disequilibrium process with stellar UV leading to the formation of solid particles. Clouds, on the other hand, are formed via equilibrium processes which involve condensation and rainout.

\cite{21GaWaMo} give an overview of the state of knowledge (as of February 2021) of exoplanet aerosols as determined from observations, modeling, and laboratory experiments.
Wavelength-dependent, and ideally also temperature-dependent, refractive indices, also known as optical constants, are required to compute the scattering and absorption coefficients of clouds and hazes (aerosols) for use in exo-planetary atmosphere models. Thermochemical data are also required for predicting which aerosol species are expected to be present at various altitudes and horizontally across a given exoplanet atmosphere. Typically the outputs from global circulation models (GCMs), such as pressure-temperature profiles as a function of longitude and latitude across the planet~\citep[see, e.g.][for details on GCMs and complex atmospheric dynamics]{21LePaHa,21ChSkTh,22ScCaDe,22SkCh,23SkNaCh,24ChSkCh} are used as a basis for kinetic cloud models which then predict which particles will condense at different locations of the atmosphere~\citep[see, e.g.][]{01AcMa,08HeAcAl,20GaThLe,21RoKeRa,22RoBaGa,helling_exoplanet_2023}. Usually some assumptions need to be made, such as initial elemental abundances available for condensing, and the strength of vertical atmospheric mixing, for example.
Due to the importance of optical and UV data for understanding aerosol properties in exoplanet atmospheres, it has been suggested by \cite{24FaWaMa} and \cite{20WaSiSt} that one should complement JWST observations with one from HST.

Similarly, planetary surfaces can be modelled by a simple reflectance model (for instance using linear mixing)~\citep{Cui_Simultaneouscharacterizationof_EaPP2018}, but also by radiative transfer modelling using optical constants, similar to the approach for Solar System planets~\citep{Hapke_Book1993, Andrieu_Radiativetransfermodel_AO2015}. 
In both contexts, already existing databases contain relevant and useful sources but in many cases new measurements would be desirable (see Section~\ref{sec:aerosols_needed}).
Here we outline some typical sources of currently available data for these different applications.

\subsection{State of the art - Data availability}

\subsubsection{Refractive Indices}\label{sec:refindices}

Codes which compute the scattering and absorption properties of a particle or a surface require the refractive index to be known. The real and imaginary part of the refractive index relates to the scattering and absorption properties of the material, respectively. 
The refractive indices of aersols are complex numbers which are intrinsically dependent on the individual compositions of the aerosols~\citep[see, for example,][]{19BiFoBa}. 
If modelling an atmospheric layer consisting of different types of aerosol materials, or modelling mixed-material aerosol particles, then the refractive indices are often mixed using effective medium theory~\citep[see, for example,][]{16MDlLi}. 

As well as there being variation with wavelength, there is also some temperature dependence with refractive indices.
This has been demonstrated by \cite{15ZeMuPo} for olivine and enstatite up to 973~K and by \cite{13ZePoMu} for corundum, spinel, and $\alpha$-quartz up to 900~K. It can be seen by such studies that the vibrational band peaks shift slightly and their shapes change with temperature, which could be enough to lead to some incorrect conclusions when trying to fit models to observed spectra. The reason for these changes is that at higher temperatures, vibrational transitions will tend to occur between higher states with slightly lower energetic differences, and the states will be broadened by vibrational mode interactions~\citep{15ZeMuPo}. This causes broader and redshifted bands with increasing temperature; see, for example, Figures 2~-~7 of \cite{15ZeMuPo}.
We note that data for high temperatures are typically not currently available for many species, and so room temperature data are often used for characterising higher temperature planetary atmospheres, potentially causing biases in analyses.

Table~\ref{tab:ref_indices} gives the wavelength and temperature range of the available refractive indices of various materials from different sources. We note this table is not exhaustive, and it is useful to search in the various databases and compilations mentioned below for additional species of interest. Studies which predict which cloud or haze species are present in various exoplanet atmospheres \citep[see, for example,][]{01AcMa,08HeAcAl,20GaThLe,22RoBaGa,helling_exoplanet_2023} are useful as references for species to include in atmospheric models.

Databases which compile the optical properties of various materials include: 

\textbf{Heidelberg-Jena-St.Petersburg-Database of Optical Constants (HJPDOC)}\footnote{\url{https://www2.mpia-hd.mpg.de/HJPDOC/}}~\citep{99HeIIKr,03JaVlHe} and 
\textbf{Database of Optical Constants for Cosmic Dust}\footnote{\url{https://www.astro.uni-jena.de/Laboratory/OCDB/}} provide optical properties (refractive index, reflectivity, transmittance) of analog materials of cosmic dust in the wavelength range from the UV to far IR mainly produced in  Astrophysical Institute and University Observatory (AIU) Jena but also from open sources, including papers, reports, dissertations, where these data were derived. Users of the database are requested to cite the original sources of data.

\textbf{Aerosol Refractive Index Archive (ARIA)}\footnote{\url{http://eodg.atm.ox.ac.uk/ARIA/}} is a comprehensive collection of refractive index datasets of typical aerosol particles covering broad spectroscopic ranges. It contains both the original experimental and interpolated (mostly from very low to room temperature) values, intended to be used in light scattering parameters by the Oxford Mie Code.\footnote{\url{https://eodg.atm.ox.ac.uk/MIE/}}

\textbf{HITRAN2020 database}\footnote{\url{https://hitran.org/aerosols/}}~\citep{2022JQSRT.27707949G} 
provides refractive indices from visible to millimeter spectral ranges for many types of species including aerosol particles, required for 
 calculation of extinction, scattering, and absorptive properties of  atmospheric particles aiming at terrestrial and planetary (low temperature) applications. 

\textbf{Optical Constants database (OCdb)}\footnote{\url{https://ocdb.smce.nasa.gov}}~\citep[see, e.g.][]{24DrGaPe}  
provides optical constants of organic refractory materials (i.e. tholins, laboratory analogs of photochemical hazes) and ices relevant to planetary and astrophysical environments from  peer-reviewed sources. The data provided mostly correspond low temperatures. 

\textbf{Open-source  database of optical constants \tt{refractiveindex.info}}\footnote{\url{https://refractiveindex.info/}}~\citep{24Polyanskiy} is another recent resource of optical data from publicly available sources such as scientific publications and material datasheets published by manufacturers.

\textbf{Texas A\&M University dust 2020 database} (TAMUdust2020)~\citep{TAMUdust2020}  compiles the optical properties of irregularly shaped aerosol particles, such as dust aerosol and volcanic ash particles, for a wide range of values of the size parameter, the index of refraction, and the degree of sphericity. The data are produced using state-of-the-art light scattering calculations and validated with laboratory/in-situ measurements. 

Original sources of refractive index data include those of works such as \cite{95HeBeMu,03JaDoMu,95DoBeHe,11ZePoMu,24DrGaPe}. A number of species are also included in Palik's handbook of optical constants~\citep{12Palik}. 

Refractive indices as a function of wavelength for different species thought to be present in exoplanetary atmospheres have been compiled in various works, such as \cite{17LaKo,17KiHe,20MiOrCh.arcis,22LiWo}. \cite{17KiHe} provide a rather extensive compilation of wavelength-dependent refractive indices (real and imaginary components; see their Table~1 for species and sources) which are stored at the associated GitHub page for the LX-MIE mie-scattering tool\footnote{\url{https://github.com/exoclime/LX-MIE}}. Similarly, the wavelength-dependent refractive indices used in \cite{15WaSi} can be downloaded from the author's webpage\footnote{\url{https://stellarplanet.org/science/condensates/}}, although some updates have become available since this early compilation.

Anisotropic crystalline particles have refractive indices which depend on the direction of propagation of incident light with respect to the particle's axes of symmetry. The complex refractive indices related to light perpendicular to the main symmetry axis are often named ordinary, while the others are labelled extraordinary. These complex indices should typically be averaged over the three ($x$, $y$, $z$) axes of the particle for use in atmospheric codes. It is not always clear if available indices have already been mixed, but they may be labelled either ordinary/extraordinary, or by their $x$, $y$, $z$ axes. Where indices have been averaged is often noted in compilations such as \cite{17KiHe} (see their Table 1). 

Although the refractive indices data in these databases exist for very broad spectroscopic ranges, from microwave to far UV, for a given species the coverage is usually rather limited, see for example Table~\ref{tab:ref_indices}. A broad spectral range coverage of the optical properties is crucial to understand the role of aerosols in (exoplanet) atmospheric retrievals via radiative transfer models. In the context of the Ariel spectroscopic range, the IR data are especially urgent, but broader ranges are also required for atmospheric chemistry models involved in the retrievals. Due to the lack of the laboratory or in-situ measurements, it is common and especially important for exoplanet atmosphere applications to use interpolations to obtain continuous data \citep[see, for example,][]{17KiHe}.

\textbf{The Granada-Amsterdam light scattering database}\footnote{\url{https://www.iaa.csic.es/scattering/list/index.html}}~\citep{12MuMoGu} contains experimental scattering matrices as functions of the scattering angle of samples of small irregular particles. These are laboratory experiments typically performed at two optical wavelengths. Laboratory data such as these are important to be able to benchmark theoretical codes, particularly those which model the scattering matrices of irregularly shaped particles. Available codes for computing complex dust particle opacities, both for spherical and fractal shapes, include optool~\citep{21DoMiTa}\footnote{\url{https://github.com/cdominik/optool.git}}, the discrete-dipole-approximation code ADDA~\citep{11YuHo}\footnote{\url{https://github.com/adda-team/adda.git}}, the T-matrix codes~\citep{96MaMi}\footnote{\url{https://www.giss.nasa.gov/staff/mmishchenko/tmatrix/}} and CORAL~\citep{23LoWaLe}\footnote{\url{https://github.com/mglodge/CORAL.git}}. Other well-used codes include PyMieScatt\footnote{\url{https://pymiescatt.readthedocs.io/en/latest/}}~\citep{18SuHeCh}, which can deal with both direct and inverse Mie calculations.

\begin{table*}
\caption{Sources used for the real ($n$) and imaginary ($k$) parts of the refractive indices of a selection of species of various materials typically used in (exo)planetary atmosphere models of clouds and hazes (aerosols). The temperature range is left blank in cases where we assume the data is not higher than room temperature.}
	\label{tab:ref_indices}
 \centering
 \resizebox{\textwidth}{!}{
	\begin{tabular}{lllccl} 
		\hline
		\rule{0pt}{3ex}Material & Form & Name & Wavelength range ($\mu$m) & Temperature range (K) & Reference \\
		\hline
  \rule{0pt}{3ex}Al$_2$O$_3$  & Crystalline & Corundum & 0.03~-~10 &  & \cite{12Palik} \\
  \rule{0pt}{3ex}Al$_2$O$_3$  & Crystalline & Corundum & 6.7~-~10,000 &  300~-~928  & \cite{13ZePoMu}$^a$ \\
  \rule{0pt}{3ex}Al$_2$O$_3$  & Amorphous & Corundum & 0.2~-~12 &    & \cite{95KoKaYa}$^a$ \\
  \rule{0pt}{3ex}Al$_2$O$_3$  & Amorphous & Corundum & 7.8~-~500 &    & \cite{97BeDoHe}$^a$ \\
    \rule{0pt}{3ex}C$_x$H$_y$N$_z$  & Solid & Tholins & 0.02~-~920 &    & \cite{84KhSaAr}$^b$ \\
    \rule{0pt}{3ex}KCl & Crystalline & Potassium chloride & 0.22~-~167 &  & \cite{87QuWi}\\
  \rule{0pt}{3ex}Fe & Metallic & Iron & 0.1~-~100,000 &  & \cite{12Palik} \\
  \rule{0pt}{3ex}Fe$_2$O$_3$ & Solid & Hematite & 0.1~-~1000 & & \cite{05Triaud}$^d$\\
  \rule{0pt}{3ex}Fe$_2$SiO$_4$ & Crystalline & Fayalite & 0.4~-~10,000 & &  Unpublished$^a$\\
  \rule{0pt}{3ex}FeO &  Amorphous & Wustite &  10~-~510 & 10~-~300 & \cite{95HeBeMu}\\
  \rule{0pt}{3ex}FeS & Crystalline & Troilite  & 20~-~480  & 10~-~300 & \cite{94PoHoBe} \\
  \rule{0pt}{3ex}MgAl$_2$O$_4$ & Crystalline & Spinel &  6.7~-~10,000 &  300~-~928 & \cite{13ZePoMu}$^a$ \\
  \rule{0pt}{3ex}Mg$_2$SiO$_4$ & Amorphous & Forsterite & 0.2~-~950 &  & \cite{03JaDoMu}\\ 
  \rule{0pt}{3ex}Mg$_2$SiO$_4$ & Crystalline & Forsterite & 2~-~100 & 50~-~295 & \cite{06SuSoTa}\\ 
  \rule{0pt}{3ex}MgO & Cubic & Magnesium oxide & 0.2~-~625 &  & \cite{12Palik} \\
  \rule{0pt}{3ex}MgSiO$_3$ & Amorphous (glass) & Enstatite  & 0.2~-~500 &  & \cite{95DoBeHe}$^e$\\
  \rule{0pt}{3ex}MgSiO$_3$ & Amorphous (sol-gel) & Enstatite  & 0.2~-~1000 &  & \cite{03JaDoMu}$^e$\\
    \rule{0pt}{3ex}MgSiO$_3$ & Crystalline & Enstatite  & 2~-~98  &  & \cite{98JaMoDo}\\
  \rule{0pt}{3ex}SiO$_2$ & Crystalline & Quartz & 0.05~-~8.4  &  & \cite{12Palik} \\
  \rule{0pt}{3ex}SiO$_2$ & Crystalline & Quartz & 6.25~-~10,000  & 300~-~928 & \cite{13ZePoMu} \\
  \rule{0pt}{3ex}SiO & Amorphous & Silicon oxide & 0.05~-~100 &  & \cite{12Palik} \\
  \rule{0pt}{3ex}TiO$_2$ &  Crystalline & Rutile & 0.4~-~10 &  & \cite{11ZePoMu}\\
    \rule{0pt}{3ex}TiO$_2$ & Crystalline  & Anatase & 0.4~-~10 &  & \cite{11ZePoMu}\\
	\hline
	\end{tabular}
 }
	\rule{0pt}{0.2ex}
	\flushleft{	\flushleft{\textit{$^a$: Available via the Database of Optical Constants for Cosmic Dust\footnote{\url{https://www.astro.uni-jena.de/Laboratory/OCDB/crsilicates.html}}}} \\
 \textit{$^b$: Available via the HITRAN2020 database\footnote{\url{https://hitran.org/aerosols/}}~\citep{2022JQSRT.27707949G}}\\
\textit{$^c$: Available via the refractiveindex.info database of optical constants\footnote{\url{https://refractiveindex.info/}}~\citep{24Polyanskiy}}\\
	\textit{$^d$: Available via the 
			Aerosol Refractive Index Archive (ARIA)\footnote{\url{http://eodg.atm.ox.ac.uk/ARIA/}}}}\\
		\flushleft{\textit{$^e$: \cite{03JaDoMu} use the sol-gel method, and \cite{95DoBeHe} a melting and quenching technique; a comparison between the two is made in \cite{03JaDoMu}.}} \\
\end{table*}

\subsubsection{Thermochemical data}

Thermochemical data is required as input to kinetic cloud models of exoplanet atmospheres, for example in the computation of bulk growth processes for a particular species as a function of pressure and temperature, which requires gas-surface reactions~\citep{21Helling}.
There are several thermochemical properties of interest for astrophysical studies, including the Gibbs free energies of formation,
the change in enthalpy, or the entropy~\citep{24LeGoSi}. Chemical and cluster data for all the species that take part in the cloud formation process
are required for complete atmospheric models. Particularly, cluster data is crucial to model the nucleation process and the formation of cloud condensation nuclei (CCN). Databases, including the NIST-JANAF Thermochemical tables~\citep{98Chase}\footnote{\url{https://janaf.nist.gov/}} and the other databases listed in Section~\ref{sec:state_eq} contain thermochemical data but rarely cluster data.
As noted by \cite{21Helling}, laboratory data are often not available, and so thermochemical data for various species can also be computed. For example, thermochemical data for TiO$_2$ nanoclusters have been recently computed by \cite{22SiGoHe,22SiGoHe_data}, and for V$_x$O$_y$ clusters by \cite{24LeGoSi}. These data are appropriate for use in modelling the formation of cloud condensation nuclei in exoplanet atmospheres. 
A variety of other sources for thermochemical data exist. For example, for TiO$_2$ data is provided by \cite{JeoChaSed2000, LeeHeGi2015,22SiGoHe_data}, with \cite{JeoChaSed2000} also providing information about other intermediate Ti$_x$O$_y$ structures, including their vibrational frequencies. For SiO and SiO$_2$ thermochemical data are available from works such as \cite{BroGo2016, FlintFort2023}. For MgO, \cite{KohlGail1997} provide thermochemical data, and \cite{ChenFelm2014} provide cluster structures and vibrational spectra for (MgO)$_x$ (but not thermochemical data directly). Other species with thermochemical data include silicon carbide~\citep{GoCrPi2017}, 
KCl~\citep{RoBe1992}, silicates~\citep{GouBrom2012}, magnesium and calcium aluminates~\citep{GoSeCa2023}, aluminium oxides~\citep{GoPlaBro2022, GoDeCri2018, lam2015, PatChang2005}, and Fe-bearing species~\citep{ChangPat2013, DingWei2009}.

Condensation curves (or vapour pressure curves), which indicate the thermal stability of a species,
are necessary inputs into codes which predict the expected cloud materials present in an exoplanet's atmosphere based on it's pressure and temperature structure, typically found via global climate models (GCMs).
Condensation curves appropriate for modelling exoplanet atmospheres up to high temperatures are available from a number of sources, with those for some typical species thought to be important in exoplanet atmospheres given in Table~\ref{tab:cond_curves}. \cite{20GaThLe}\footnote{\url{https://github.com/natashabatalha/virga/blob/master/virga/pvaps.py}} give expressions for condensation curves from a number of different sources, including some of those mentioned in Table~\ref{tab:cond_curves}.

For cooler atmospheres, such as those in our solar system, ices are formed when gaseous species become supersaturated. The altitudes where the saturation occurs is a function of the mixing ratio of the species in question and the atmospheric temperature. Species dependent condensation curves will therefore vary with season and latitude, for example in the case of benzene in Titan, the latitudinal dependence is particularly evident in the fall at the poles where the decrease in stratospheric temperature is more dramatic and leads to an increase in the mole fractions. \cite{17Barth} extrapolated high-temperature vapor pressure data of benzene (C$_6$H$_6$) to lower temperatures to calculate benzene condensation curves. Their results predicted benzene and HCN to condense at near identical altitudes for the temperature profiles. However, \cite{21DuIrBa} measured the equilibrium vapor pressure of pure crystalline benzene (C$_6$H$_6$) at temperatures relevant to Titan. Their measurements indicated that the prior extrapolations from literature values underestimated the vapor pressure. Feeding their laboratory data into a microphysical model, \cite{21DuIrBa} showed that C$_6$H$_6$ polar cloud particles form first at higher altitudes and continue to exist at all latitudes down to the saturation point of HCN at slightly lower altitudes where co-condensation of C$_6$H$_6$ and HCN is possible. This is because the larger measured vapor pressure of C$_6$H$_6$ allows for larger ice particles that continue to grow deeper down in the troposphere, beyond the condensation point of HCN. It is therefore important for accurate condensation curve data to be available across a wide range of temperatures for a large number of species relevant to (exo)planetary and solar system atmospheres.

\begin{table*}
\caption{Example of sources used for vapour pressure/condensation curves of various materials typically thought to be important in exoplanet atmospheres, particularly those at higher temperatures such as hot Jupiters.}
 \centering
	\begin{tabular}{lll} 
		\hline
		\rule{0pt}{3ex}Material  & Reference & Note \\
		\hline
  \rule{0pt}{3ex}Al$_2$O$_3$  & \cite{16WaViLe} & Following the approach of \cite{10ViLoFe}  \\
    \rule{0pt}{3ex}CaTiO$_3$  & \cite{16WaViLe} \\
    \rule{0pt}{3ex}KCl &  \cite{12MoFoMa}\\
  \rule{0pt}{3ex}Fe & \cite{10ViLoFe}  \\
  \rule{0pt}{3ex}Fe$_2$O$_3$ & \cite{90ShHu} & Coefficients for a large number of condensed species are given in Table 2b \\
  \rule{0pt}{3ex}Fe$_2$SiO$_4$ & \cite{90ShHu} & \\
  \rule{0pt}{3ex}FeO & \cite{90ShHu} & \\
  \rule{0pt}{3ex}FeS & \cite{18WoHeHu.exo} & Equation inside GGchem source code \\
  \rule{0pt}{3ex}MgAl$_2$O$_4$ & \cite{18WoHeHu.exo} & Equation inside GGchem source code \\
  \rule{0pt}{3ex}Mg$_2$SiO$_4$ & \cite{10ViLoFe} \\ 
  \rule{0pt}{3ex}MgSiO$_3$ & \cite{10ViLoFe} \\
  \rule{0pt}{3ex}SiO$_2$ & \cite{23GrLeWa}\\
  \rule{0pt}{3ex}SiO & \cite{13GaWePu} & Older data from \cite{06NuFe,60Schick} \\
  \rule{0pt}{3ex}TiO$_2$ & \cite{04WoHe} &\cite{18PoZhGa} give this formula rewritten in different units\\
	\hline
	\end{tabular}
	\label{tab:cond_curves}
\end{table*}

Notable sources for condensation curve formulae which can be used in atmospheric models include \cite{06ViLoFe}, who give approximations for condensation curves of various sulphur- and phosphorus-bearing species, and \cite{10ViLoFe} for iron- silicon- and magnesium-bearing species. \cite{12MoFoMa} compute condensation curves for Cr, MnS, Na$_2$S, ZnS, and KCl, for use in characterising brown dwarf atmospheres, based on the thermochemical models of \cite{06ViLoFe,10ViLoFe}. \cite{16WaViLe} give expressions for equilibrium condensation temperatures as functions of pressure and metallicity for various Al- and Ti-bearing species. Appendix A of \cite{01AcMa} provides relations for saturation vapour pressures of a few species, such as NH$_3$, H$_2$O, Fe, and MgSiO$_3$ (enstatite). For NH$_3$ they fit measurements tabulated in the CRC Handbook of Chemistry and Physics~\citep{71Weast}. For H$_2$O they make use of data from \cite{81Buck}, and from \cite{76BaLe} for Fe and enstatite. 

Section 3.2 of \cite{18WoHeHu.exo}, along with their supplementary document~\citep{17WoHeTu}, give a detailed comparison of condensed phase data from either the NIST-JANAF database~\citep{98Chase} or the geophysical SUPCRTBL database\footnote{\url{https://models.earth.indiana.edu/supcrtbl.php}}~\citep{92JoOeHe,16ZiZhLu}. They offer comparisons of 121 species, and find good agreement for all species apart from anorthite, CaAl$_2$Si$_2$O$_8$. These data are used in \cite{18WoHeHu.exo} to compute vapour pressure curves, with an example for Fe shown in their Figure D.1.

Figure 4 of \cite{13GaWePu} illustrates the importance of using the most up-to-date measurements for vapour pressure curves, as they compare their experimentally determined data for SiO with data from \cite{06NuFe} and \cite{60Schick}. There is a considerable difference between the curves shown for SiO, with the newer measured data allowing SiO to be present in solid form up to higher temperatures than the older data. It is recommended that any users of thermochemical data such as condensation curves cite the sources of their data to allow for comparisons and for re-analysis in the case of newer data being produced. A thorough investigation into the reliability of the data used for condensation curves of species important for characterising (exo)planetary atmospheres such as those mentioned here, including additional newer measurements, would be a worthwhile endeavour, useful for assessing which species are expected in the atmospheres of current and future observations.

\subsubsection{Measured cloud and haze production in the laboratory}\label{sec:lab_cloud}

Atmospheric chamber experiments can be used to simulate the atmospheres of Solar System bodies, in order to study haze formation, and measure transmittance and reflectance spectra. This has been done for Titan~\citep{22HeHoRa,04ImKhEl}, Triton~\citep{22MoHoHe}, and early Earth~\citep{09DeTrPa,18HoHeUg}; in such studies many haze properties can be measured, such as production rate, bulk composition, molecular composition, and their transmission and reflectance from the optical to the near-infrared. The latter is done with Fourier Transform Infrared (FTIR) Spectroscopy. In the case of \cite{22HeHoRa}, they provide a set of optical constants of Titan haze analogs in the wavelength range 0.4~-~3.5~$\mu$m. Similar atmospheric chamber experiments have been used to study the formation of water ice clouds on Mars~\citep{11PhJoMa}, 
organic aerosols in anoxic and oxic atmospheres of Earth-like exoplanets~\citep{18GaCaVr}, precursors to haze formation in cool exoplanet atmospheres~\citep{19HeHoLe.exo}, and to explore haze production in super-Earth and mini-Neptune atmospheres~\citep{18HoHeLe}.

Several groups contribute to databases of refractive indices using different laboratory setups, such as COSmIC at NASA Ames~\citep{23ScRoRa}, PAMPRE at LATMOS~\citep{18GaCaVr,24DrGaPe}, and John Hopkins University~\citep{22HeHoRa}. Previous work on Titan haze analogs revealed that the setup conditions likely affect the analog's composition and thus their refractive indices~\citep[see, e.g.,][]{15BrMuCo}. Further laboratory experiments are important to fully understand the composition of different solid particles and their resulting optical properties.

\subsubsection{Surface reflectance}

For terrestrial exoplanets and exomoons, the bottom boundary layer of the atmosphere must be considered as the surface. Table \ref{tab:reflectance_database} summarises the most important databases of surface reflectance in Planetary Science. They encompass Earth natural rocks analogues, but also synthetic materials in the lab (such as ices), meteorites and in-situ samples measured back in Earth, such as the Moon or Ryugu asteroid. 
The reflectance spectrum of Phobos was measured from 0.4~-~4.75~$\mu$m by \cite{23WaGaPo}.
The dataset from in-situ measurements of reflectance in planetary science is usually scattered and unfortunately, such a database is missing.

Reflectance measurements can be done over several wavelength ranges, but also for different observation geometries (incidence and emergence angles), leading to BiDirectionnal Reflectance Distribution Function (BRDF). Laboratory measurements can also be done for various surface properties (such as grain size, grain shape, compaction, roughness) but also various kinds of mixtures (such as linear geographic mixture, intimate mixture, intra-grain mixture).

The databases which contain information on observation geometries and the nature of the samples do not always contain optical constants, as can be seen by Table~\ref{tab:reflectance_database}. Additionally some databases contain characterisation of the samples using other methods of investigations (Raman spectroscopy, petrological description, X-ray diffraction, Laser Induced Breakdown Spectroscopy, X-ray fluorescence or electron microprobe analysis). The SSHADE (Solid Spectroscopy Hosting Architecture of Databases and Expertise)~\citep{schmitt2018sshade} database regroups several databases. It provides spectral and photometric data obtained by various spectroscopic techniques including reflectance spectroscopy. The measured samples include ices, pure minerals, rocks, organic and carbonaceous materials as well as liquids. They are either synthetic, natural samples from Earth or other planetary bodies such as (micro-)meteorites or lunar soils. Additional characterisations of the samples are not systematically done but optical constants are often available and observation geometries are generally given. The USGS-Speclab \citep{kokaly2017usgs} database regroups spectra measured with laboratory, field, and airborne spectrometers, acquired on samples of specific minerals, plants, chemical compounds, and man-made materials. It also contains mathematically computed mixtures. 
The RELAB database regroups spectroscopic data acquired on pure minerals, natural and synthetic samples. 
The ECOSTRESS database \citep{ECOSTRESS} regroups data acquired on natural and man-made materials. 
The Mineral and Rock Sample Database University of Winnipeg  \citep{PSF} contains data acquired on pure minerals and natural or synthetic mixtures. 
Finally, the PTAL \citep[Planetary Terrestrial Analogue Library,][]{dypvik2021planetary} regroups data acquired on whole rocks. 
Additional characterisation are systematically done by multiple techniques.

\begin{table*}
    \centering
     \caption{Sources for the reflectance data of various materials typically used in (exo)planetary surface models.}
    \begin{tabular}{lccl}
    \hline 
\rule{0pt}{3ex} Name & Reflectance & Optical constants & Link\\
    \hline 
    SSHADE & X & X & \url{https://www.sshade.eu/}\\
    USGS-Speclab & X & &\url{https://crustal.usgs.gov/speclab/}\\
    RELAB & X &  &\url{https://sites.brown.edu/relab/relab-spectral-database/}\\
    ECOSTRESS & X &  &\url{https://speclib.jpl.nasa.gov}\\
    Mineral and Rock Sample  & \multirow{ 2}{*}{X} &  &\multirow{ 2}{*}{\url{https://www.uwinnipeg.ca/c-tape/sample-database.html}}\\
    Database University of Winnipeg &  &  &\\
    PTAL & X &  &\url{http://erica.uva.es/PTAL/}\\
\hline
\normalsize
    \end{tabular}
   
    \label{tab:reflectance_database}
\end{table*}

\subsection{Data used by retrieval codes}

There are a number of exoplanetary atmospheric modelling and retrieval codes which include the effects of solid-state aerosol species in their radiative transfer calculations, and thus require wavelength-dependent, and ideally also temperature-dependent, refractive index data~\citep[e.g.][]{15WaSi,17PiMa,20MiOrCh.arcis,23MaItAl,23TaPa}. The LRS instrument onboard JWST, covering $\sim$~5~-~12.5~$\mu$m,  allows for observations of vibrational-mode aerosol signatures which are commonly found in this wavelength region; for example quartz clouds were inferred in hot Jupiter WASP-17~b by \cite{23GrLeWa}~\citep[using optical constants of crystalline SiO$_2$ at 928~K from][]{13ZePoMu} and silicate clouds (MgSiO$_3$, SiO$_2$ and SiO) have been inferred in hot Neptune WASP-107~b by \cite{24DyMiDe}. 

Many retrieval codes have their own built-in library of refractive indices, the option to directly include pre-computed cloud opacities, or both. See for example Table 1 of \cite{20MiOrCh.arcis} for species included in atmospheric modelling and retrieval code ARCiS, with the option to read in your own refractive index data and use the code to compute either spherical or irregularly shaped particles~\citep{05MiHoKo} of a defined particle size distribution. petitRADTRANS~\citep{19MoWaBo.petitRT} also has a list of built-in cloud opacities\footnote{\url{https://petitradtrans.readthedocs.io/en/latest/content/available_opacities.html}}, with the option to add more. There are also available add-ons which can be coupled with existing retrieval codes. For example, \textit{YunMa} is an exoplanet cloud simulation and retrieval package~\citep{23MaItAl} which can be coupled to the TauREx3 platform to retrieve microphysical cloud properties in exoplanet atmospheres. 

Typically in cases of databases and compilations, the original citation for the measured or computed refractive indices is mentioned. It is important these citations are also included in any atmospheric modelling or retrieval works, as there may be future updates for particular species. Works such as \cite{15ZeMuPo} highlight the importance of this; they show temperature-dependence has an impact on the refractive indices and therefore the spectral features which are included in atmospheric models for enstatite and olivine. 

The refractive indices determined by using different techniques can differ from one another; for example in the case of amorphous MgSiO$_3$ (see Table~\ref{tab:ref_indices}) \cite{03JaDoMu} use the sol-gel method, and \cite{95DoBeHe} use melting and quenching technique; a comparison between the two is made in Figure 4 of \cite{03JaDoMu}.

Surfaces are often neglected but some retrievals propose to include surface reflectance measurements by linear mixing~\citep{Cui_Simultaneouscharacterizationof_EaPP2018}. More advanced radiative transfer modelling will be used, based on optical constants, for instance using the Hapke model~\citep{Hapke_Book1993}. Surface reflection is also an important component of studies such as \cite{08Stam,17FaRoSt,22TrSt}, who consider the reflection spectra of a planet's atmosphere and surface (if there is one), including polarisation state; these studies are particularly useful for modelling the observations of future instruments such as the Habitable Worlds Observatory (HWO)~\citep{23VaGeBo}.

\subsection{What's being worked on?}

Ongoing laboratory experiments~\citep[see, for example, setups described in works such as][]{20DePeCl,24DrGaPe,24HeRaMo} are essential for measuring new refractive indices, exploring photochemical pathways, and verifying theoretical calculations related to aerosols in planetary and exoplanet atmospheres. \cite{24HeRaMo}, for example, recently measured the density and optical properties of organic haze analogues generated in water-rich exoplanet atmosphere experiments, for use in characterising observations from JWST. \cite{21GaWaMo}, who give an overview of the state of knowledge (as of February 2021) of aerosols for exoplanet atmospheres, highlight the importance of ongoing laboratory experiments and development of theoretical methods. A recent review by \cite{24ScDrWo} further highlights the importance of measured optical constants of laboratory aerosols for interpreting JWST observations. 

As mentioned in Section~\ref{sec:lab_cloud}, there are several groups with different laboratory setups who are working on measuring the optical properties of different aerosol analogs. Collaborations are currently ongoing between LATMOS~\citep{18GaCaVr,24DrGaPe} and NASA Ames~\citep{23ScRoRa} to compare optical properties found with different laboratory setups. Such collaborations will provide more complete data sets and help better understand implications for future observations. Work is ongoing to combine different laboratory spectroscopic measurements in order to provide continuous data of refractive indices across a broad spectral range covering the ranges of JWST and Ariel; see, for example, \cite{24HeRaMo} and \cite{24DrGaPe} for current examples of this.

For condensation curve data, work is ongoing from groups such as those at NASA Ames, where three groups are working independently to address the different properties and optical constants required for accurate condensation curves.

\subsection{Data needs: What's missing and urgent?}\label{sec:aerosols_needed}

\begin{enumerate}
  \item \textbf{Wavelength coverage:} as shown in Table~\ref{tab:ref_indices}, the available data for a number of aerosols do not have the wavelength-coverage required to cover the observational range of Ariel (0.5~-~7.8~$\mu$m) and/or other telescopes related to exoplanet atmospheres such as JWST ($\sim$~0.6~-~28.5~$\mu$m, with the LRS instrument covering $\sim$~5~-~13~$\mu$m). We also note that refractive indices are important for modelling reflectance spectra of exoplanets and their atmospheres, including albedo in the visible region~\citep[see, for example,][]{23KrLePa,23ChStHe}. 
  
  \item \textbf{Temperature-dependence:} As mentioned above, the main spectral features of aerosols which could be observed in exoplanet atmospheres using JWST, for example, may be shifted at high-temperatures in comparison to room-temperature~\citep{13ZePoMu,15ZeMuPo}, making high-temperature measurements of refractive indices of high priority. \cite{13ZePoMu,15ZeMuPo} demonstrate the temperature-dependence of optical properties (i.e. refractive indices) of aerosol particles such as enstatite and olivine which are expected in typical hot gas giant exoplanets. \cite{15ZeMuPo} demonstrate the vibrational bands of olivine and enstatite becoming broader and redshifted with increasing temperature; see, for example, their Figures 2~-~7. It is assumed a similar effect will be found in other species, however there are no available studies on temperature dependence for the vast majority of refractive indices and most are therefore only available for terrestrial temperatures. 
Obtaining newer high temperature data for a wide range of other species important in higher temperature atmospheres would greatly aid in the analysis of observed spectra from missions such as JWST and Ariel. 
Reflectance spectra and refractive indices above 500~-~600~K are also important for surfaces, but at such high temperatures are usually not available.

  For ices, important for characterising cooler solar system bodies, three types of temperature effects are usually considered \citep{98ScQuTr}: (i) Phase transitions upon heating, irreversible for some cases, can lead to profound changes in spectral features. (2) Thermal annealing, also associated with irreversible reorganisation of amorphous to more ordered structure resulting in a progressive shift and narrowing of the bands. (3) Thermal contraction of the crystal and  phonon-phonon interactions can lead to a change of the spectroscopic bands with a complex temperature behaviour depending on many factors including the solid structure, phase etc, with the bandwidth generally increasing on heating. This change is reversible upon temperature. 

For cooler Solar system conditions, the temperature dependence of the refractive index have been studied, e.g. by \citet{08WaBr,00BiLuPe}.
The temperature dependence of the refractive index of H$_2$O has been investigated recently by \cite{22HeDiWa} between 30~K and 160~K (see their Figure 7). These results for water and other small molecules are available in the Leiden Ice Database for Astrochemistry - LIDA~\citep{22RoRaOl}. From what we know about ices, there is a clear temperature dependence which for H$_2$O results in a roughly 10\% increase from 30~K to 160~K~\citep{22HeDiWa}. 

Like for the majority of the data types considered in this work, more laboratory studies, especially at warmer conditions, are needed for refractive indices.   
   
  \item \textbf{A guide on what tools to use:} computing cloud opacities from refractive index data requires some additional assumptions, including the shape of the cloud particles, and whether the particles are mixed composition. For example, Mie computations are fast but only applicable for spherical particles, while the discrete dipole approximation (DDA)~\citep{94DrFl} or $T$-matrix method~\citep{96MaMi} are very accurate for non-spherical particles which can be composed of multiple materials, but also computationally expensive. There are some methods which lie between these extremes, allowing for the computation of scattering and absorption coefficients for non-spherical particle shapes, at a lower cost but typically also lower accuracy that DDA. For example the Distribution of Hollow Spheres (DHS) method~\citep{05MiHoKo} is thought to be a good compromise between accuracy and speed, with the option to choose a degree of irregularity of the particles. The DHS method has been used in studies such as \cite{24DyMiDe}. There is also the modified mean field theory (MMF) method, designed for computing opacities of dust aggregates~\citep{18TaTa}. Photochemical hazes are well described by mean field approximations, as the solid particles produced by photochemistry coagulate at relatively high pressures in the atmosphere and form fractal aggregates of spherical particles~\citep{97BoRaCa}. The Fractal Meanfield Scattering Code of \cite{97BoRaCa}, available on GitHub\footnote{\url{https://github.com/storyofthewolf/fractal_optics_coreshell.git}}, can be used to get the optical properties of fractal aggregates using refractive indices and fractal dimension as inputs.
  A recent study by \cite{23LoWaLe} investigates the differences which result in the use of the Mie, MMF, and DDA approaches, and demonstrates that using a lower resolution for DDA can also be a good compromise in terms of accuracy and speed. There has been a call amongst some users of atmospheric modelling and retrieval codes for some simple tools to create and read-in cloud opacities. There are two approaches here: either the code reads in refractive indices and particle sizes, and uses Mie theory or similar (DHS is available as part of the ARCiS~\citep{20MiOrCh.arcis} code, for example) to compute opacities on the fly; or pre-computed grids of opacities for various species, particle sizes, and shapes, can be read in. Codes such as YunMa~\citep{23MaItAl} assume spherical particle sizes. For surfaces, simple linear mixing of pure component spectra can be easily surpassed by more advanced radiative transfer models, such as Hapke's~\citep{Hapke_Book1993}. The effect of the surface texture (compacity, roughness, grain size) should be also taken into account if the actual data contains sufficient information to retrieve these parameters.
\item \textbf{Amorphous vs crystalline forms:} there is some discussion in the literature over whether crystalline or amorphous structures are expected under certain situations. 
It is thought that condensates are typically crystalline if their condensation timescale is faster than their annealing timescale, and amorphous if the opposite is true. This is because in the former case the condensates can rearrange quickly enough to form the crystalline structure, but if the condensation occurs much faster then the formed condensate will be amorphous~\citep{99GaSe,20HeWoHe}. \cite{23GrLeWa} demonstrate the impact of choosing either amorphous or crystalline structures for their refractive indices of solid-state SiO$_2$, with some resulting small differences between their retrieved spectra of WASP-17~b. It is therefore important for refractive indices for both crystalline and amorphous forms to be available for models. \cite{21Helling} calls for opacity data for condensates across large wavelengths ranges and crystalline species, along with opacity modelling for non-spherical and charged cloud particles.
\item \textbf{Additional species:} it has been noted by members of the community that sulphuric acid-based aerosols are of interest but are missing from the list of available aerosol refractive indices. 
\item \textbf{Condensation curves:} these are required for all species in order to correctly predict their presence in exoplanet atmospheres, over a large range of pressures and temperatures. As mentioned above, Figure 4 of \cite{13GaWePu} illustrates the considerable difference between condensation curves of SiO using different measured data, with the newer measurements allowing SiO to be present in solid form up to higher temperatures than the older ones. This highlights the importance of laboratory data for cloud modelling.
\item \textbf{In-situ measurements of reflectance:} as noted in this section, these data are available from different sources (see Table~\ref{tab:reflectance_database}), but not available in one easily accessible database.
\item \textbf{Cloud nucleation parameters:} constraints on the surface tension, contact angles and similar parameters are required for the description of cloud nucleation~\citep{24ArLa}. These are typically used even for non-molecular condensates, which are characterised by the absence of the corresponding condensing species in the gas phase e.g. MgSiO$_3$ and Na$_2$S~\citep{24ArLa}, thus a better understanding of the heterogeneous mechanisms at play for these cases is also needed. 
\item \textbf{Laboratory data to aid understanding of the coupling of haze and clouds:} it is becoming more evident with the latest observations that the coupling of haze and clouds is important~\citep[see, for example,][]{24ArLa}. To better understand this coupling, experiments that investigate the potential of haze particles acting as nucleation sites for the condensation of anticipated condensates in exoplanet atmospheres are needed.
\end{enumerate}

\section{Atmospheric Chemistry}\label{sec:chemistry}

Data required for atmospheric chemistry models appropriate for exoplanet atmospheres primarily include thermochemical data for equilibrium networks, and collision and reaction rates for kinetic networks. 

\subsubsection{Equilibrium chemistry}

There are a variety of codes designed to model chemical equilibrium in atmospheres, including exoplanet atmospheres. For example, ACE~\citep{12AgVeIr,20AgMaAn}, FastChem\footnote{\url{https://github.com/exoclime/FastChem}}~\citep{18StKiPa,22StKiPa}, and GGchem\footnote{\url{https://github.com/pw31/GGchem}}~\citep{18WoHeHu.exo} all work by minimising the Gibbs free energy of the system for a given pressure and temperature. The minimisation of Gibbs free energy requires thermochemical data as input.
Many codes make use of the algorithm of \cite{14GoMc} to compute the free energy minimisation. 

Here we give a little more context to what data are required for codes such as these. Each species is characterised by two sets of thermochemical coefficients, that can be in a 7-format or in a 9-format. The 9-format, being a more recent one, permits more accurate calculations. There is a set for the ``high-temperature'' range and another one for the ``low-temperature'' range.
For each of these ranges, thermochemical properties of each species in the Normal Conditions Enthalpy ($h^0$) and Entropy ($s^0$) can be calculated thanks to the NASA polynomials. For the 7-format, the expressions are:
\begin{equation}
    \frac{h^0(T)}{RT} = a_1 + \frac{a_2T}{2} +\frac{a_3T^2}{3} + \frac{a_4T^3}{4} + \frac{a_5T^4}{5} + \frac{a_6}{T}  
\end{equation}

\begin{equation}
    \frac{s^0(T)}{R} = a_1 \mathrm{ln}T  + a_2T +\frac{a_3T^2}{2} + \frac{a_4T^3}{3} + \frac{a_5T^4}{4} + a_7.
\end{equation}
The polynomials corresponding to the 9-format are slightly different: 
\begin{multline}
    \frac{h^0(T)}{RT} = -a_1 T^{-2} + a_2 T^{-1} \mathrm{ln}T + a_3 +  \frac{a_4T}{2} \\+  \frac{a_5T^2}{3} +   \frac{a_6T^3}{4} +  \frac{a_7T^4}{5} + \frac{b_1}{T}
\end{multline}

\begin{multline}
    \frac{s^0(T)}{R} = - \frac{a_1T^{-2}}{2}  - a_2 T^{-1} +  a_3 \mathrm{ln}T + a_4T \\ +   \frac{a_5T^2}{2} +  \frac{a_6T^3}{3} +  \frac{a_7T^4}{4} + b_2.
\end{multline}
With these two quantities calculated, the Gibbs Free Energy of the system can be calculated:

\begin{equation}
    G_{sys} = \sum_l^L (h^0_l(T) - Ts^0_l(T) + RT \mathrm{ln}\frac{P}{P^0} + RT \mathrm{ln}N_l) \times N_l
\end{equation}

The sets of $N_l$ that will minimise this equation (the Gibbs Free Energy of the system) will correspond to the composition at thermochemical equilibrium. 
Some databases which gather equilibrium thermodynamic coefficients for large sets of species are listed in Section~\ref{sec:state_eq}.

\subsubsection{Disequilibrium chemistry}

In addition to equilibrium chemistry, some codes include disequilibrium chemistry processes (i.e. quenching induced by vertical mixing, photodissociation, condensation and evaporation, surface deposition, outgassing and atmospheric escape) that control the atmospheric chemical composition \citep[i.e.][]{07Munoz,2011moses_diseq, 12VeHeAg, 20VeCaBo, 2016tremblin_atmo, 16RiHe,2017tsai,22AlChVe}.
These codes require data adapted to the type of exoplanets studied. For instance, for warm exoplanets, such as hot Jupiters, data which are validated/measured/calculated at high temperatures are necessary. The data are: chemical schemes, that is to say a list of hundreds to thousands of reactions, with the associated reaction rates, but also physico-chemical data for photodissociations (UV absorption cross-sections and quantum yields). Thermodynamic data are also necessary to calculate reverse reactions rates.

\subsection{State of the art - Data availability}

\subsubsection{Equilibrium chemistry}\label{sec:state_eq}

The C/H/O/N (CHON) chemical species included in ACE~\citep{12AgVeIr,20AgMaAn} are mainly sourced from \cite{14GoMc}, \citet{04AtBaCo,06AtBaCo,07AtBaCo} and \citet{10BoHeFo}, with some constants computed using THERGAS~\citep{95MuMiSc}. ACE is commonly used in conjunction with chemical kinetic models by \citet{12VeHeAg,20VeCaBo} and in the more recently re-written Python version FRECKLL \citep{22AlChVe}. FastChem~\citep{18StKiPa,22StKiPa} sources their thermochemical data from the NIST-JANAF database~\citep{98Chase}\footnote{\url{https://janaf.nist.gov/janbanr.html}}. A version of FastChem has recently been released which includes equilibrium condensation and rainout~\citep{KiStPa}, and comes with a python interface.
GGchem~\citep{18WoHeHu.exo} uses multiple sources of thermochemical data, with new species continuously added to the data files on the GitHub page. The sources for their chemical data comes from \citep{16BaCoxx}, NIST-JANAF \citep{98Chase} and \citet{73Tsuji}. Appendix C of \cite{18WoHeHu.exo} includes a comparison of the molecular equilibrium constants collected from various sources; the full comparison data can be found at \cite{17WoMiHu}. A graphical comparison of mineral Gibbs free Energy data can also be found at \cite{17WoHeTu}. Other sources of thermochemical data include the Burcat database~\citep{05BuRu}\footnote{\url{http://garfield.chem.elte.hu/Burcat/burcat.html}} and \cite{90ShHu}.  Recently, \citet{jt899} provided NASA polynomials for 464 molecules with an emphasis on those species which are important
for exoplanets.

\subsubsection{Disequilibrium chemistry}

Concerning chemical schemes, one major difficulty is to get data corresponding to the temperature of exoplanets. Due to long-standing studies of Solar System bodies, most of the chemical schemes developed for planetology are valid for low-temperature medium (e.g. \citet{Atreya1994, Moses2005,Moses2018, Dob2014,Dob2016}). However, in the combustion domain, many studies regarding chemical kinetics at high temperatures have been performed in the last few decades and can be relevant for modeling the atmospheres of exoplanets (e.g. \citet{cathonnet1982,keromnes2013, Burke2016}). Based on this fact,\cite{12VeHeAg, 15VeHeAg, 20VeCaBo} and \cite{2023veillet} have developed and updated reliable chemical schemes validated through combustion experiments for CHON species. These schemes have been validated on a large range of temperatures and pressures, typically 500--2500~K and 10$^{-6}$--100~bar, thanks to comparisons with many experimental datasets \citep{2023veillet}. This range of validation is probably adequate, as the chemical composition of the upper part of the atmosphere, which might in certain cases be at lower temperatures than 500 K, is mainly governed by photolysis processes and/or vertical mixing-induced quenching.
Even if those have not been validated with the same methodology as the aforementioned ones, chemical schemes with sulphur have also been developed and presented in various studies, such as \cite{16ZaMaMo}, \cite{2021tsai}, and \cite{21HoRiSh}.
The network of \cite{2021tsai} was recently used to contextualise the photochemically produced SO$_2$ feature detection in the atmosphere of hot Jupiter WASP-39~b~\citep{23TsLePo.wasp39b}. Other disequilibrium chemical schemes tailored for exoplanets include the STAND network~\citep{16RiHe}, which includes effects of UV photochemistry, cosmic ray chemistry and lightning-driven chemistry, and the network of \cite{10MoViKe} which uses thermochemical/photochemical kinetics and transport models. ExoREL~\citep{19Hu,21Hu} is a code to compute disk-averaged reflected-light spectra at any planetary phase, taking into account the radiative properties of clouds and their feedback on  planetary albedo. 

 Databases such as the JPL Chemical Kinetics and Photochemical Data database\footnote{\url{https://jpldataeval.jpl.nasa.gov/index.html}}, the KInetic Database for Astrochemistry (KIDA) database\footnote{\url{https://kida.astrochem-tools.org/}}~\citep{12WaHeLo}, the NIST chemical kinetic database\footnote{\url{https://kinetics.nist.gov/kinetics/}}, and the UMIST Database for Astrochemistry (UDfA)\footnote{\url{http://udfa.ajmarkwick.net}}~\citep{UDfA} are all sources of kinetic and photochemical data. KIDA provides evaluations of reaction rates and uncertainties, and contains chemical schemes used in planetology. Concerning UV photochemical data, very few data exist at high temperatures. \cite{2013venotCO2, 2018venotCO2} studied experimentally the thermal dependency of carbon dioxide and its impact on atmospheric composition and  \citet{2023sf2a.conf..377F} 
 studied the one of \ce{C2H2}.  \citet{15GrFaCl.H2S} measured UV absorption cross-sections of sulphur-containing compounds at elevated temperatures. See Section~\ref{sec:UV} for details on UV photodissociation and photoabsorption cross-sections.

\subsection{Data used by retrieval codes}

\subsubsection{Equilibrium chemistry}

Many of the aforementioned equilibrium chemistry codes are included in various atmospheric modelling and retrieval codes, such as TauREx3~\citep{21AlChWa} where ACE is included as standard with the option to include others using the plugin feature of TauREx~3.1, and ARCiS~\citep{18OrMi.arcis,20MiOrCh.arcis}, where GGchem~\citep{18WoHeHu.exo} is fully integrated. FastChem~\citep{18StKiPa,22StKiPa} is the main chemical model used in the HELIOS radiative transfer code \citep{HELIOS,20KiHeOr}. Other codes such as petitRADTRANS~\citep{19MoWaBo.petitRT} and Exo-REM~\citep{15BaBeBo} include self-written chemical routines. For example, easyCHEM is a self-written Gibbs free energy minimizer which is included in petitRADTRANS and described in \cite{17MoBoBo}. \cite{22AlChVe} compares computed atmospheric spectra of a hot gas giant exoplanet using some of the available equilibrium chemistry codes plugged into TauREx3; ACE~\citep{12AgVeIr,20AgMaAn}, FastChem~\citep{18StKiPa}, GGchem~\citep{18WoHeHu.exo}, and also GGchem with condensation included. \cite{24YaPlBo} explore the reliability of 1D equilibrium chemistry models to predict the composition of 3D exo-atmospheres with simulated transmission spectra of Ariel and JWST observations. They use the ACE chemical scheme in the TauREx retrieval code. ATMO~\citep{ATMO,2016tremblin_atmo} is a radiative/convective equilibrium code which is coupled to the CHNO-based chemical network of \cite{12VeHeAg}.

\subsubsection{Disequilibrium chemistry}

In addition to equilibrium chemistry, some atmospheric retrieval codes explicitly parameterise the disequilibrium chemistry process of quenching. 
For example, \cite{21KaMi.arcis} outlines the disequilibrium chemistry scheme included in atmospheric retrieval and modelling code ARCiS~\citep{18OrMi.arcis,20MiOrCh.arcis}, which includes vertical mixing of CH$_4$, CO, H$_2$O, NH$_3$, N$_2$, and CO$_2$. ARCiS uses thermochemical data collated as part of GGchem~\citep[mentioned above;][]{18WoHeHu.exo}, along with some extra species sourced from \cite{05BuRu} and \cite{05RuPivo}.
 Recently, a step forward has been made with TauREx3 using the kinetic code FRECKLL (Full and Reduced Exoplanet Chemical Kinetics distiLLed) as a plugin \citep{22AlChVe}, in order to perform retrievals taking into account chemical kinetics. The kinetic network included in FRECKLL is based on the chemical scheme of \cite{20VeCaBo}. VULCAN is another open source chemical kinetics code for exoplanetary atmospheres, written in python~\citep{2017tsai}. Their paper describes the rate coefficients and thermodynamic data used in the network. ATMO~\citep{ATMO,2016tremblin_atmo}
 includes a kinetic chemical network, as well as the equilibrium chemistry network described above.
HyDRA~\citep{17GaMa} is a disequilibrium retrieval framework for thermal emission spectra of exoplanetary atmospheres.

\subsection{What's being worked on?}

A chemical scheme validated through combustion experiment coupling CHON species to sulphur species is currently in progress and should be published in 2024 (Veillet et al., in prep). There is currently work underway to include P and S in CHON chemical schemes, using ab initio methods. Sticking coefficients of gas phase molecules onto solid atmospheric haze are currently in progress and should be published in 2024 (Perrin et al. in prep).  Several UV absorption cross sections (\ce{NH3}, HCN, \ce{SO2}) are currently under study at LISA with their own UV platform and will be published in the coming years (Collado et al. in prep).

\subsection{Data needs: What's missing and urgent?}

\begin{enumerate}
  \item \textbf{Rocky exoplanets:} data for rocky exoplanets require chemical networks that include many different species, which can be much more complex than what is usually the case for hydrogen-dominated atmospheres. These species include KOH, NaOH, SiO$_2$, CaOH, FeO, MgO, MgOH, as outlined in works such as \cite{17TeYuxxi} and \cite{20HeWoHe} 
    \item \textbf{Reaction rates for various different molecules:} \cite{21Helling} call for more thermochemical data, particularly reaction rates, for elements other than CHON(S). This has been pointed out by others, for example for phosphine~\citep{21BaPeSe}, which is applicable to Venus as well as exoplanet atmospheres; for chlorine-bearing species which would be helpful for Mars; and species such as TiO and VO for high-temperature exoplanet atmospheres. Other elements which ideally would be included in chemical schemes include Si, Mg, Na, K, Ca, and Al~\citep{2019astro2020T.146F}.
    \item \textbf{Reaction rates for ions:} ion chemistry may affect the deeper layers of the atmosphere through, for example, the formation of haze. The ionosphere is also relevant because it connects with the upper atmospheres and therefore with the long-term evolution of the planet's composition.

     \item \textbf{Reaction rate coefficients at a variety of pressures and temperatures:} Reaction rates must therefore be known at temperatures ranging from $\sim$~30~K to above 3000~K, and because the deep atmospheric layers are chemically mixed with the layers probed by spectroscopic observations, at pressures up to about $\sim$~100~bar.
     \item \textbf{Condensation:} some codes, such as GGchem~\citep{18WoHeHu.exo} and the latest version of FastChem~\citep{KiStPa}, already include condensation. Expanding more codes to explicitly include condensation and including additional elements in chemical networks will be useful to fully characterise where the elements of an atmosphere will go under chemical equilibrium conditions.
     \item \textbf{Gas-particle processes:} when aerosol particles are present in the atmosphere, gas-particle chemistry has been shown to influence not only the particles, but also the atmospheric content~\citep[][and Kn{\'i}{\v z}ek et al. in prep.]{23CaReMo,23PeKnLa}. For that purpose, sticking coefficients of gas components on solid particles are largely missing.  
\end{enumerate}

\section{UV photodissociation/photoabsorption data}\label{sec:UV}

The need for a description of molecular absorption in the ultraviolet (UV) region is motivated mainly by the following two reasons: 1) direct observations in the UV, which are not in the Ariel instrumental range but are of relevance to a number of existing, past, or potential future instruments; the WFC3/UVIS instrument on board the Hubble Space Telescope (HST) (200~-~1000~$\mu$m), the Galaxy Evolution Explorer (GALEX) (135~-~280~nm)~\citep{20ViNaMa}, the Far Ultraviolet Spectroscopic Explorer (FUSE) (90.5~–~119.5~nm)~\citep{04MoMcKr}, Cassini Ultra-violet Imaging Spectrograph (UVIS) (55.8-190 nm)~\citep{04EsBaCo}, and the Habitable Worlds Observatory (HWO) future mission concept; and 2) utilisation in atmospheric chemistry calculations for considering the chemical consequences of photodissociation, see Section~\ref{sec:chemistry}.  For example, improving the cross-section data of H$_2$O in the the vacuum ultraviolet (VUV, $\sim$100~-~200~nm) leads to significant changes in the outcome of photochemistry: for an abiotic habitable planet with an anoxic, CO$_2$–N$_2$ atmosphere orbiting a Sun-like star, it enhances the OH production which in turn suppresses a broad range of species in anoxic atmospheres \citep{20RaScHa}, including H$_2$ and CH$_4$. This is just one mechanism, clearly showing the sensitivity of the models to UV cross-sections and thus highlighting the critical need for laboratory data used photochemical models, whether from measurements or calculations. 
The impact of the strong variation of CO$_2$ absorption cross-sections with temperature on abundance and photodissociation rates of many species (CO$2$, CH$_4$, NH$_3$) was demonstrated by \citet{2018venotCO2}. 

The special but typical feature of molecular spectra in the UV/VUV wavelength region is their continuum. Their featureless character is due to the dissociative effects in the high energy/low wavelength region. It is this type of spectra produced by transitions to or from states above a dissociation limit and caused by the dissociation that we mainly discuss in this section. Photoabsorption includes both bound–bound  transitions (these appear as either sharp lines or continuum) and bound–free photodissociation (continuum) transitions, see, e.g., \citet{21PeYuTe,23TePeZh}. Bound-bound transitions which result in sharp lines are often included as part of a line list which extends into the UV region (see Section~\ref{sec:linelists}).
These types of spectra can be completely featureless for transitions to the continuum (from bound states to fully unbound, or dissociative states, e.g. HF above 100-160 nm \citep{22PeTeYu}) or mostly consisting of sharp lines (from bound to bound or quasi-bound states, e.g. MgH in the A-X band at 0.5~$\mu$m \citep{jt858}), or anything in between \citep{22PeTeYu}.  Both continuum absorption and photodissociation need to be represented as cross-sections rather than the lines of a typical line list, as they are continuum processes.

This creates challenges not only for the production of UV spectral data but also for its provision and format. For the featureless UV spectra, it is natural to provide data in the form of temperature dependent cross-sections. This is the format of the MPI-Mainz UV/VIS Spectral Atlas\footnote{\url{https://www.uv-vis-spectral-atlas-mainz.org/uvvis/}}~\citep{13KeMoSa}
and Leiden\footnote{\url{https://home.strw.leidenuniv.nl/~ewine/photo/index.html}}~\citep{Leiden, Leiden_2023} databases.
 In cases where the cross-sections are dominated by the continuum, the pressure effect is considered to be negligible, but it is noteworthy that in cross-sections dominated by resonances to Rydberg and valence states, high pressures can significantly affect the absorption cross-sections. Such effects have been observed in NO where transitions to Rydberg states were significantly broadened and shifted up until 1000~bar where the Rydberg states have all but vanished from the spectrum \citep{miladi_pressure_1975, miladi_pressure_1978}. However, most UV spectra comprise a mixture of the bound-bound transitions (sharper lines) and continuum.

\subsection{State of the art - Data availability}

One of the main sources of photodissociation cross-sections of molecules is the Leiden database\footnote{\url{https://home.strw.leidenuniv.nl/~ewine/photo/index.html}} which provides photodissociation and photoionisation cross-sections for 116 atoms and molecules of astrochemical interest collected from the literature, both experimental and theoretical \citep{Leiden, Leiden_2023}. The photodissociation thresholds and threshold products provided by this database are summarised in Table~\ref{tab:Leiden:cross}, and the provided photoionisation thresholds summarised in Table~\ref{tab:Leiden:ion} for molecules at low (interstellar/terrestrial) temperatures. The PHIDRATES website\footnote{\url{https://phidrates.space.swri.edu/}} provides a set of tools to calculate photoionisation, photodissociation, and photodissociative ionisation rate coefficients~\citep{92HuKeLy}. There is also the MPI-Mainz UV/VIS Spectral Atlas\footnote{\url{https://www.uv-vis-spectral-atlas-mainz.org/uvvis/}}~\citep{13KeMoSa}, where experimental temperature dependent cross-sections for a larger number of molecules are provided. We note, however, that there are some limitations in terms of temperature-dependence and wavelength coverage of the data in cross-section databases such as these, so it is worth checking the range of temperatures covered by the cross-sections for each species of interest.

\begin{table*}
 \caption{\label{tab:Leiden:cross}Photodissociation thresholds and threshold products in the Leiden database~\citep{Leiden,Leiden_2023}.}
    \centering
    \begin{tabular}{llcllcllc}
    \hline
           & Photodis.      & Thresh &          & Photodis.  & Thresh &         & Photodis & Thresh   \\
Species    & products       & Photodis & Species  & products   & Photodis & Species & products & Photodis   \\
\hline
AlH        & Al + H             & 392     & CO$_2$        & CO + O             & 227       & NH$_2$CHO   & NH$_3$ + CO   &      208          \\
C$_2$         & C + C         &    193 & CS         & C + S             & 168       & NH$_3$      & NH + H$_2$    &      301   \\
C$_2$H        & C$_2$ + H             & 253     & CS$_2$        & CS + S             & 278       & NO       & N + O      &      191   \\
C$_2$H$_2$       & C$_2$H + H          &    217 & H$_2$       & H + H      &      274       & NO$_2$      & NO + O     &      398         \\
C$_2$H$_3$        & H$_2$CC + H         &    1240 & H$_2^+$      & H + H$^+$     &      290 & NaCl    & Na + Cl  &      293      \\
C$_2$H$_4$       & C$_2$H$_2$ + H$_2$        &    258 & H$_2$CO     & H$_2$ + CO    &      361 & NaH     & Na + H   &      606   \\
C$_2$H$_5$       & C$_2$H$_4$ + H      &    260 & H$_2$CS     & H$_2$ + CS    &      313 & O$_2$      & O + O    &      242   \\
C$_2$H$_5$OH     & C$_2$H$_5$O + H      &    208 & H$_2$O      & H + OH     &      242 & O$_2^+$     & O + O$^+$   &      186   \\
C$_2$H$_6$       & C$_2$H$_5$ + H       &    287 & H$_2$O$_2$     & OH + OH    &      556 & O$_3$      & O$_2$ + O   &     1180   \\
C$_3$         & C$_2$ + C         &    268 & H$_2$S      & H + SH     &      318 & OCS     & CO + S   &      280   \\
C$_3$H$_3$     & C$_3$H$_2$+ H      &    300 & HC$_3$H     & l-C$_3$H + H  &      400 & OH      & O + H    &      279   \\
C$_3$H$_7$OH     & C$_3$H$_7$O + H          &    208 & HC$_3$N     & H + C$_3$N    &      244 & OH$^+$     & O$^+$ + H   &      247         \\
CH         & C + H         &    358 & HCN      & H + CN     &      243 & PH      & P + H    &      398   \\
CH$^+$        & C + H$^+$          &    303 & HCO      & H + CO     &     2037 & PH$^+$     & P$^+$ + H   &      369   \\
CH$_2$        & C + H        &    417 & HCO$^+$     & H + CO$^+$    &      145 & S$_2$      & S + S    &      283         \\
CH$_2^+$       & CH$^+$ + H        &    204 & HCOOH      & H + HCOO     &      333 & SH      & S + H    &      345         \\
CH$_3$        & CH + H$_2$      &    271 & HCl      & H + Cl     &      279 & SH$^+$     & S$^+$ + H   &      320   \\
CH$_3$CHO     & HCO + CH$_3$      &    342 & HCl$^+$     & H + Cl$^+$    &      267 & SO      & S + O    &      233   \\
CH$_3$OCH$_3$      & CH$_3$O + CH$_3$      &    196 & HF       & H + F      &    211.4 & SO2     & O + SO   &      219   \\
CH$_3$OCHO      & HCO + CH$_3$       &    248 & HNC        & H + CN         & 165       & SiH     & Si + H   &      417   \\
CH$_3$CN      & CH$_2$CN + H       &    308 & HNCO       & NH + CO         & 333       & SiH$^+$    & Si$^+$ + H  &      391   \\
CH$_3$NH$_2$     & CH$_3$NH + H              &    244 & HO$_2$      & H + O$_2$     &      476 & SiO     & Si + O   &      150         \\
CH$_3$OH      & CH$_3$O + H       &   280 & LiH      & Li + H     &      579       & c-C$_3$H   & C$_3$ + H   &      289   \\
CH$_3$SH      & CH$_3$S + H          &    331 & MgH      & Mg + H     &      623 & c-C$_3$H$_2$  & c-C$_3$H + H &      284  \\
CH$_4$        & CH$_3$ + H          &    277 & N$_2$       & N + N      &      127       & SiO     & Si + O   &      150   \\
CH$_4^+$       & CH$_3^+$ + H         &    1031 & N2O      & N$_2$ + O     &      735       & l-C$_3$H   & C$_3$ + H   &      379         \\
CN         & C + N         &    160 & NH       & N + H      &      362 & l-C$_3$H$_2$  & l-C$_3$H + H &      320         \\
CO         & C + O          &    110 & NH$^+$      & N$^+$ + H     &      281 & l-C$_4$    & C$_3$ + C   &      263   \\
CO$^+$        & C$^+$ + O         &    149 & NH$_2$      & NH + H     &      314 & l-C$_4$H   & C$_4$ + H   &      267   \\
--        & --             & --     & --      & --     &      -- & l-C$_5$H   & C$_5$ + H   &      348   \\
\hline

\end{tabular}
   
\end{table*}

\begin{table*}
 \caption{\label{tab:Leiden:ion}Ionisation thresholds in the Leiden database~\citep{Leiden, Leiden_2023}.}
    \centering
    \begin{tabular}{lclclclc}
    \hline
           & Thresh &           & Thresh &           & Thresh &          & Thresh   \\
Species    & Photoion & Species & Photoion & Species & Photoion & Species & Photoion\\
\hline
Al         & 207     & CH$_3$OH       & 113       & HCO      & 152  &      O$_2$      & 103       \\
AlH         & 156     & CH$_3$SH       & 131       & HCOOH      & 110    &    O$_3$      & 99       \\
C         & 110     & CH$_4$       & 98       & HCl      & 97    &    OCS     & 111       \\
C$_2$         & 102     & CN       & 87       & HF      & 77.5   &     OH      & 95       \\
C$_2$H         & 107     & CO       & 88       & HNC      & 103     &   P     & 118       \\
C$_2$H$^-$         & 412     & CO$_2$       & 90       & HNCO      & 108  &      PH      & 122       \\
C$_2$H$_2$         & 109     & CS       & 110       & HO$_2$      & 109    &   Rb         & 297       \\
C$_2$H$_3$         & 147     & CS$_2$       & 123       & K      & 286   &      S         & 120       \\
C$_2$H$_4$         & 118     & Ca       & 203       & Li      & 230       & S$_2$         & 133      \\
C$_2$H$_5$         & 153     & Ca$^+$       & 104      & LiH      & 152      &  SH         & 119        \\
C$_2$H$_5$OH         & 119     & Cl       & 96        & Mg      & 162   &     SO         & 120      \\
C$_2$H$_6$         & 108     & Co       & 158       & Mn      & 167      &  SO$_2$         & 100       \\
C$_3$         & 102     & Cr       & 183       & N      & 85      &  Si         & 152       \\
C$_3$H$_3$         & 143     & Fe       & 158       & N$_2$      & 79 &       SiH         & 156       \\
C$_3$H$_7$OH         & 123     & H       & 91       & N$_2$O      & 96   &     SiO         & 108       \\
C$_2$H$^-$         & 348     & H$^-$       & 1470       & NH      & 92    &     Ti         & 182       \\
C$_6$H$^-$         & 327     & H$_2$       & 82       & NH$_2$      & 111  &     Zn         & 132       \\
CH         & 117     & H$_2$CO       & 114       & NH$_2$CHO      & 121      &  c-C$_3$H         & 129       \\
CH$_2$         & 119     & H$_2$CS       & 135       & NH$_3$      & 122   &     c-C$_3$H$_2$         & 136       \\
CH$_3$         & 126     & H$_2$O       & 98       & NO      & 134    &    l-C$_3$H         & 136       \\
CH$_3$CHO         & 121     & H$_2$O$_2$       & 117       & NO$_2$      & 127  &      l-C$_3$H$_2$         & 119      \\
CH$_3$OCH$_3$         & 127     & H$_2$S       & 119       & Na      & 241 &       l-C$_4$         & 116       \\
CH$_3$OCHO         & 113     & HC$_3$H       & 139       & NaH      & 179   &    l-C$_4$H         & 129         \\
CH$_3$CN         & 102     & HC$_3$N       & 107       & Ni      & 162   &     l-C$_5$H         & 168      \\
CH$_3$NH$_2$         & 135     & HCN       & 91       & O      & 91 &       --         & --       \\
\hline
\end{tabular}
   
\end{table*}

In the past few years, there have been multiple VUV cross-sections (photoabsorption and photoionisation) measured at the VUV DESIRS beamline at the SOLEIL synchrotron facility in France~\citep{nahon_2012}. Absolute photoabsorption cross-sections have been measured using a Fourier-transform absorption cell~\citep{de_oliveira_high-resolution_2016} where multiple species have been measured such as OH~\citep{heays_high-resolution_2018}, CO~\citep{hakalla_vis_2016, lemaire_atlas_2018}, S$_2$~\citep{stark_fourier-transform-spectroscopic_2018}, SO~\citep{heays_ultraviolet_2023}, and others. Absolute photoionisation cross-sections of radical species have been measured with the double-imaging photoelectron photoion (i2PEPICO) instrument DELICIOUS3~\citep{garcia_delicious_2013} where the radicals were produced continuously in a stable manner with a flow tube~\citep{garcia_synchrotron-based_2015}. This has allowed the measurements of photoionisation cross setions of species such as OH~\citep{harper_quantifying_2019}, NH$_2$~\citep{harper_photoionization_2021}, SH~\citep{hrodmarsson_absolute_2019}, and others. In the frame of the EXACT project\footnote{\url{https://www.anr-exact.cnrs.fr/}} VUV absorption cross-sections are also measured experimentally. The first species studied was carbon dioxide~\citep{2013venotCO2, 2018venotCO2}.

Regarding computations of photoionisation or photoabsorption, the XCHEM code has recently been used to compute the photoionisation of H$_2$O~\citep{23FeBoGo}, and also the valence-shell photoionisation of CO$_2$~\citep{24PrGoBe}. Photoabsorption cross-sections for atmospheric volatile organic compounds (VOCs) have been computed by \cite{20PrIbMa,22PrMaHu}, using a combination of time-dependent and time-independent methods. \cite{23NdQuDa} explore the temperature dependence of the electronic absorption spectrum of NO$_2$ up to 2200~K, down to a wavelength of 0.25~$\mu$m (up to 40,000~cm$^{-1}$) by combining \textit{ab initio} methods with high-resolution experiments. \cite{20XuBaDr} utilise machine learning methods for the computation of absorption cross-sections in the visible and UV wavelength region, including benzene. Machine learning algorithms for the computation of electronically excited states are also utilised by \cite{21SrStSl} and \cite{23SrSl}. 

Collections of theoretical UV cross-sections of PAHs are available online\footnote{\url{ http://astrochemistry.ca.astro.it/database/pahs.html }}~\citep{07MaJoMu} and measurements of the cross-sections of cationic PAHs~\citep{16ZhRoJo, 20WeJoGi} have been used recently by \cite{22BeFoJa} to show the importance of up-to-date experimental measurements to understand the contribution of the ionisation of PAH cations to a variety of environments.

The ExoMol database has started providing photoabsorption and photodissociation data, see \citet{23TePeZh}. This will include UV line lists (bound-bound transitions only), UV photoabsorption cross-sections (bound-continuum only) and separately full temperature dependent photoabsorption cross-sections (all together). The data are produced with essentially the same computational ExoMol-methodology used for production of line lists, see Section~\ref{sec:ExoMol}, i.e. a mixture of first-principle quantum-mechanical calculations and empirical refinement, see \citet{21PeYuTe}. As in the case of the line lists, the advantage of theoretical spectra is in the completeness at higher temperatures.
ExoMol also provides hosting for experimental VUV/UV cross-sections of molecules, see \citet{jtVUV}. The recent AloHa AlH line list gives an example of the treatment of mixed spectral lines and continuum absorption in UV \citep{jt922} while \citet{22PeTeYu} gives an example of photodissociation cross-sections.

Absolute VUV branching ratios (sometimes known as quantum yields) of molecules are generally not very well known, in part due to the experimental difficulties of measuring them. Rotationally resolved absolute photodissociation branching ratios of small molecules like CO \citep{20RaScHa,20GaSoJa,13GaSoCh}, CO$_2$ \citep{14LuChGa}, N$_2$ \citep{16SoGaCh,17ShYiGa}, and others have been obtained using two tunable VUV laser systems. Other advancements have been in the use of VUV free electron lasers such as the one in the Dalian Light Source in China, where VUV photodissociation branching ratios have been measured for species such as H$_2$S \citep{20ZhZhHa}, CS$_2$, and OCS along with even more recent results of H$_2$O \citep{23ChAsYu,24WoBaZa}. 
There are more recent measurements in the context of exoplanet atmospheres that are noteworthy such as recent measurements of ethane (C$_2$H$_6$) \citep{20ChYaCh} considering its relevance to both Hycean atmospheres and Enceladus. As the sizes of molecules grow, the fragmentation patterns become more difficult to disentangle. In the case of C$_2$H$_6$, there are seventeen different fragmentation channels available at the Ly$\alpha$ limit, each other with different sets of fragmentation products which can be formed in various electronically excited states.

Regarding the electronically excited species, to a zeroth order approximation the electronic fluorescence proceeds faster than the collision frequency in the atmosphere in question. The validity of this approximation is dependent on the pressure and temperature of the relevant environments but typically non-dissociative fluorescence lifetimes of small molecules are of the order of one to a few microseconds. For environments where collision rates and these fluorescence rates are comparable, collisions with electronically excited species should not be neglected because they could lead to different photochemical paradigms that are currently neglected in most atmospheric models. For example, when H$_2$O is photodissociated from the D state, OH is formed in the electronically excited A state which has a fluorescent lifetime of 1 microsecond \citep{24LaKl}.

\subsection{What's being worked on?}

\begin{enumerate}
     \item ExoMol is working on producing temperature-dependent UV photoabsorption and photodissociation data for a variety of molecules including OH, SO and HCN.
\item Recently, acetylene has been measured \citep{2023sf2a.conf..377F} and more data are to  be published soon (Fleury et al. in prep), as part of the EXACT project\footnote{\url{https://www.anr-exact.cnrs.fr/}}, where UV absorption cross-sections are measured experimentally. The EXACT project aims to measure UV absorption cross-sections at high temperatures for many species, using their experimental setup at LISA.
\item VUV photoabsorption cross-sections and VUV Photoionisation cross-sections are currently being studied for a number of molecules at the DESIRS beamline at the SOLEIL synchrotron facility in France. Current efforts involve sulphur-containing species; photoabsorption cross-section of CS and C$_2$ will be measured and the photoionisation cross-sections of sulphur containing radicals such as CCS will be investigated.
\item PAHs: similar to the recent experimental work of \cite{24RoGaHr} on the cyanonaphthalenes, the photoionisation of a large sample of neutral PAHs of varying sizes and symmetries are being investigated (Hrodmarsson et al. in prep) to improve previous experimental data on PAH photoabsorption data from the 1990’s~\citep{90VeLeHe, 94ToLeJo,96JoBaLe,97JoRuBa,99JoBaLe} currently in use to describe the contribution of neutral PAHs to the gas heating in the ISM.
At the LISA laboratory, UV absorption cross sections for several molecules of atmospheric importance are currently under study. A future instrumental direction at LISA is to construct a double VUV laser system with a VMI spectrometer that will be dedicated to measuring VUV branching ratios of molecules of interest to exoplanet atmospheres. 

\end{enumerate}

\subsection{Data needs: What's missing and urgent?}

\begin{enumerate}
     \item \textbf{Photoabsorption cross-sections at high temperatures:} \cite{2019astro2020T.146F} (a collaborative Astro2020 Science White Paper) expressed the need for photoabsorption cross-sections measurements and their temperature dependency in the VUV wavelength range (115~-~230~nm) for molecules such as N$_2$, O$_2$, O$_3$, H$_2$O, CO, CO$_2$, CH$_4$, NH$_3$,
     TiO, VO, HCN, C$_2$H$_2$, H$_2$S, PH$_3$.
     \item \textbf{Branching ratios:} Despite a huge progress in laboratory provision of branching ratios, e.g. \citet{20buSaAb}, more data for atmospheric studies is needed~\citep{19JeVaAu,20GaSoJa}. 
     \item \textbf{Non-LTE:} most photodissociation cross sections come from the study of molecules in thermal equilibrium. However, there it is generally accepted that the upper atmospheres of most observable exoplanets are not thermal and so Non-LTE cross sections will be needed.

\end{enumerate}

\section{Data Standards}\label{sec:data-standards}

\subsection{Principles of and need for data standards}

Standards for the description of atomic and molecular data are crucial for their effective use in the study of exoplanet atmospheres. The successful adoption of a set of standards for the description of species, quantum states and spectroscopic processes facilitates interoperability between simulation codes, data exchange between researchers, the long-term preservation of data and automatic processing of data, for example through Application Programming Interfaces (APIs).

In any discussion about data standards a distinction must be made between metadata and data formats. All datasets are associated, implicitly or explicitly, with some set of conventions for describing metadata: the context, applicability, provenance, physical units and additional information which give meaning to the raw numbers of the data themselves. The format in which these numerical data are provided may vary between databases (depending, for example, on the size of the data sets). However, an interoperable ecosystem of databases must adopt standardised, unambiguous and machine-readable metadata conventions to allow comparison and aggregation across data sets.

Modern database services aspire to conform to so-called FAIR principles of data curation \citep{16FAIRDATA}; in brief, FAIR data are: 

\begin{itemize}
    \item Findable: associated with globally unique and persistent identifiers; registered in a searchable, online resource.
    \item Accessible: retrievable through a standardised protocol.
    \item Interoperable: associated with metadata in a formal, well-described representation.
    \item Reusable: clearly-licensed and documented, with detailed provenance and rich metadata attributes providing context to the data.
\end{itemize}

In the context of atomic and molecular data, the adoption of these principles can be realised as follows

\begin{itemize}
\item Atomic and molecular species and states should be specified using common identifiers.
\item Data sets, such as line lists and cross-sections, should be assigned globally unique identifiers and be locatable through these identifier using online services.
\item Where data sets are updated, older versions should be permanently available, identified and time-stamped.
\item All numerical data should be associated with clear metadata describing its format, physical units and limits of applicability (for example, valid temperature or pressure range).
\item Digital Object Identifiers (DOIs) should be attached to data sets to permanently identify the origin of the data; typically such DOIs will refer to the data's associated publication, such as a journal article.
\end{itemize}

\subsection{Focus on Interoperability: VAMDC and other services}\label{sec:VAMDC}

The Virtual Atomic and Molecular Data Centres (VAMDC) consortium \citep{wwwVAMDC,2020AtomsVAMDC} is a group of research institutes that share a common technical and political framework for the distribution and curation of atomic and molecular data. The goal of the consortium is to build a software infrastructure allowing the linking and searching of multiple databases through a common data format (``XSAMS'') and query language (``VSS2''). The VAMDC has built a platform (the ``portal'') for the simultaneous search of, at the time of writing, 38 databases and the aggregation of matching data sets. It now is concerned with further developing data standards, promoting collaboration in data production and evaluation, and in broadening its scope beyond spectroscopy and plasma physics. As machine learning matures and becomes more widely-adopted, the VAMDC is also exploring the application of artificial intelligence to its data services, to facilitate, for example, data transformation and visualisation and to assist with the validation and aggregation of data.

An up-to-date summary of the current status of the VAMDC and its database nodes can be found in the report of its most recent annual meeting, held as a Technical Meeting of the International Atomic Energy Agency in November 2023 \citep{23CCN-VAMDC}.

\subsection{What's being worked on?}

The PyValem software library \citep{wwwPyValem} provides a framework of Python routines for parsing standardised, text-based representations of atomic and molecular species and states; the syntax of these representations is largely adopted by the HITRAN and ExoMol databases, and is fully implemented in the CollisionDB database \citep{24CollisionDB, wwwCollisionDB} of plasma collisional processes. In addition to providing a common language for referring to species and states, this library can also perform a degree of validation on its input, for example by ensuring that quantum numbers are valid and reactions conserve stoichiometry and charge.

Several of the resources described in this paper expose an API for the automated querying and retrieval of atomic and molecular data; in addition to the VAMDC portal service outlined above, there are well-documented APIs for the HITRAN, ExoMol and CollisionDB databases; the software tools for interacting with these APIs are either already available or under active development.

The MAESTRO database, introduced in Section~\ref{sec:opacity_codes}, is under ongoing development. It aims to create one uniform and standardised database for opacities useful in the characterisation of exoplanet atmospheres, and to aid in the establishment of community standards for computing opacity data~\citep[see, e.g.,][]{24GhBaCh}.

\subsection{Data standards needs: What's missing and urgent?}

At the time of writing there are several competing standards for describing atomic and molecular species; whilst some of these are widely-adopted in some areas of chemistry and physics (for example, SMILES \citep{SMILES}, or InChI / InChIKey strings \citep{InChI-InChIKey}), there is no entirely satisfactory and easy-to-use standard for small molecules, particularly metal hydrides, in the context of astrophysical spectroscopy.

As described above, the line lists required of even relatively small molecules for spectroscopic modelling at elevated temperatures can become very large, with over $10^{10}$ individual transitions contributing the observable spectra. Therefore, database services supporting the asynchronous retrieval of large data files are needed. New software tools are required for the visualisation, validation, extraction and, potentially, reduction of such data sets. These tools should allow for the interconversion between popular data formats (XSAMS, HITRAN, ExoMol, and so on) and must be tested and maintained on a regular basis. In addition to improving usability, these tools can also perform the important function of validating the integrity of the data by checking for format consistency and data completeness.

A further issue is the complexity of the quantitative description of line broadening effects: this necessitates a flexible data model that can be applied to multiple databases; at present different approaches are taken by those databases that contain broadening data sets.

\section{Conclusion}\label{sec:conclusion}

In this white paper we have focused on a number of different data types: molecular and atomic line lists (Section~\ref{sec:linelists}), molecular and atomic line shapes 
(Section~\ref{sec:lineshapes}), computed cross-sections and opacities (Section~\ref{sec:opacities}), collision induced absorption (CIA) and other continuum data (Section~\ref{sec:CIA}), data for aerosols (Section~\ref{sec:aerosols}), data for atmospheric chemistry models (Section~\ref{sec:chemistry}), and data for UV photodissociation or photoabsorption (Section~\ref{sec:UV}). Needs for data standards, metadata, existing rules and associated tools were discussed in Section~\ref{sec:data-standards}. For each data type we have focused on: data availability, what data is typically used by atmospheric modelling and retrieval codes, what is missing and urgent, and what is currently being worked on. Of course the information provided in this paper is not comprehensive, but we have tried to give an overview for each data type for both data users and data providers. 

We also note that there are some important data topics relating to exoplanet atmospheres which we do not address in this particular paper. For example, accurate data regarding stellar spectra are important for photochemistry (see Section~\ref{sec:chemistry}) and UV photodissociation processes (see Section~\ref{sec:UV}). The secondary eclipse measurement of a transiting exoplanet gives information on the emission and reflection spectra. These measurements are presented as a ratio of planetary flux to stellar flux, and therefore also require an accurate stellar flux model.
We note there are some resources for theoretical models of stellar spectra such as the Kurucz-Castelli grids of models\footnote{\url{http://kurucz.harvard.edu/grids.html}}~\citep{02HeKuVe,04CaKu}, models from the SVO (Spanish Virtual Observatory) \footnote{\url{http://svo2.cab.inta-csic.es/theory/newov2/index.php}}, which provides data for 70 collections of theoretical spectra and observational templates, and the PHOENIX stellar models\footnote{\url{ https://phoenix.astro.physik.uni-goettingen.de/}}~\citep{13HuWeDr}. The Virtual Planet Laboratory regroups data from various publications\footnote{\url{ https://vpl.uw.edu/models/spectral-database-tools/}}, and there are additional sources of stellar spectra available from the MUSCLES survey of 11 low-mass, planet-hosting stars\footnote{\url{https://archive.stsci.edu/prepds/muscles/}}~\citep{16FrPaYo}.
We refer to works such as \cite{17GaMa}, \cite{20ShLaRe} and \citet{rzad009} for further information on the topic of stellar spectra for use in characterising exoplanet atmospheres.

Although the primary focus of the working group who initiated this white paper is for data needs appropriate for the Ariel space mission, we do discuss data for e.g. spectral regions which fall outside the observing window of Ariel. Such regions can be very useful for theoretical models which inform predictions for target planets for the Ariel mission, or for observations which are complementary to those planned by Ariel. Many of these data needs will be shared by all those in the exoplanet atmosphere community, not only those working directly on preparing for the Ariel space mission.

\subsection{Ariel-data GitHub platform}\label{sec:github}

The description of the data associated with this paper and the Ariel mission in general is available at GitHub via the project \url{https://github.com/Ariel-data}. The goal of this platform is to provide a go-to place both for the data-users and data-providers, for the users to highlight their data needs and make requests, and for the data providers to link to the available data. As an open access tool, GitHub provides huge advantages of forming direct dialogues between these sides of the community, where the modelers can make their shopping-lists for the data producers to start planning their calculations or experiments, even for those who are currently not directly involved in the Ariel consortium or in the field of exoplanetary science in general. 

We invite everyone to test and make use of this system.

\section{Data availability}

The data discussed in this paper will be available for access at GitHub via the project \url{https://github.com/Ariel-data}, which was created as a platform to bridge the two communities, data-providers and data-users. As an open platform, for the users, it provides an efficient way to communicate their needs, while the data providers can use it to make their work more visible. Any data-related GitHub ``issues'' raised for the attention of the Ariel Working-Group will be visible to the entire community of the lab data providers, from experimentalists to theoreticians.

\section{Acknowledgements} 

We thank the comprehensive efforts of the Reviewers for a number of constructive comments through the paper. The ExoMol project is supported by the European Research Council (ERC) under the European Union's Horizon 2020 research and innovation programme through Advance Grant numbers 883830 (ExoMolHD). \\
K.L.C was funded by UK Research and Innovation (UKRI) under the UK government’s Horizon Europe funding guarantee as part of an ERC Starter Grant [grant number EP/Y006313/1].\\
S.Y. acknowledges funding by STFC Projects No. ST/Y001508/1. \\
S.R. acknowledges funding by the Belgian Science Policy Office (BELSPO) through the FED-tWIN program (Prf-2019-077 - RT-MOLEXO) and through financial and contractual support coordinated by the ESA Prodex Office (PEA 4000137943, 4000128137).\\
A.F. acknowledges funding by the Knut and Alice Wallenberg Foundation (KAW 2020.0303) and the Swedish Research Council (2020-00238).\\
This research was supported by the Excellence Cluster ORIGINS which is funded by the Deutsche Forschungsgemeinschaft (DFG, German Research Foundation) under Germany's Excellence Strategy - EXC-2094 - 390783311.\\
A.V.N. and O.E. acknowledge the support from the Russian Scientific Foundation (RSF, No. 22-42-09022).\\
M.R. acknowledges support from the French ANR TEMMEX project (Grant 21-CE30-0053-01) and from the Romeo computer center of Reims Champagne-Ardenne.\\
F.S. acknowledges support from the ``Institut National des Sciences de l’Univers'' (INSU), the ``Centre National de la Recherche Scientifique'' (CNRS) and ``Centre National d’Etudes Spatiales'' (CNES) through the ``Programme National de Plan\'{e}tologie''.\\
T. Zi acknowledges NVIDIA Academic Hardware Grant Program for the use of the Titan V GPU card and the support by the CHEOPS ASI-INAF agreement n. 2019-29-HH.0 and the Italian MUR Departments of Excellence grant 2023-2027 ``Quantum Frontiers''.\\
Ch. H and H.L.M. acknowledge funding from the European Union H2020-MSCA-ITN-2019 under grant agreement no. 860470 (CHAMELEON).\\
O.V. acknowledges funding from the ANR project ‘EXACT’ (ANR- 21-CE49-0008-01) and  the Centre National d’\'{E}tudes Spatiales (CNES). This work was also supported by CNES, focused on Ariel, through `EXACT'.  \\
J.K.B. is supported by an STFC Ernest Rutherford Fellowship, grant number ST/T004479/1. \\
A.M. (Alessandra Migliorini) acknowledges funding from the Italian Space Agency (ASI) contract with the National Institute for Astrophysics (INAF) n. 2018-22-HH.0.1-2020.\\
I.S. acknowledges the support by DFG (DFG, German Research Foundation) – project No. SA 4483/1–1.\\
M.R. acknowledges the support by the DFG  priority program SPP 1992 ``Exploring the Diversity of Extrasolar Planets'' (DFG PR 36 24602/41).\\
A.K. (Anton\'{i}n Kn\'{i}\v{z}ek) acknowledges support from grant no. 24-12656K by the Czech Science Foundation and the ESA Prodex project under PEA 4000129979.\\
I.E.G and R.J.H acknowledge funding support through NASA grant 80NSSC23K1596 and NASA PDART grant 80NSSC24K0080.\\
Z.M. acknowledges, through Centro de Qu\'{i}mica Estrutural, the financial support of Funda\c{c}\~{a}o para a Ci\^{e}ncia e Tecnologia (FCT) for the projects UIDB/00100/2020 and UIDP/00100/2020, and through Institute of Molecular Sciences, the financial support of FCT for the project LA/P/0056/2020.\\
A.B. was supported by the Italian Space Agency (ASI) with \textit{Ariel} grant n. 2021.5.HH.0.\\
N.C. thanks the European Research Council for funding via the ERC OxyPlanets project (grant agreement No. 101053033). \\

\bibliographystyle{rasti}
\bibliography{journals_astro,linelists,exoplanets,CIA,Bibliography_LW,programs,ARIEL_database,jtj,exomol}

\end{document}